\documentclass{aa}  
\usepackage{changepage}
\usepackage{tabularx,multirow,booktabs,blindtext}
\usepackage{adjustbox}
\usepackage{graphicx}
\usepackage{subfig}
\usepackage{placeins} 
\usepackage{url}
\usepackage[titletoc,toc,title]{appendix}
\usepackage{titlesec}
\usepackage{chngcntr}
\usepackage{rotating}
\usepackage{txfonts}
\usepackage{graphicx,rotating}
\usepackage{amsmath} 
\usepackage{float}
\usepackage{makecell}
\usepackage{siunitx}
\usepackage{hyperref}
\usepackage{subfig}
\usepackage{txfonts}
\usepackage[T1]{fontenc}
\usepackage{epsfig}
\usepackage{natbib}
\usepackage{color}
\usepackage{amssymb}  
\usepackage{tabularx}
\usepackage{comment}
\usepackage{longtable}
\usepackage{ltablex}
\usepackage{caption, booktabs}

\DeclareGraphicsExtensions{.pdf,.png,.jpg}
\bibpunct{(}{)}{;}{a}{}{,}

\usepackage{tablefootnote}
\usepackage{threeparttable}

\begin{document}

\title{Galaxy clusters in the VIDEO fields: Detection and characterisation in the context of MOONRISE}

   \author{P. Galois{$^{1}$}\fnmsep\thanks{e-mail: pierre.galois@oca.eu} 
        \and C. Benoist{$^{1}$}
        \and G. Castignani{$^{2}$}
        \and M. Maturi{$^{3,4}$}
        \and S. Maurogordato{$^{1}$}
        \and P. Jablonka{$^{5,6}$}
        \and Y.~M.~Bahé{$^{5,7}$}
        \and D.~J. McLeod{$^{8}$}
        \and R.~A.~A. Bowler{$^{9}$}
        \and O. Cucciati{$^{2}$}
        \and L. Moscardini{$^{10,2,11}$}
        \and M. Magliocchetti{$^{12}$}
        \and M. Radovich{$^{7}$}
        \and M. Cirasuolo{$^{13}$}
        \and G. Covone{$^{14,15,16}$}
        \and F. Cullen{$^{8}$}
        \and C.~T. Donnan{$^{8}$}
        \and H. Flores{$^{6}$}
        \and W. Hartley{$^{17}$}
        \and R. Maiolino{$^{18,19,20}$}
        \and R.~J. McLure{$^{8}$}
        \and N. Napolitano{$^{14,21,22}$}
        \and M. Paolillo{$^{14,15,16}$}
        \and M. Vaccari{$^{23,24,25}$}
        }   
       
        \institute{\centering \textit{(Affiliations can be found after the references)}}
        \date{Received January 23, 2026; accepted August 5, 2026}

  \abstract{
    The identification and characterisation of galaxy clusters at different cosmic times are fundamental to constrain cosmological scenarios and to understand the joint evolution of structure formation and galaxies in dense environments. In this work, we analyse the cluster content of the $\sim 4.5 \text{ deg}^{2}$ XMM-LSS and CDFS VIDEO fields, which are expected to be partially covered by the upcoming MOONRISE survey. Using AMICO and WaZP photometric redshift-based cluster finders, we constructed a sample of $519$ cluster candidates detected by both finders in the redshift range $z = 0.1-3$, including $74$ detections at $z > 1.5$. 
    For all detections, we identifed the brightest central galaxy (BCG) and computed a list of probabilistic cluster memberships. Our photometric redshift measurements of the clusters agree well with spectroscopic redshifts from the literature, when available. From ancillary spectroscopic data, we assign $z_\text{spec}$ measurements to $116$ cluster candidates based on their spectroscopic members and to $204$ based on their likely BCGs. We also show that candidates containing radio-loud members were efficiently recovered using the prior-based cluster finder PPM. 
    We perform a preliminary analysis of the galaxy content of these candidates, focusing on the red-sequence components of their apparent colour-magnitude diagram. By comparing their mean colours with models of galaxy evolution, we show that this population is consistent with a model of passive evolution with a formation at high redshift, and is already in place at $z = 1.5-2.0$. Finally, our cluster sample was used to 
    evaluate how these clusters would be detected and characterised, according to various MOONRISE strategies. We show that cluster spectroscopic confirmation and characterisation could be efficiently achieved up to $z\sim1.7$ even with the shallowest survey strategy. This brings unprecedented insight into the physical properties of high-redshift galaxy clusters and into galaxy formation in dense environments.
    }
    

\titlerunning{Galaxy clusters in the VIDEO fields: detection and characterisation in the context of MOONRISE}
\authorrunning{P. Galois et al.}
    
\maketitle
\nolinenumbers

\section{Introduction}

\noindent Galaxy clusters are the most massive gravitationally bound and virialised structures in the Universe. They are excellent tracers of the matter density in the cosmic web, providing precious insight into cosmological models. Moreover, they are natural laboratories for studying how galaxies evolve within dense environments over cosmic times. Consequently, systematic studies of these structures over a wide redshift range provide valuable insight into the processes shaping structure formation and into how these processes have impacted galaxy evolution throughout cosmic history.

Studies at low redshift have demonstrated that galaxy cluster environments exert a strong influence on galaxy properties relative to field environments. Several mechanisms are suggested
to explain this effect: the hot intracluster medium strips infalling galaxies of their cool gas components as they enter the densest regions of clusters (ram-pressure stripping, \citealt{Gunn_1972}), repeated high-speed encounters disturb low-mass galaxies (\citealt{Li_2012}), while major merging within infalling groups shapes and pre-processes the galaxies (\citealt{Dressler_1997}, \citealt{van_Dokkum_1999}). As a result, galaxies in clusters tend to be redder and more quiescent, and, as opposed to the field, they present themselves as lenticulars and ellipticals.
The large fraction of passive galaxies present in cluster cores at low and intermediate redshifts populates a well-defined 'red-sequence' in the colour-magnitude diagram (\citealt{Lidman2004, Lidman2008}, \citealt{Lopez-Cruz_2004}, \citealt{Mei2009}).
The slope and scatter of this relation, as well as its evolution with redshift, have been widely used to constrain models of galaxy formation and evolution (\citealt{Bower1992}, \citealt{deLucia2007}, \citealt{Kodama2007}, \citealt{Stott_2009}, \citealt{Valentinuzzi_2011}).

However, at redshifts greater than one, consensus is not yet achieved. Several analyses show strong star formation in cluster cores (e.g. \citealt{Brodwin2013}, \citealt{Wang2016_T}).
In contrast, some clusters appear to be dominated by red and quiescent galaxies (e.g. \citealt{Kubo_2021}, \citealt{Andreon_2014}, \citealt{Strazzulo_2019}, \citealt{Noordeh_2021}), even in clusters (or protoclusters) at redshift $z \sim 2$  (\citealt{Kodama2007}, \citealt{Willis2020}, \citealt{McConachie22}). Moreover, some of them already exhibit an in-place or nearly formed brightest cluster galaxy (BCG) beyond $z = 1$ (e.g: \citealt{Chu_2021}, \citealt{Shi_2023}). Some also show the presence of an intra-cluster light component detected up to $z \sim 2$ (\citealt{Joo_2023}, \citealt{Werner_2023}), suggesting the complex history of their galaxy populations. 
The lack of large and representative samples of clusters beyond $z=1$  currently limits our ability to place strong constraints on the various scenarios of galaxy formation and transformation, and on the efficiency of the associated processes.

Galaxy clusters can be detected in different ways. For example, they can be traced via the X-ray emission of their ionised gas components (e.g. \citealt{Pacaud_2016}, \citealt{Bulbul_2024}), with millimetre observations via the Sunyaev-Zel'dovich effect (e.g. \citealt{Bleem_2015}, \citealt{Ade_2016}), through their weak-lensing signal (e.g. \citealt{Miyazaki2002, Miyazaki2007}, \citealt{Gavazzi2007}, \citealt{Shan_2012}, \citealt{Leroy2023}), or by the direct identification of galaxy overdensities. 
For the latter, different techniques have been used. They can be based on various cluster features used to limit projection effects, such as the identification of red-sequence galaxies (\citealt{Gladders_2000}, \citealt{Rykoff_2014}).
These approaches have been shown to be highly effective at redshifts below one, where the red sequence is well established, but they may fail to detect higher redshift clusters that lack a prominent red sequence. 

With the advent of multi-band photometric surveys, an alternative technique is to detect overdensities of galaxies in photometric redshift space, independent of any prior assumptions on the galaxy population. To be effective, this approach requires high-quality photometric redshifts to minimise contamination from projection effects, especially for the identification of low-mass systems. 
This led to a multitude of group and cluster catalogues built from a number of surveys: 
\citealt{Adami2010} from the Canada-France-Hawaii Telescope
Legacy Survey (CFHTLS), 
\citealt{Oguri2018} from the Hyper Suprime-Camera (HSC) Subaru Strategic Program, \citealt{Thongkham2024} from the Massive and Distant Clusters of WISE Survey 2 (MaDCoWS2),
\citealt{Maturi_2019, Maturi2025} from the Kilo-Degree Survey (KiDS)
, 
\citealt{WAZP} and \citealt{Benoist_2025} from Dark energy Survey (DES),
\citealt{Doubrawa_2024} from the Javalambre-Physics of the Accelerating Universe Astrophysical Survey (J-PAS). 
In parallel, group and cluster catalogues have also been obtained from spectroscopic surveys, such as the Sloan Digital Sky
Survey (SDSS; e.g. \citealt{Miller_2005}, \citealt{Yang_2007}, \citealt{Tempel_2012}) or the Galaxy And Mass Assembly (GAMA; \citealt{GAMA}), or with a combined approach based on both photometric and spectroscopic redshifts such as the Observations of Redshift Evolution in Large-Scale Environments survey (ORELSE; \citealt{Hung2020}).

These large-scale photometric and spectroscopic surveys revealed thousands of clusters up to $z=1$, allowing a deep understanding of cluster properties in the past $7$ Gyr of the history of the Universe. However, they lack the depth required to extensively identify and study clusters beyond $z = 1-1.5$. Consequently, research for high-redshift clusters is still limited to individual follow-up cases or within deep but small fields, constructed for this purpose. One of these fields is the COSMOS field, which has been widely studied at different wavelengths, allowing the identification of groups and clusters of galaxies over a wide range of redshifts (e.g. \citealt{Finoguenov_2007}, \citealt{Epinat_2024}, \citealt{Toni_2024, Toni_2025a}). However, the relatively small covered area ($\sim2 \text{ deg}^2$) restricts its statistical power to systems of lower mass, preventing general conclusions to be drawn about the properties of clusters and their evolution over time. The landscape is set to change in the near future with the advent of wide-field, deep optical, and infrared surveys such as the Euclid Deep and Wide Surveys (\citealt{Mellier2025}, \citealt{EWS}. See \citealt{Bhargava_2025} for cluster detection in the Euclid Q1 release).

On the other side, next-generation wide-field, high-multiplex spectrographs (WEAVE, 4MOST, PSF, MOONS) are also expected to transform our understanding of galaxy clusters and their role within the cosmic web.
In particular, the MOONS instrument (\citealt{Cirasuolo2020}) offers a remarkable combination of large multiplexing capability, high sensitivity, and broad simultaneous spectral coverage (spanning from the optical to near-infrared bands), over a wide field.
The Main MOONS GTO Extragalactic Survey (MOONRISE, \citealt{MOONRISE})  will obtain key spectroscopic information for $ \sim 500000$ galaxies in the redshift range $[0.8, 2.6]$. Its spectroscopic capabilities will be used to get a full three-dimensional characterisation of the galaxy distribution over $\sim 11$ Gyr, to understand the assembly history of galaxy clusters and their connection to the cosmic web, and test how environment affects the properties of their members galaxies.

The VIDEO XMM-LSS and CDFS fields (\citealt{Jarvis_2013}) are, along the COSMOS field and the Euclid Q1 fields (\citealt{Euclid_Q1_fields}), among the deepest extragalactic fields currently available, with a wealth of available photometric and spectroscopic data up to $z = 1$ and beyond. In this paper, we discuss the high redshift cluster content of these two fields obtained by using two different photometric redshift-based detection algorithms. From the two sets of detections, we built a robust joint sample of cluster candidates in the redshift range $0.1-3$. We achieved a first validation of some of these cluster candidates using ancillary spectroscopic data and by comparing with spectroscopically confirmed clusters from the literature. Then, we performed a preliminary analysis of the properties of the galaxies within these candidates by constructing their apparent Colour-Magnitude diagram, and by comparing with models of galaxy evolution. Finally, these candidates were used as a benchmark to test the capabilities of the MOONRISE survey for cluster science. 

This paper is organised as follows. In Sect.~\ref{section:data} we describe the content of the datasets used in this work and explain how they have been constructed. In Sect.~\ref{section:cluster_finder}, we introduce and describe the two cluster finder algorithms used in this work. Section~\ref{section:GS_construction} is dedicated to describing the construction of a subsample of robust detections. Then, in Sect.~\ref{section:GS_analysis}, we perform a first validation of this particular subset of detections using ancillary spectroscopic data and previously confirmed galaxy clusters. In Sect.~\ref{section:red-sequence}, we focus on the identification of the red-sequence components of our candidates within their apparent Colour-Magnitude diagram, and compare with models of galaxy evolution to put constraints on the properties of this particular subpopulation of galaxies. Section~\ref{section:MOONRISE} is dedicated to testing the ability of the different MOONRISE survey strategies to observe and characterise these systems, and finally, we discuss these results and draw our conclusion in Sect.~\ref{section:conclusion}. Throughout this paper, we assume a flat $\Lambda$CDM cosmology with $H_0 = 70\text{ km}\text{s}^{-1}\,\text{Mpc}^{-1}$, $\Omega_\text{m} = 0.30$ and $\Omega_\Lambda = 0.70$. All the quantities in pc correspond to physical lengths.

\begin{figure*}
    \centering
    \captionsetup{labelfont=bf}
    \includegraphics[width=17cm]{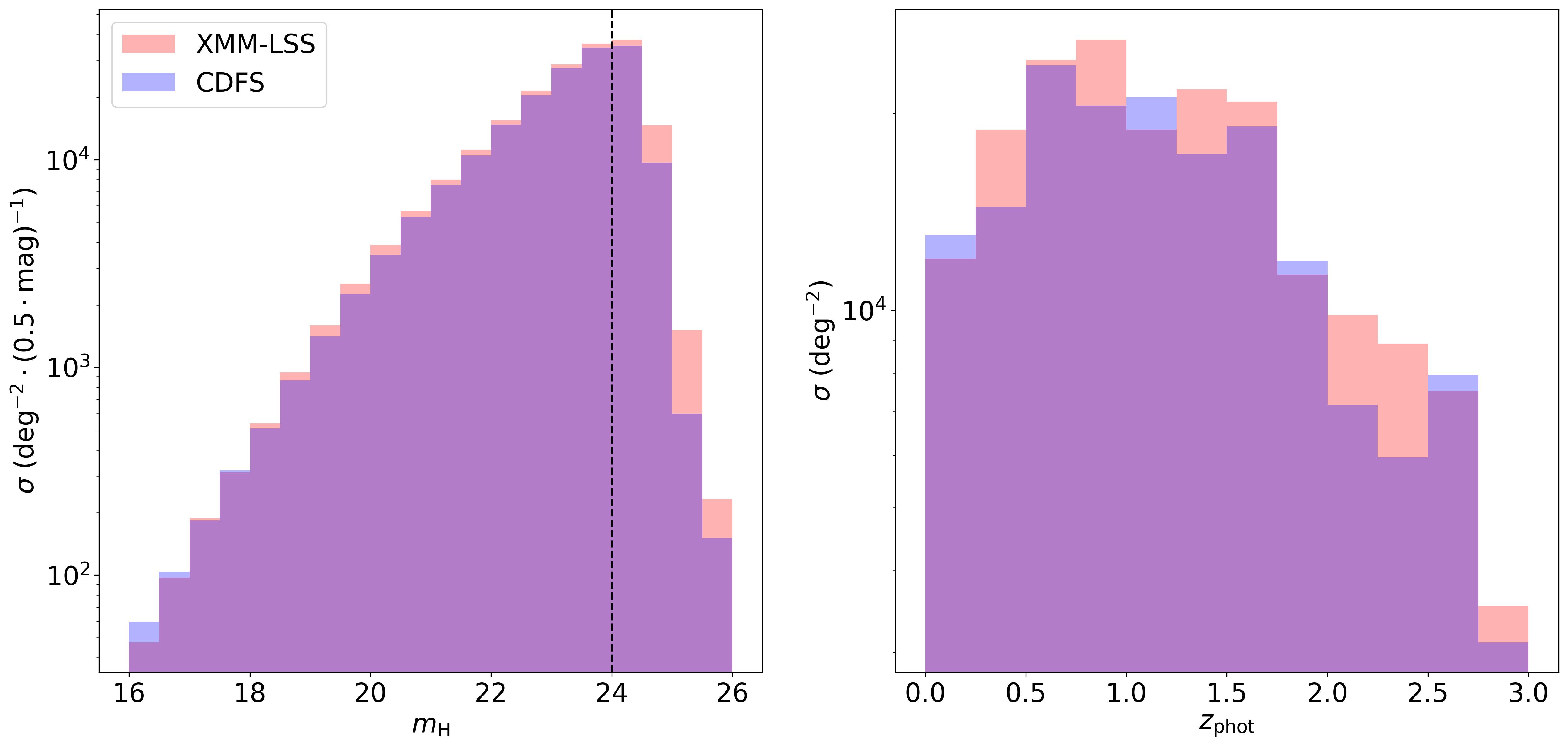}   
    \caption{\textit{Left}: VISTA H-magnitude galaxy counts for the XMM-LSS and CDFS fields. The vertical dashed line represent the $m_\text{H} = 24$ completeness limit. \textit{Right}: Distribution of the photometric redshifts of galaxies in the XMM-LSS and CDFS within the redshift range $z_\text{phot} \in [0, 3]$.}
    \label{figure:mag_counting}   
\end{figure*}

\section{Data}
\label{section:data}

\subsection{The VIDEO catalogues of galaxies}
\label{subsection:video_cat}

As part of the ongoing preparation for the MOONRISE GTO programme (\citealt{Cirasuolo2020}, \citealt{MOONRISE}), near-infrared selected catalogues spanning the VIDEO XMM-LSS and CDFS footprints have been produced. The details of these catalogues will be provided in future work (McLeod et al., in prep). 
We subsequently summarise their construction here.

Each of the VIDEO \citep{Jarvis_2013} XMM-LSS and CDFS fields spans approximately $\simeq 4.5 \text{ deg}^2$. For both of these fields, McLeod et al. (in prep) uses the available Y, J, H, and K$_{s}$ imaging from VIDEO DR4. The CDFS photometry includes UV/optical \textit{u}-, \textit{g}-, \textit{r}-, and \textit{i}-band data taken from the VLT Survey Telescope VOICE survey \citep{Vaccari_2016}. Available HSC imaging in \textit{g}-, \textit{r}-, \textit{i}-, and \textit{z}-band are also included, although we note that the $r$-band coverage is available only over a $\simeq2 \text{ deg}^2$ subset of the area (\citealt{Ni_2019,Ni_2021}). Finally, in the mid-infrared, IRAC 3.6$\mu$m and 4.5$\mu$m imaging publicly released from the Cosmic Dawn Survey \citep{Moneti2022} are included.

The XMM-LSS field consists of three sub-fields with varying UV to mid-infrared coverage. In particular, the UltraDeep Survey (UDS) covering a $\simeq 0.8 \text{ deg}^2$ region benefits from significantly deeper coverage in J, H, and K from UKIDSS \citep{Lawrence2007} UDS DR11 (Almaini et al. in prep). UV coverage mainly comes from CFHTLS T0007 \citep{Hudelot_2012} $u$-band imaging, with additional CFHT MegaCam $u$-band imaging from MUSUBI \citep{Mehta2018,Wang_2022} over the UDS footprint. In the optical, the XMM-LSS field is covered by HSC \textit{g}-, \textit{r}-, \textit{i}-, \textit{z}-, and \textit{y}-band imaging, as well as by HSC narrow-band NB816 and NB921 filters, as part of the second data release from the Subaru Strategic Program (SSP; \citealt{Aihara_2019}). In the CFHTLS-D1 sub-field, there is additional CFHT \textit{g}-, \textit{r}- \textit{i}-, and \textit{z}-band imaging from the aforementioned CFHTLS T0007 release (\citealt{Hudelot_2012}), while in UDS Subaru SuprimeCam (SSC) imaging in B, V, R, \textit{i}, and \textit{z}$^{\prime}$ (\citealt{Furusawa_2008}) are included. The SSC images used were those prepared as ancillary data products for the UDS DR11 data release by C. Simpson (Gemini). Finally, the more recent $z^{\prime}_\mathrm{new}$ imaging with improved red wavelength sensitivity from \citet{Furusawa_2016} are used. In the mid-infrared, the XMM-LSS benefits from IRAC 3.6$\mu$m and 4.5$\mu$m imaging from the SERVS programme \citep{Mauduit_2012}. For the deeper coverage in the UDS, McLeod et al. (in prep) utilises the \citet{Mehta2018} release which includes the SPLASH (PI Capak, see \citealt{Mehta2018}), SERVS (\citealt{Mauduit_2012}), SEDS (\citealt{Ashby2013}), and S-CANDELS \citep{Ashby2015} datasets.

Prior to catalogue construction, all imaging was PSF-homogenised to a common Moffat profile of FWHM 0.95$^{\prime\prime}$, with the exception of the CDFS VOICE \textit{u}, HSC \textit{g}, and HSC \textit{r} imaging, which were found to have FWHM$\gtrsim0.95^{\prime\prime}$. The catalogue construction follows the methods employed in \citealt{McLeod2021}. \textsc{Source-Extractor} (\citealt{Bertin1996}) was run in dual-image mode, with the VIDEO H-band sets as the detection image. For the UDS, a UKIRT WFCAM H-band catalogue was also produced, in order to leverage the superior depth in that region. Photometry was measured in 2-arcsec diameter apertures, with the exception of the lower resolution IRAC imaging. To measure IRAC photometry, the deconfusion software \textsc{TPHOT} (\citealt{Merlin_2015}) was used. \textsc{TPHOT} takes the positional and surface brightness information for objects from their higher resolution images (i.e. the detection H-band) as an input, along with a transfer kernel between the higher and lower resolution PSFs, in order to fit model fluxes for these same objects in the lower resolution IRAC imaging.

\begin{figure*}
    \centering
    \captionsetup{labelfont=bf}
    \includegraphics[width=17cm]{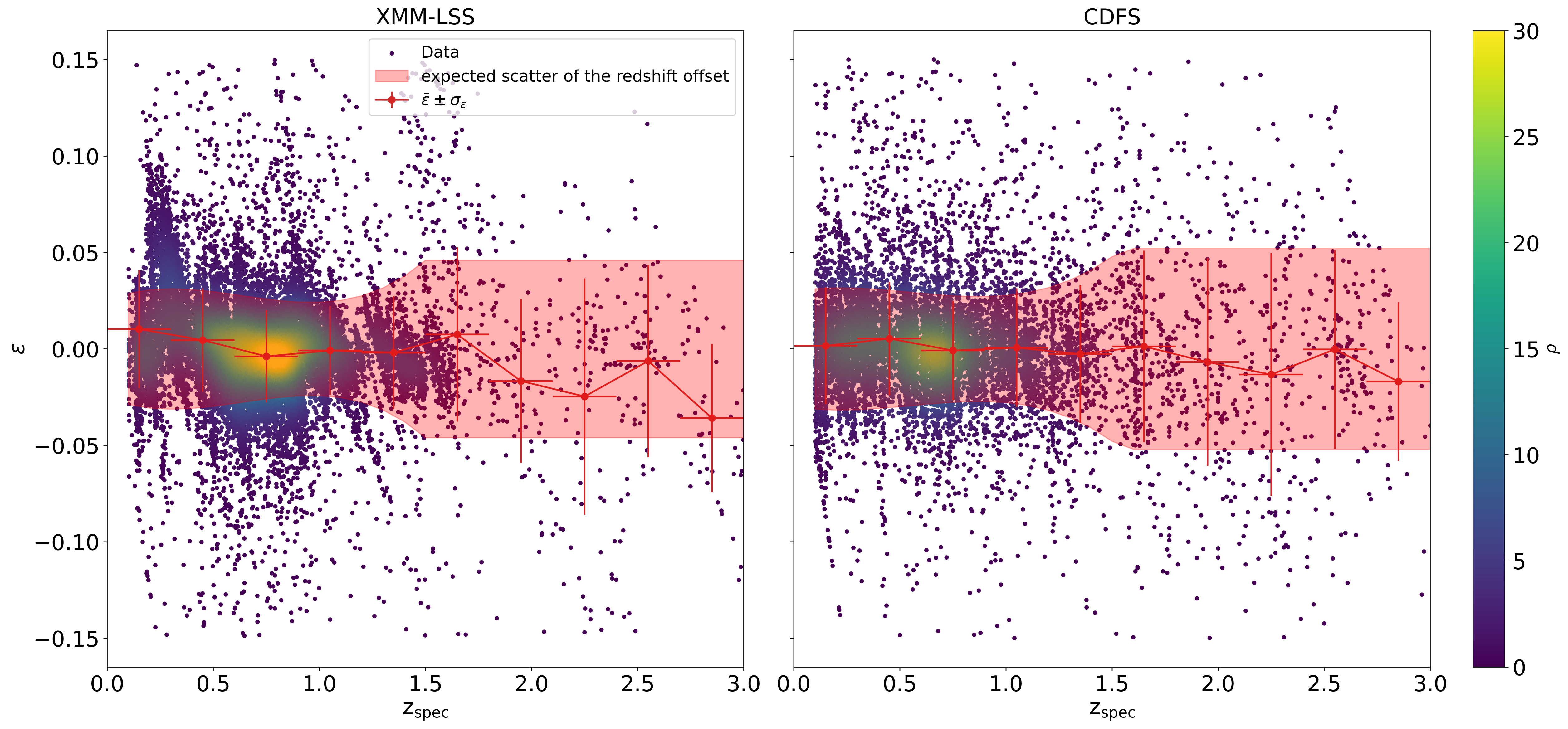}    
    \caption{Evolution of the mean bias $\bar{\varepsilon}$ in bins of $z_\text{spec}$ in the XMM-LSS (left panel) and the CDFS (right panel). The y-axis error bars represent the standard deviation $\sigma_\epsilon$ computed in the corresponding redshift bin. The x-axis error bars represent the bin widths. All data points are shown in the background, with density reconstructed using a Gaussian kernel. The filled red area represents the estimated scatter of the redshift offset, derived from polynomial fitting of $\sigma_\epsilon$. We assume it constant beyond $z=1.5$, due to the lack of available spectroscopic data to extrapolate it properly. The colour maps represent the amplitude of the probability density function of the points distribution.}
    \label{figure:z_error}   
\end{figure*}

\begin{figure}
    \centering
    \captionsetup{labelfont=bf}
    \includegraphics[width=0.48\textwidth]{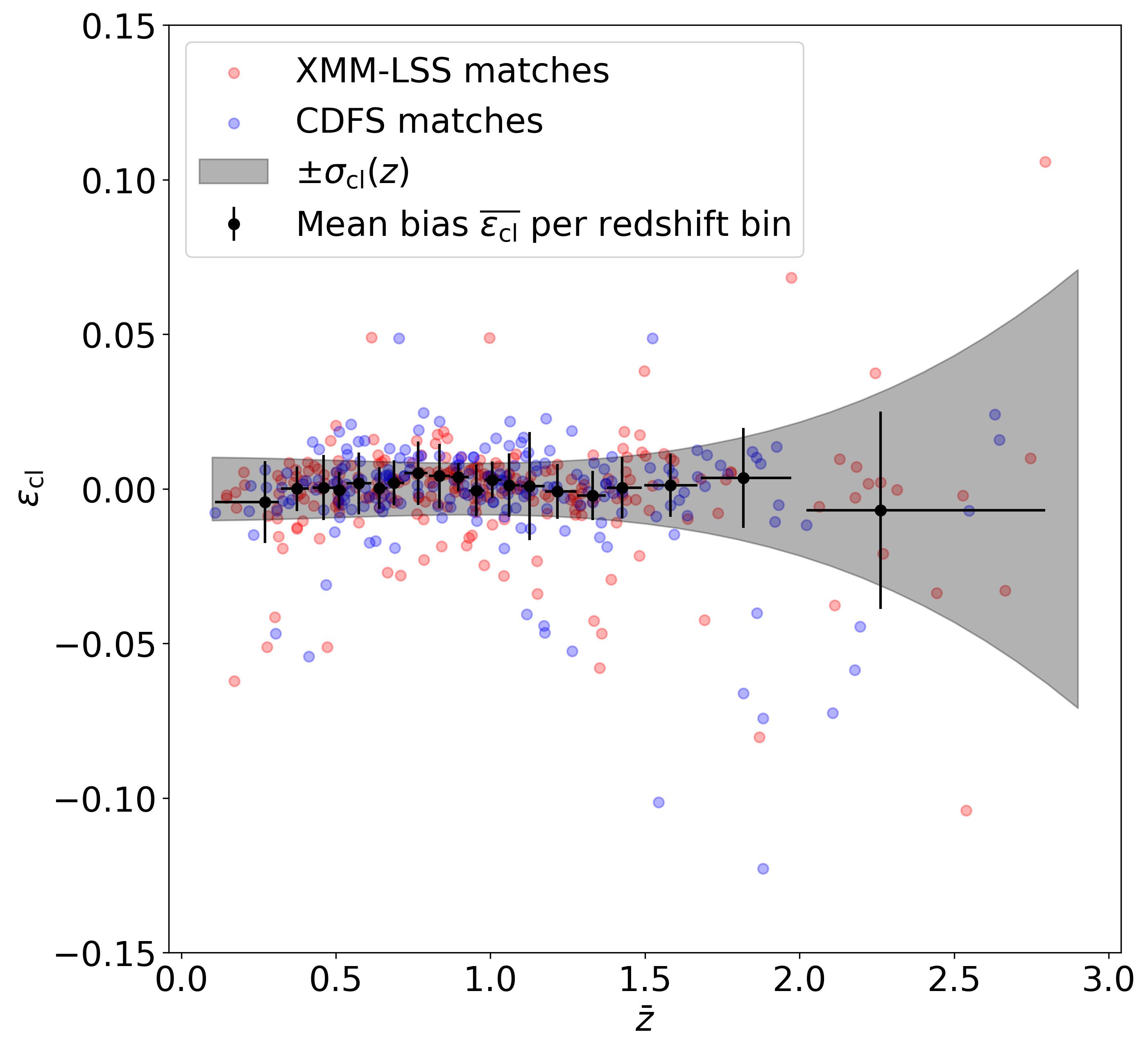}    
    \caption{Evolution of the redshift scatter $\varepsilon_\text{cl}$ between the AMICO and WaZP matched detections for a projected physical separation criterion of $R = 150$~kpc. The black dots represent the mean bias computed per bins of redshift, with the associated bins width and standard deviation as x and y-error bars, respectively. The grey envelops represent the estimated dispersion $\sigma_\text{cl}(z)$ of the offset between the AMICO and WaZP $z_\text{phot}$ for a pair of detections at a given redshift. We limit the y-axis to the range $[-0.15, +0.15]$ for the purpose of readability, five matches from the CDFS and four from the XMM-LSS are outside this range. The last bin is larger than the others to take into account the lower density of points at these redshifts.}
    \label{figure:bias_std_matching}   
\end{figure}

\begin{figure}
    \centering
    \captionsetup{labelfont=bf}
    \includegraphics[width=0.48\textwidth]{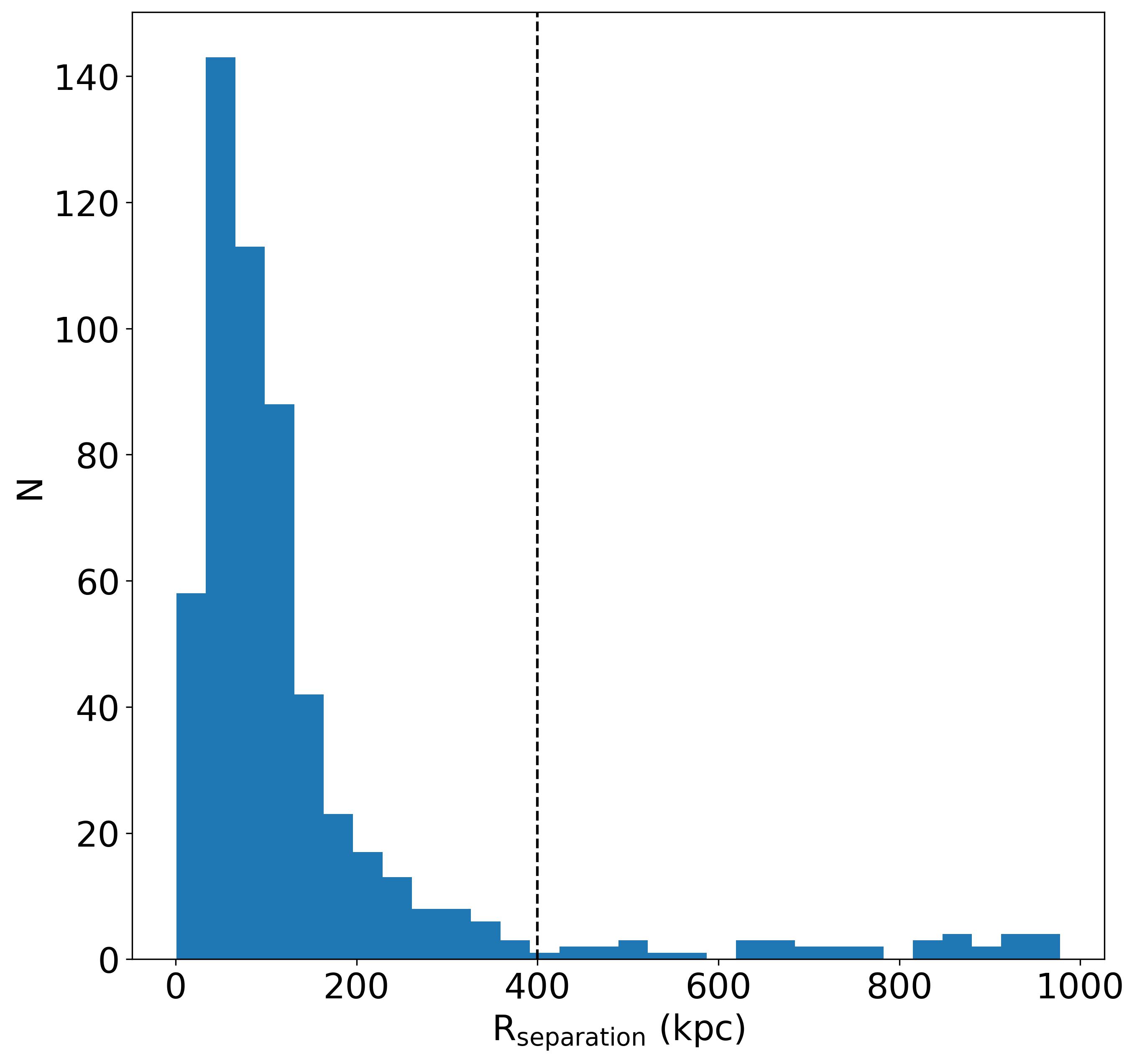}    
    \caption{Angular separation distribution of the AMICO-WaZP pairs of detections for a $R = 1.0$\,Mpc matching radius at the $2\sigma(z)$ level. The vast majority of the distribution is contained within $R = 200$~kpc ($\sim 83~\%$), with a few pairs separated by more than $400$~kpc ($\sim 7~\%$ of the pairs, in both fields). The dashed vertical line represent the $400$ kpc matching criterion we derived from this figure.}
    \label{figure:ang_sep_matching}   
\end{figure}

\begin{figure*}[t]
    \centering
    \captionsetup{format=plain}
    \captionsetup{labelfont=bf}
    \includegraphics[width=4.5cm]{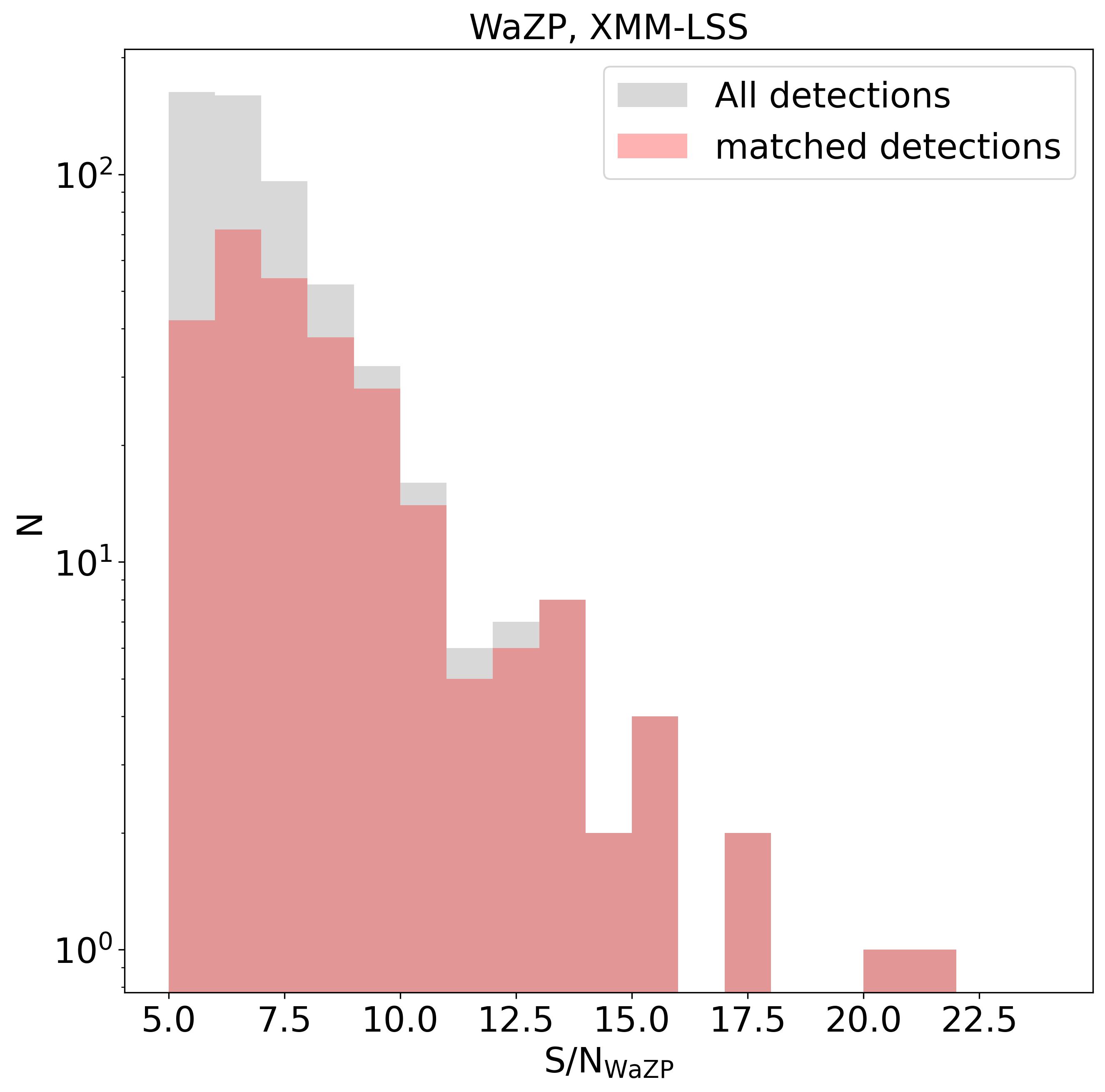}
    \includegraphics[width=4.5cm]{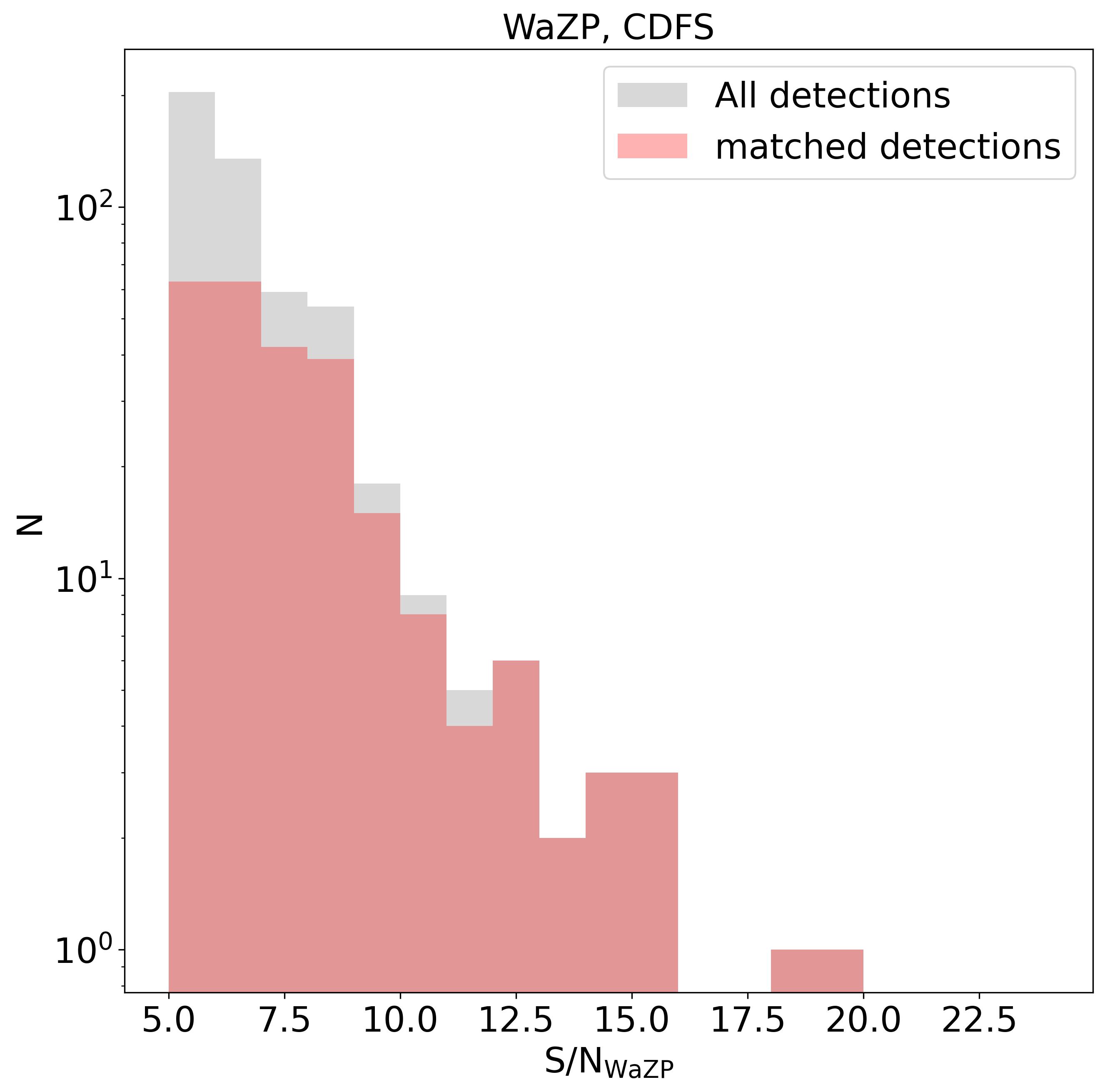}   
    \includegraphics[width=4.5cm]{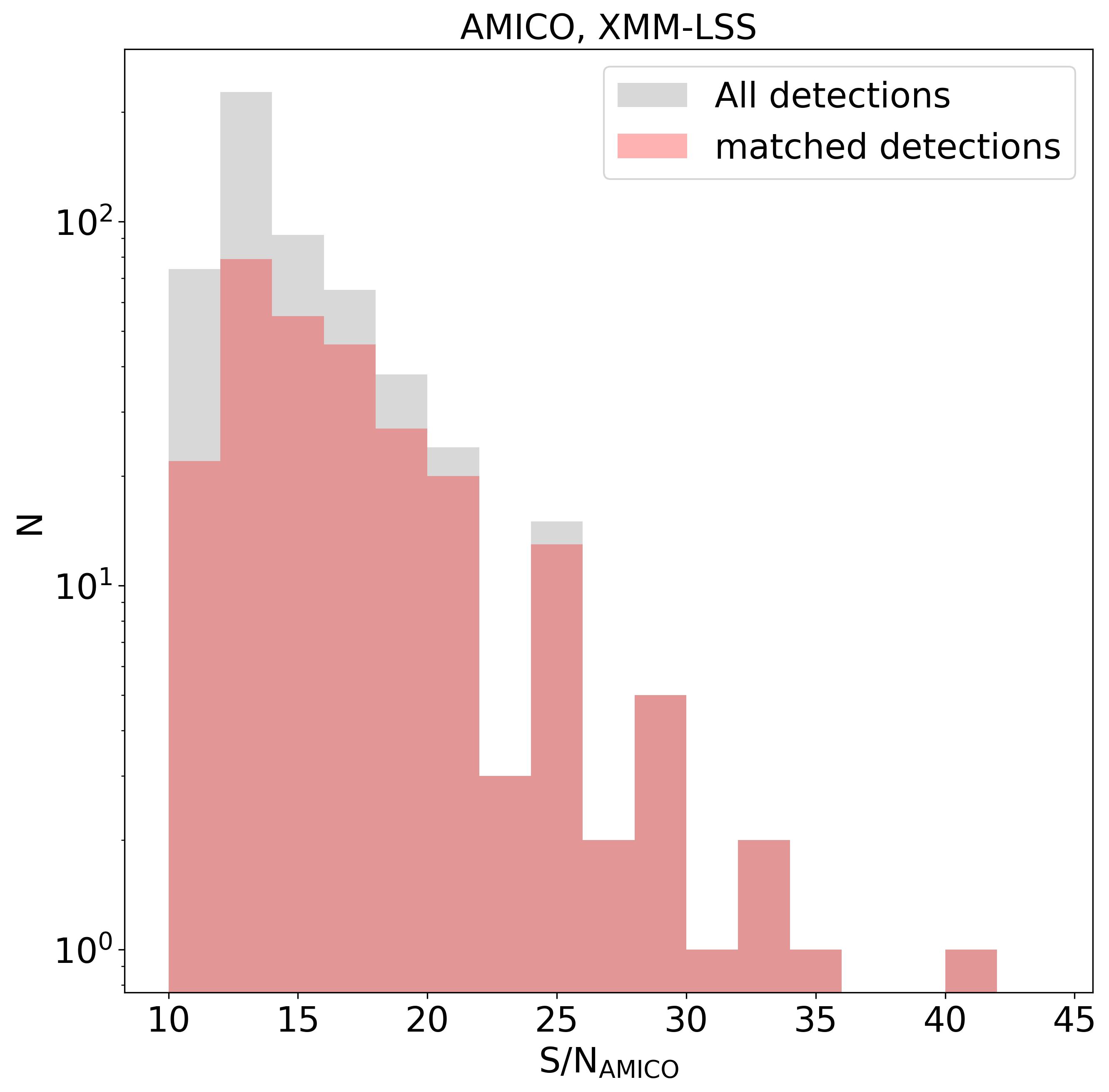}
    \includegraphics[width=4.5cm]{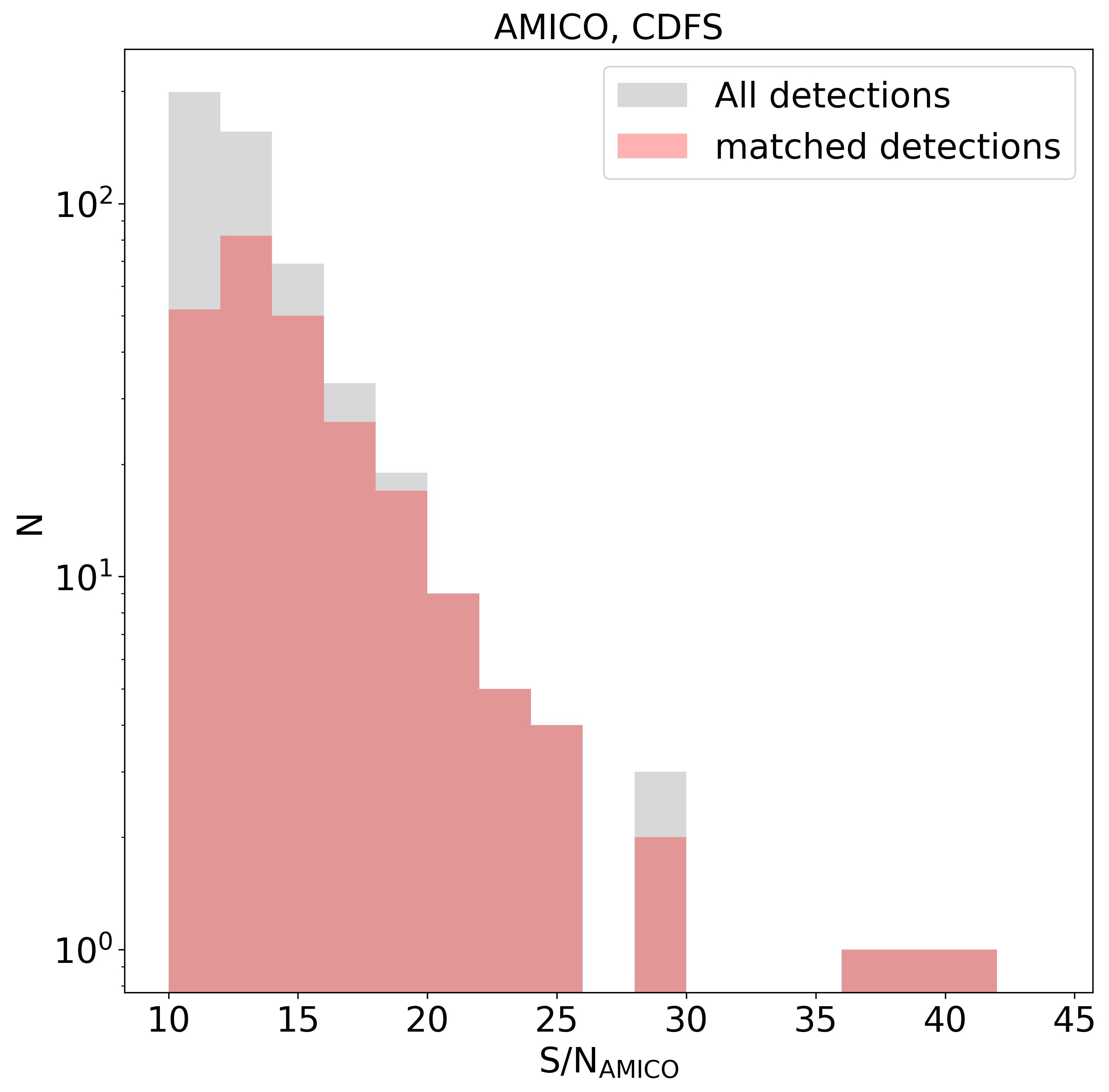}

    \caption{Distribution of the WaZP and AMICO SNR of the matched cluster candidates compared to all the detections in the XMM-LSS and CDFS.}
    \label{figure:S/N_matches_vs_all}   
\end{figure*}

\begin{figure}[t]
    \centering
    \captionsetup{labelfont=bf}
    \includegraphics[width=0.48\textwidth]{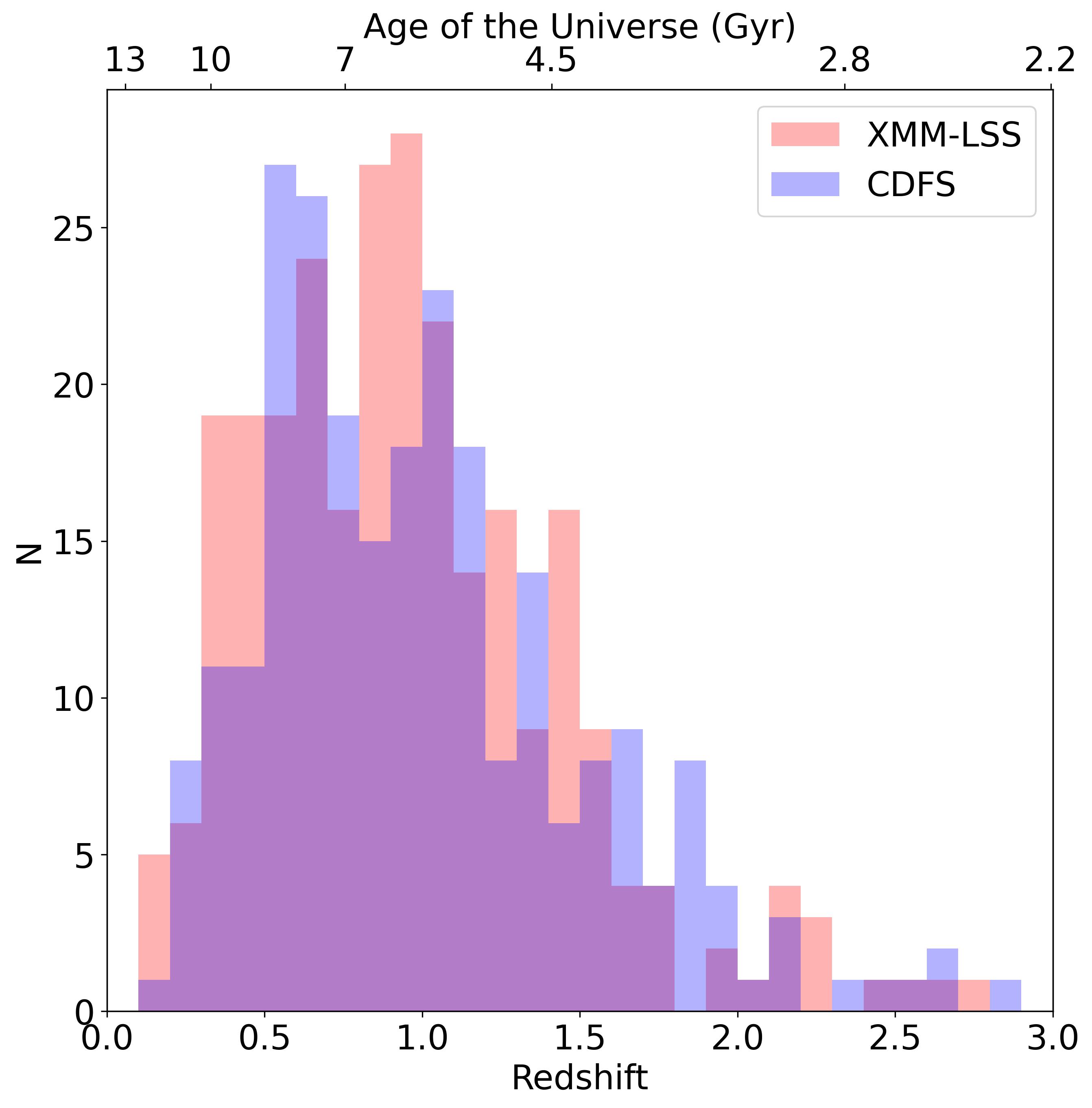} 
    \caption{Distribution of the photometric redshifts of the AMICO-WaZP joint cluster candidates in the XMM-LSS (red) and CDFS (blue). These distributions appear similar in both fields.}
    \label{figure:GS_zdistri_CF}   
\end{figure}

The final working catalogues used in this study was limited to regions that have an optimal coverage at different wavelengths to ensure a better homogeneity of the derived photometric redshifts in the two fields. In particular, some regions not covered neither by the $r$- or $i$-bands were excluded. 
In addition, several series of masks were applied to limit the number of false positive sources or sources with shallow photometry (essentially at the edges). Based on the Gaia DR3 catalogue \citep{gaiadr3}, stars were masked by disks with magnitude-dependent radii following an empirical relation. Brightest ($m_r \le 15$) foreground galaxies were also masked. Finally, several image artefacts that caused the concentration of spurious detections in space and redshift were masked by hand. The resulting effective areas are 4.55~deg² for the XMM-LSS field and 4.11~deg² for the CDFS field. The normalised galaxy number counts in bins of VIDEO H-band magnitude ($m_\text{H}$ in the following) are shown in left panel of Fig.~\ref{figure:mag_counting}. They are consistent and show a very similar depth, both complete up to $m_\text{H} = 24$. At this magnitude, we obtain galaxy densities of 38.0 and 36.1~arcmin$^{-2}$ for the XMM-LSS and CDFS fields, respectively.

\subsection{The photometric redshifts}
\label{subsection:zphot_cat}

\noindent Photometric redshifts $z_\mathrm{phot}$ were determined for sources by taking the median $z_\text{phot}$ from seven different SED fitting runs, a method proven to be effective at recovering more accurate $z_\text{phot}$ than one run using one code with its systematics, alone (see, for example, \citealt{Dahlen_2013}). Five of these used the template fitting code \textsc{LePhare} \citep{Arnouts2011}, including BC03 \citep{BC03}, three alternate PEGASE2 \citep{Fioc1999} runs that use different star-formation histories and metallicities, and the COSMOS \citep{Ilbert2009} SED libraries. The remaining two runs were performed with EAZY \citep{Brammer_2008}, using the default PCA templates and the PEGASE2 libraries. 
Once the median $z_\text{med}$ of all $z_\text{phot}$ values returned by these codes was found, the photometric redshift was fixed to this value.

In the following, we will only consider galaxies within the redshift range $z_\text{phot} \in [0, 3]$. The right panel of Fig.~\ref{figure:mag_counting} shows the redshift distribution of galaxies in the redshift range $[0, 3]$. Overall, galaxies appear to be distributed in redshift in a similar way in both fields.

\setlength{\tabcolsep}{6pt}
\renewcommand{\arraystretch}{1.5}
\begin{table}[t!]
\captionsetup{labelfont=bf}
\adjustboxset{width=\textwidth}
\centering
\caption{\label{table:zspec_survey_xmm} List of spectroscopic surveys and number of $z_\text{spec}$ used for the XMM-LSS galaxies. }
\begin{tabular}{ccc} 
 \hline\hline
 \makecell{Survey \\ name} & \makecell{Number \\ of galaxies} & Reference\\
 \hline
  DESI-DR1 & 10399 & \cite{DESI-DR1} \\ 
  GAMA-DR3 & 2158 & \cite{GAMA_DR3} \\ 
  SDSS-DR16 & 2652 & \cite{SDSS_DR16} \\ 
  UDSz & 1330 & \makecell{\cite{UDSz1} \\ \cite{UDSz2}} \\
  VIPERS-PDR2 & 11753 & \cite{VIPERS_PDR2} \\
  \hline
\end{tabular}

\vspace{0.5cm}
\caption{\label{table:zspec_survey_cdfs} List of spectroscopic surveys and number of $z_\text{spec}$ used for the CDFS galaxies. }
\begin{tabular}{ccc} 
 \hline\hline
 \makecell{Survey \\ name} & \makecell{Number \\ of galaxies} & Reference\\
 \hline
 2dF & 185 & \cite{2dF} \\ 
 2dFLenS & 23 & \cite{2dFlenS}\\ 
 3D-HST & 3724 & \makecell{\cite{3DHST_1} \\ \cite{3DHST_2}}\\ 
 ACES & 3730 & \cite{ACES}\\ 
 ATLAS & 367 & \cite{ATLAS}\\ 
 OzDES & 3271 & \cite{Lidman_2020} \\
 FIGS & 63 & \cite{ELG_FIGS}\\ 
 MUSE-Wide & 671 & \cite{MUSE}\\ 
 VANDELS & 93 & \cite{VANDELS}\\
 VUDS & 45 & \cite{VUDS}\\
 VVDS & 195 & \cite{VVDS}\\ \hline
\end{tabular}
\end{table}

\begin{figure*}[t]
    \centering
    \captionsetup{format=plain}
    \captionsetup{labelfont=bf}
    \includegraphics[width=17cm]{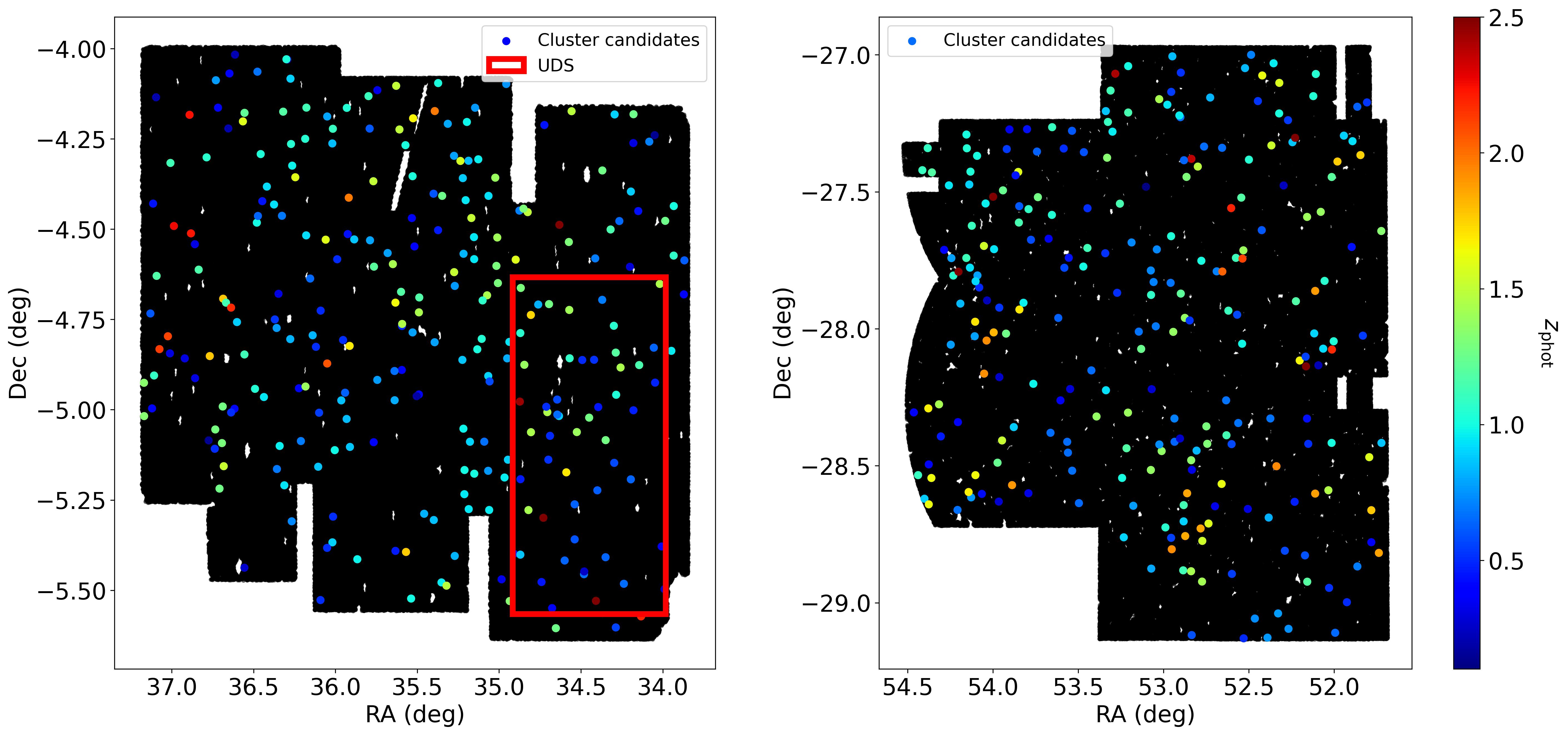}      
    \caption{Spatial distribution of the cluster candidates within the XMM-LSS (left) and the CDFS (right). The black backgrounds represent the footprints of both fields. The colours indicate the mean photometric redshift of the AMICO-WaZP joint detections. The red rectangle marks the limit of the footprint of the UDS subfield, the region which will be covered by MOONRISE in the XMM-LSS.}
    \label{figure:GS_spatial_distri}   
\end{figure*}

\subsection{The spectroscopic redshifts}
\label{subsection:zspec_cat}

\noindent In addition to the photometric redshift catalogues, we have access to catalogues of spectroscopic redshifts ($z_\text{spec}$) for both fields, compiled from various public surveys. Spectroscopic information was associated to the photometric catalogues within a $1$~arcsec radius, resulting in $28292$ matches in the XMM-LSS and $12367$ in the CDFS. For this matching, only highly reliable spectroscopic redshifts were considered, based on the quality flags provided by each survey. Table\,\ref{table:zspec_survey_xmm} and Table~\ref{table:zspec_survey_cdfs} list the different surveys used in this study and the associated number of unique counterparts for the XMM-LSS and CDFS galaxies within the photometric redshift range $z = 0.1-3$.
The available spectroscopic data allow us to sample well large domains of magnitude and redshifts. However, as we show in Fig.~\ref{figure:zspec_mag_xmm} and Fig.~\ref{figure:zspec_mag_cdfs}, some domains remain poorly covered, especially at redshifts beyond 1.5 or at magnitudes $m_H \gtrsim  23$. Our capacity to evaluate the quality of photometric redshifts in these domains will therefore be limited.

\subsection{Quality assessment of the photometric redshift}
\label{subsection:zphot_quality}

\noindent To quantify the quality of photometric redshifts, we computed the deviation of $z_\text{phot}$ relative to $z_\text{spec}$ measurements for the same galaxy by introducing the offset $\varepsilon$ defined as:  

\begin{equation}
    \varepsilon = \frac{z_\text{phot}-z_\text{spec}}{1+z_\text{spec}}.
    \label{equation:zbias}
\end{equation}
In this work, we used $z_\text{spec}$ as the reference measurement of galaxy redshifts, with errors considered negligible compared to those from $z_\text{phot}$. Following \citealt{Ilbert_2006}, $\sim 5\%$ of the galaxies with $\vert \varepsilon \vert > 0.15$ were considered as catastrophic errors. The remaining $95\%$ of galaxies are shown in Fig.~\ref{figure:z_error}. They were used to estimate 
the mean bias, $\bar{\varepsilon}$, and the associated standard deviation, $\sigma_\epsilon$, computed in bins of $z_\text{spec}$.

Up to $z \sim 1.5$, statistics remain well controlled, with a mean bias amplitude smaller than $0.01$ and a maximal scatter of $\sigma_\varepsilon(z) \sim 0.03$ in both fields. However, at higher redshift, because of the lack of spectroscopic data, $z_\text{phot}$ metrics cannot be accurately estimated. 
We derived the scatter $\sigma_\epsilon(z)$ in the range $z = 0.1 - 1.5$ in both fields via a polynomial fit, and assumed the scatter to be constant beyond this upper limit, as we cannot constrain it properly at these redshifts. These are shown as the red envelopes in Fig.~\ref{figure:z_error}.
These functions will be used in the following to estimate the error on the galaxy photometric redshift at a given redshift for the corresponding field when needed.

\section{Detection of galaxy clusters}
\label{section:cluster_finder}
\noindent In this section, we describe the two cluster finders used in this work: AMICO and WaZP. 
Both algorithms have been widely used over the last few years, proving themselves highly efficient at identifying cluster and group candidates over a wide range of redshift (e.g. see \citealt{Maturi_2019, Maturi_23, Maturi2025}, \citealt{Toni_2025a}, \citealt{Bhargava_2025} for AMICO, \citealt{WAZP}, \citealt{Benoist_2025} for WaZP).  

Both cluster finders weight galaxies depending on their magnitudes relative to a reference magnitude that varies with redshift. The goal was to build a cluster richness estimator that would provide the same richness to a cluster that would be seen at different redshifts. Thus, ensuring a constant comoving galaxy density is essential for enabling a consistent comparison of the signal-to-noise ratio (SNR or S/N) and richness of clusters detected at different redshifts.
This reference magnitude was chosen here to be the expected knee of the Luminosity Function, $m^*(z)$. To trace its evolution, we computed the flux variation of a passively evolving population formed in a single burst at high redshift and calibrated with known clusters at low redshift \citep[e.g.][]{2006ApJ...650L..99L}. In this study, we derived ${\rm m}_H^{\star}(z)$ from the passive evolution of a burst galaxy with a formation redshift $z_{form}=5$ taken from the PEGASE2 library \citep[\texttt{burst\_sc86\_zo.sed},][]{Fio97}.

\subsection{AMICO}

The Adaptive Match Identifier of Clustered Objects (AMICO, \citealt{Bellagamba_2018}, \citealt{Maturi_2019}) is designed to detect galaxy clusters in photometric catalogues using the linear optimal matched-filter technique described in \citealt{Maturi_05}. In this framework, the data are modelled as $D(x) = S(x) + N(x)$ where the cluster signal is given by $S(x) = A\cdot M_c(\mathbf{x})$.
Here, $A$ is the cluster amplitude and $M_c$ is a model describing the observable properties of clusters members, $\mathbf{x}$, which in this case include right ascension, declination, photometric redshift with its associated uncertainty, and H-band magnitude. The amplitude, $A$, which scales with cluster richness, is obtained through the convolution of the data with an optimal filter, as detailed in \citealt{Bellagamba_2018} and \citealt{Maturi_2019}. The noise term, $N$, captures the contribution of noise measured directly from the redshift and magnitude distributions of the galaxies, assumed to be uniform and produced by random Poissonian galaxy counts.

The filtering kernel is automatically derived via a constrained minimisation procedure, producing an unbiased minimum-variance estimate of the cluster amplitude $A$. AMICO scans the sky with an angular and redshift resolution of $0.3$ arcmin and $0.01$, respectively, and iteratively identifies the cluster candidates by locating the most significant signal-to-noise peaks produced by the filter. The scheme is used iteratively to allow for better identification of blended detections. 

The probability of the $i$-th galaxy being a member of the $j$-th detection is estimated as
\begin{equation}
P_i(j) = P_{f,i} \frac{A_j M_{c,j}(r_{ij},m_i) p_i(z_j)}{A_j M_{c,j}(r_{ij},m_i) + N(m_i,z_j)} \quad .
\end{equation}
Here, $r_{ij}$ is the angular separation between detection and galaxy, and $P_{f,i} = 1 - \sum_k P_i(k)$ is the probability
of the i-$th$ galaxy to belong to the field. These quantities are defined iteratively during the detection process starting with $P_{f,i}=1$. The intrinsic richness of the $j$-th detection is defined as $\lambda_{*j} = \sum_i P_i(j)$, where the sum runs over all galaxies within $R_{200}$ of the template cluster model and magnitudes smaller than $m_*+1.5$ evaluated at the detection redshift. Here, $R_{200}$ refers to the radius of a sphere of mean density equal to $200$ times the critical density of the Universe, and is empirically derived from numerical simulations for a cluster of typical mass $ M\sim 10^{14}h^{-1} \text{M}_\odot$, at a given redshift.

\subsection{WaZP}

The Wavelet Z-Photometric (WaZP) cluster finder (\citealt{WAZP}, \citealt{Benoist_2025}) is designed for the optical detection of galaxy clusters from multi-wavelength photometric imaging galaxy surveys. It is based on the search for projected overdensities of galaxies in photometric redshift space. It does not make any assumption about the existence of a red-sequence and relies only weakly on the cluster radial profile and on the luminosity function (redshift-dependent faintest magnitude). A complete description of the WaZP detection method is presented in \citealp{WAZP} and \citealt{Benoist_2025}. Here, we provide a summary of the main steps.

Overdensity peaks are initially extracted from pixelised wavelet-based density maps, constructed from the galaxy catalogue in various photometric redshift slices. In each slice, galaxies are weighted based on their redshift probability distribution function (here, Gaussian). Then, the peaks of successive slices are merged and the peak with the largest SNR enters the final list of clusters. The accuracy of the cluster position on the sky is limited by the resolution of the wavelet maps chosen here as $1/16$\,Mpc.

Then, following the prescription of \citet{Castignani_2016}, cluster membership probabilities are computed. They depend on galaxy cluster-centric distances, magnitudes, and photometric redshifts.
Finally, cluster members are used to jointly estimate the cluster richness and radius.
The richness is the sum of the membership probabilities within a radius that corresponds to an overdensity of $200$ times the mean number density of the galaxy background (similar to \citealp{hansen2005}). Here, for the richness computation, only members brighter than ${\rm m}^{*}(z_{\rm cluster}) + 1.5$ are considered.

\begin{figure*}[t]
    \centering
    \captionsetup{format=plain}
    \captionsetup{labelfont=bf}
    \includegraphics[width=5.66cm]{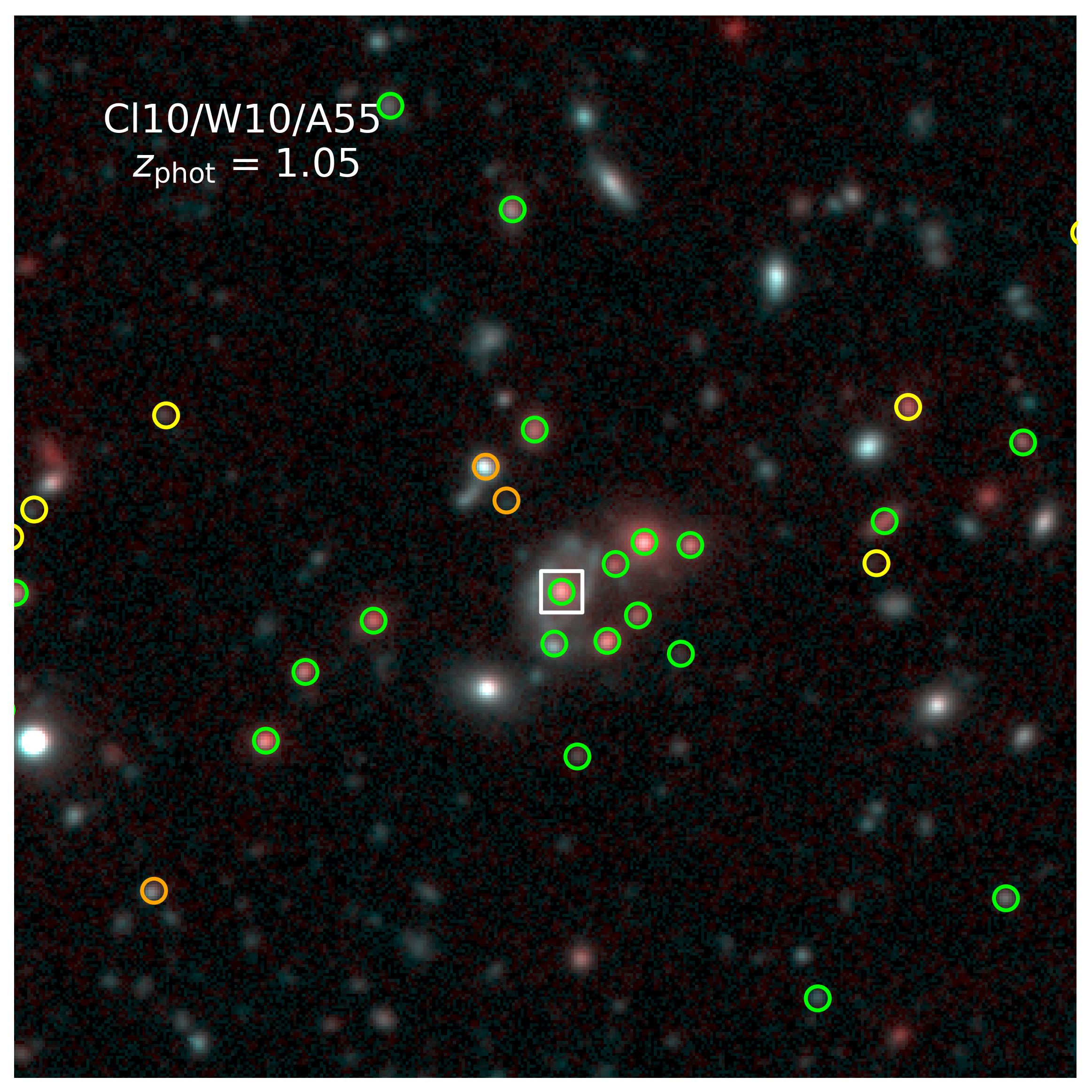}    
    \includegraphics[width=5.66cm]{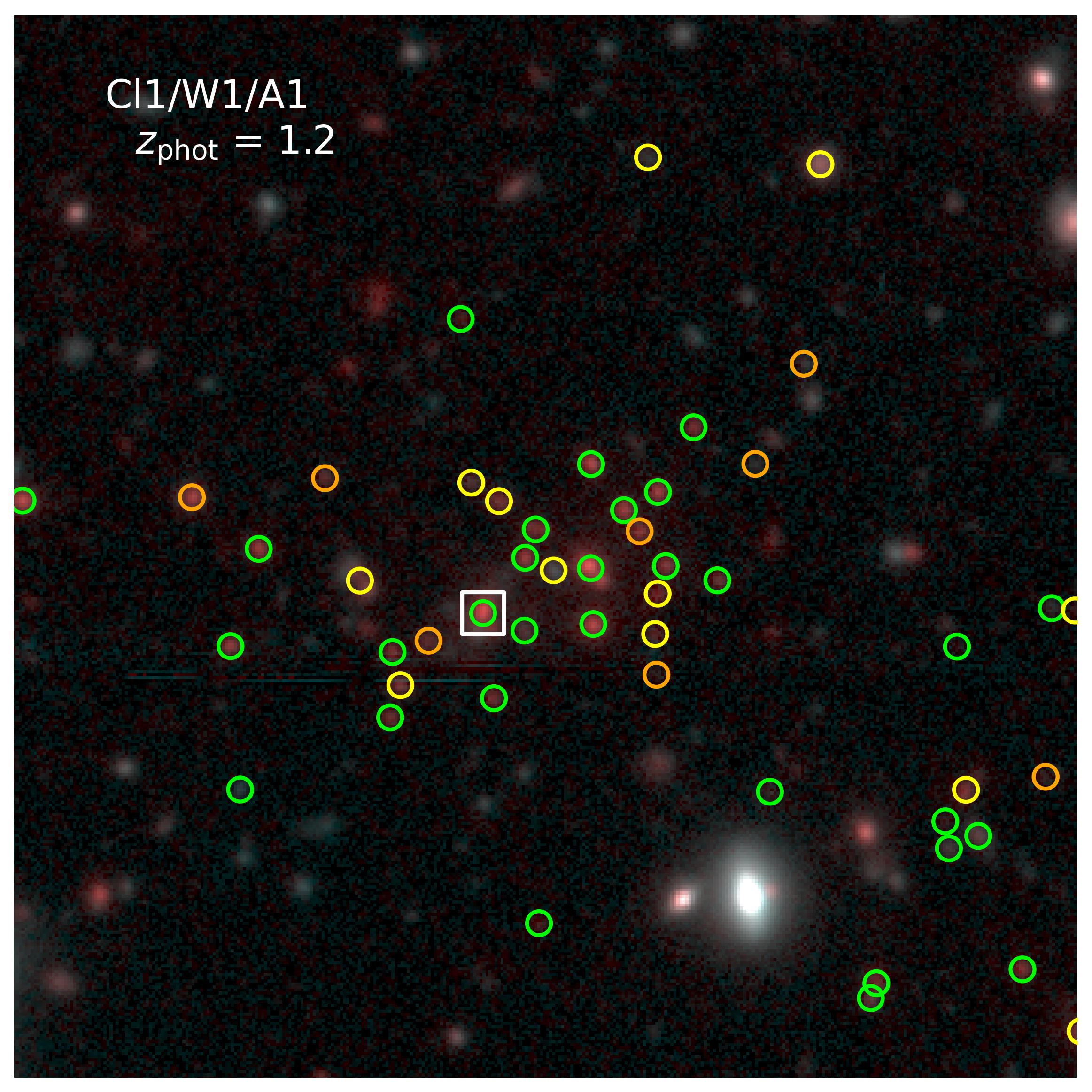}
    \includegraphics[width=5.66cm]{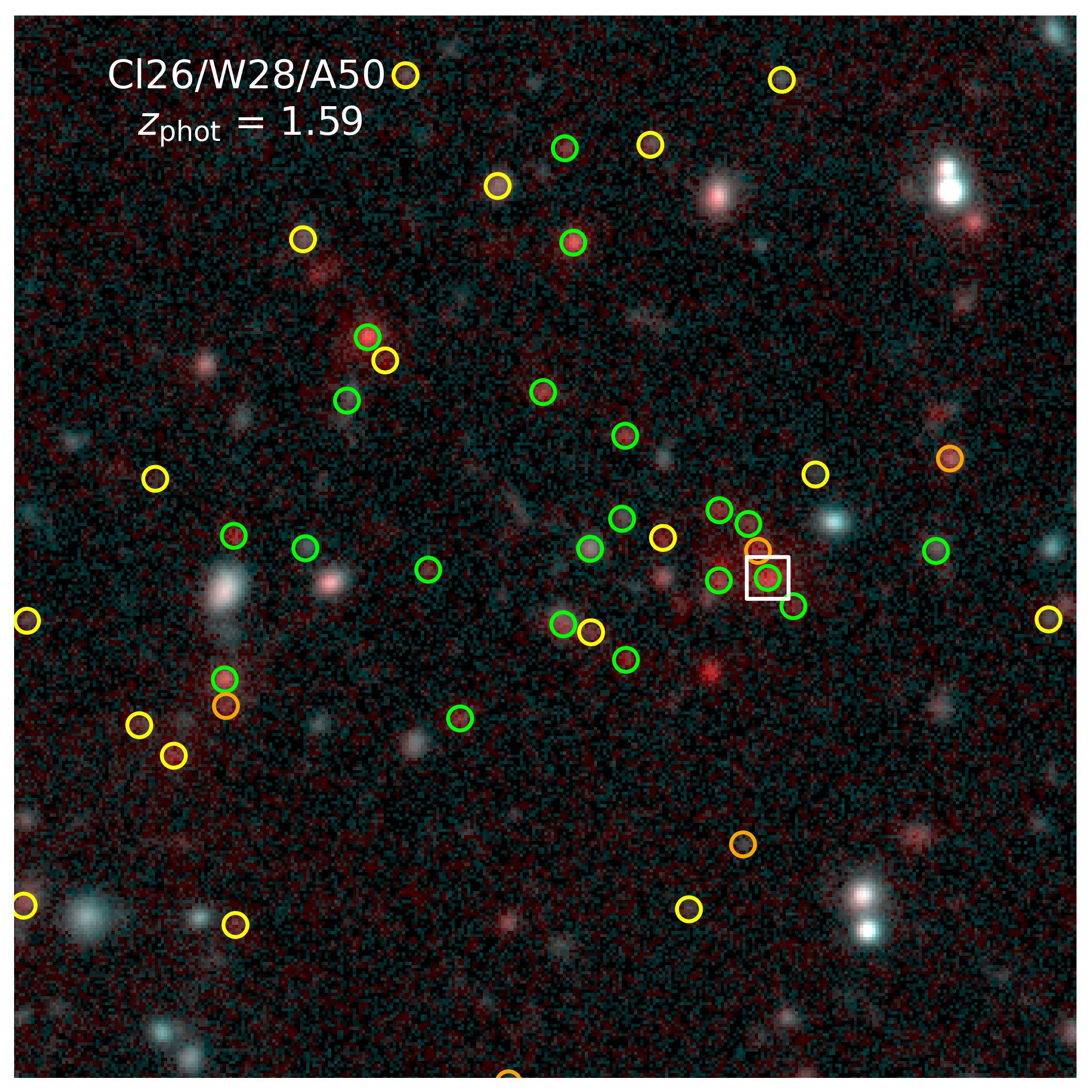}    

    \includegraphics[width=5.66cm]{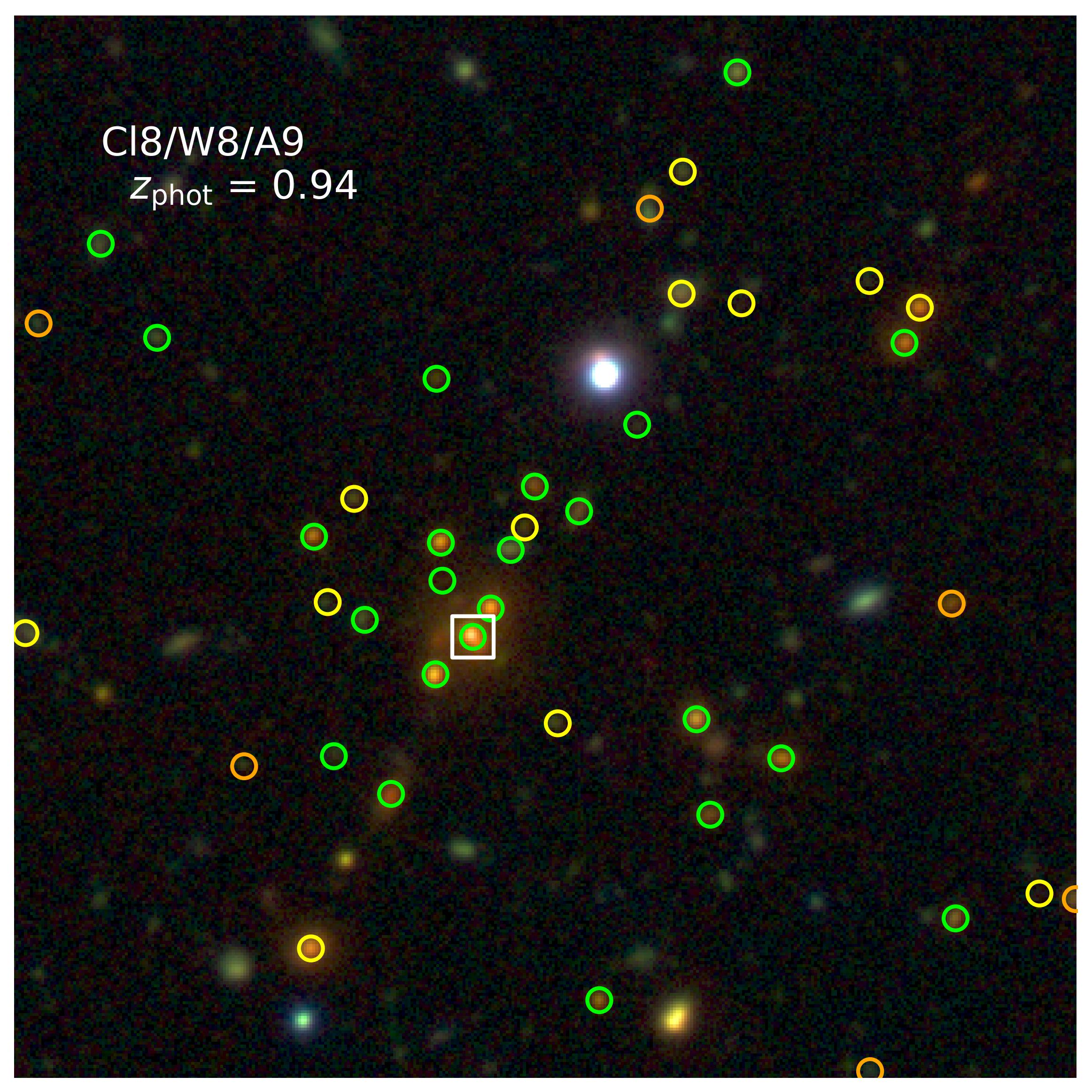}    
    \includegraphics[width=5.66cm]{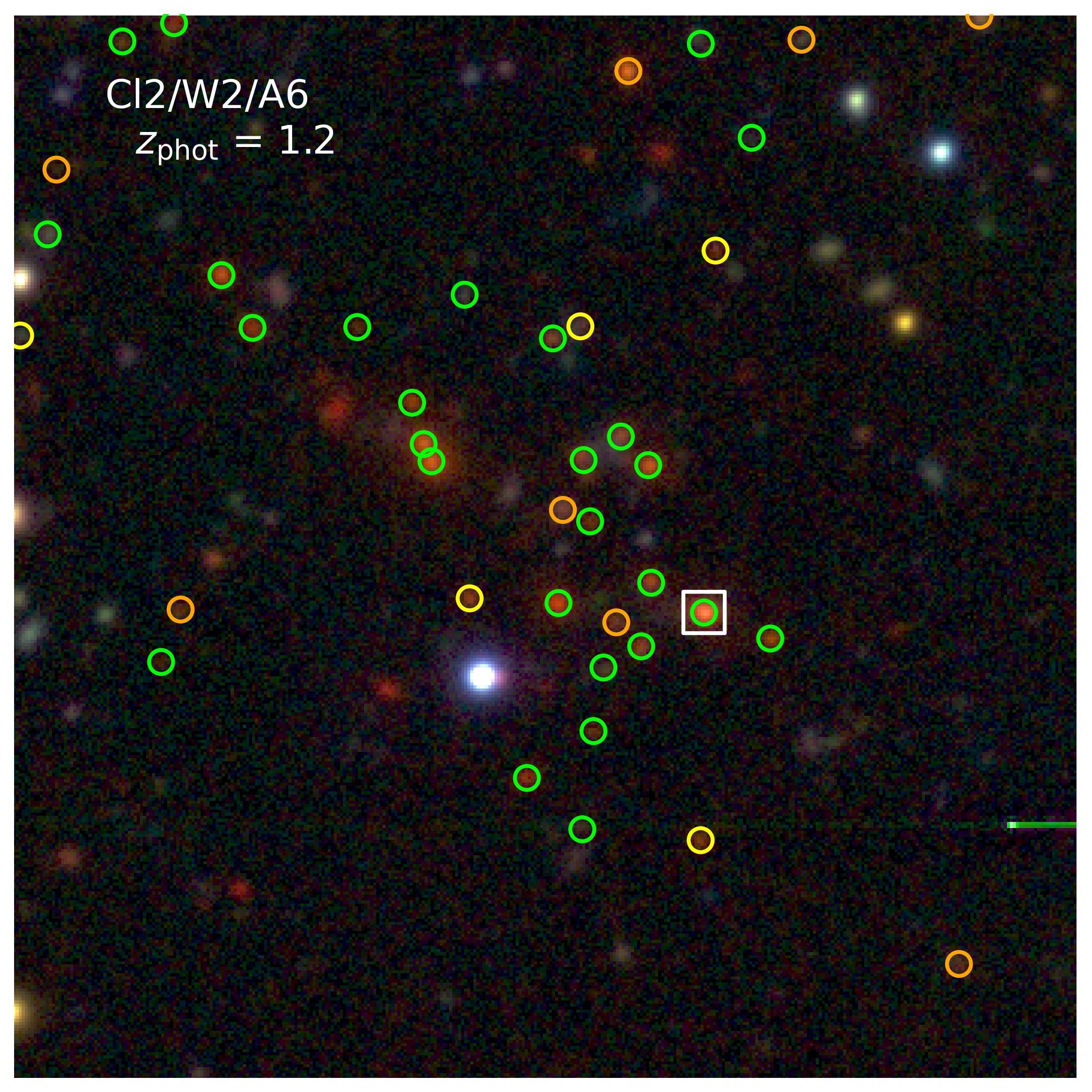}    
    \includegraphics[width=5.66cm]{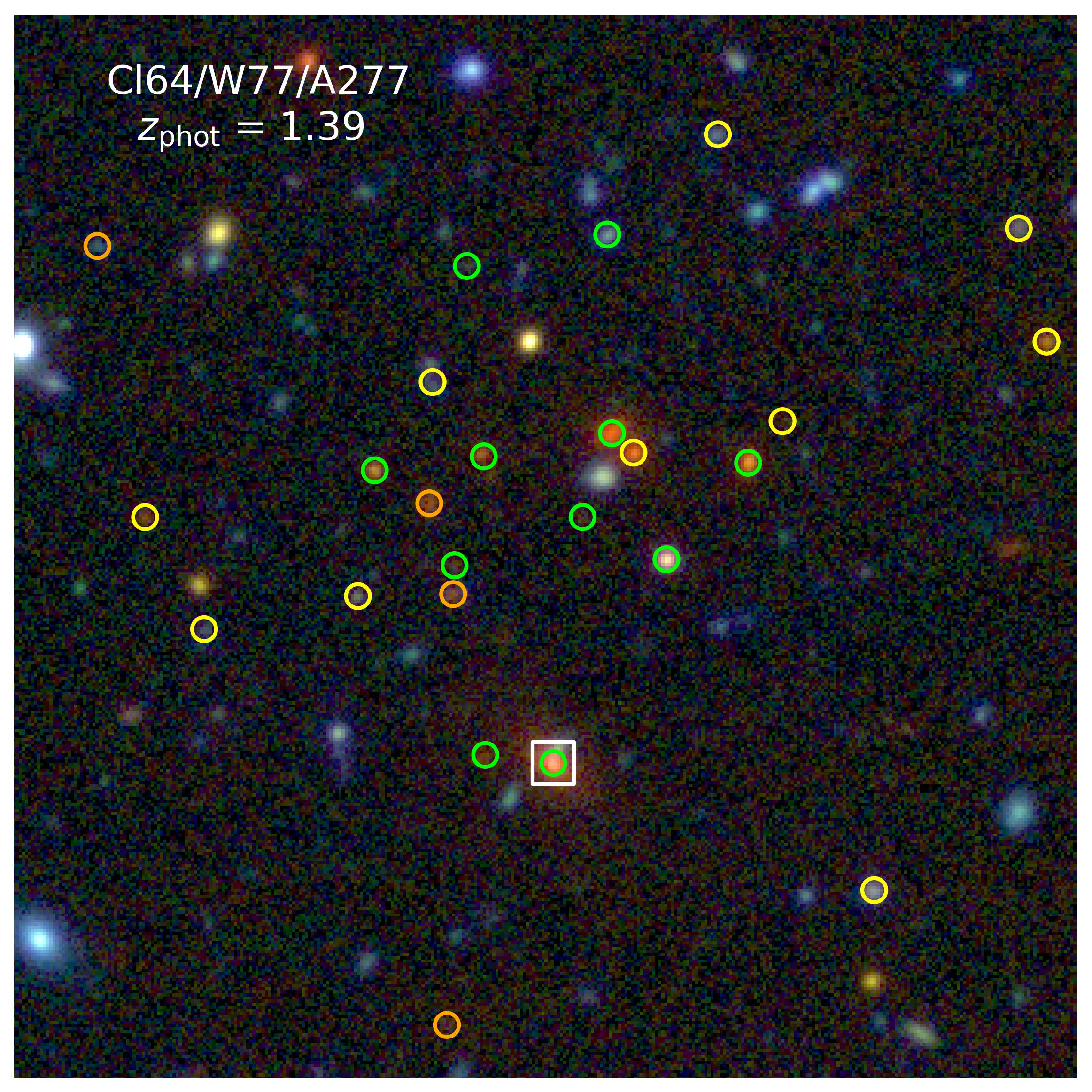}
    
    \caption{Coloured cutouts of three XMM-LSS (top) and three CDFS (bottom) cluster candidates from the joint catalogue in the redshift range $z_\text{phot, joint} \sim 1-1.6$. All images are built using HSC R, I, and VIDEO H bands. The field of view is $0.6 \times 0.6\,\text{Mpc}^2$. We display the position of the members with membership probabilities of 0.75-1.0, 0.5-0.75, and 0.25-0.5 by green, yellow, and orange circles, respectively. The location of the BCG candidate is represented by a white square (see Sect.\ref{subsection:BCG}). The images are centred at the joint cluster position. The full membership extends beyond the FoV.}
    \label{figure:RGB_medium_z}   
\end{figure*}

\begin{figure*}[t]
    \centering
    \captionsetup{format=plain}
    \captionsetup{labelfont=bf}
    \includegraphics[width=4.25cm]{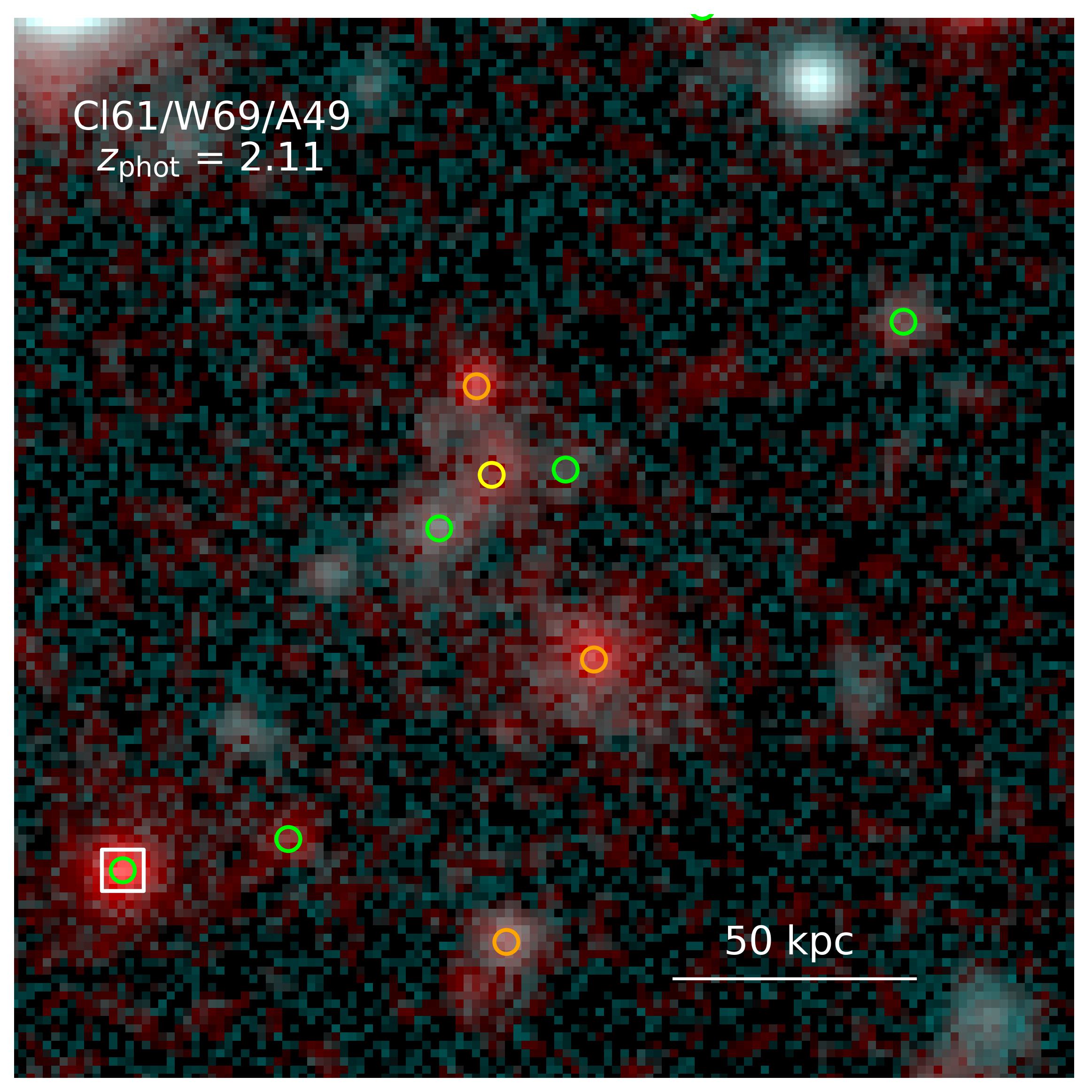}    
    \includegraphics[width=4.25cm]{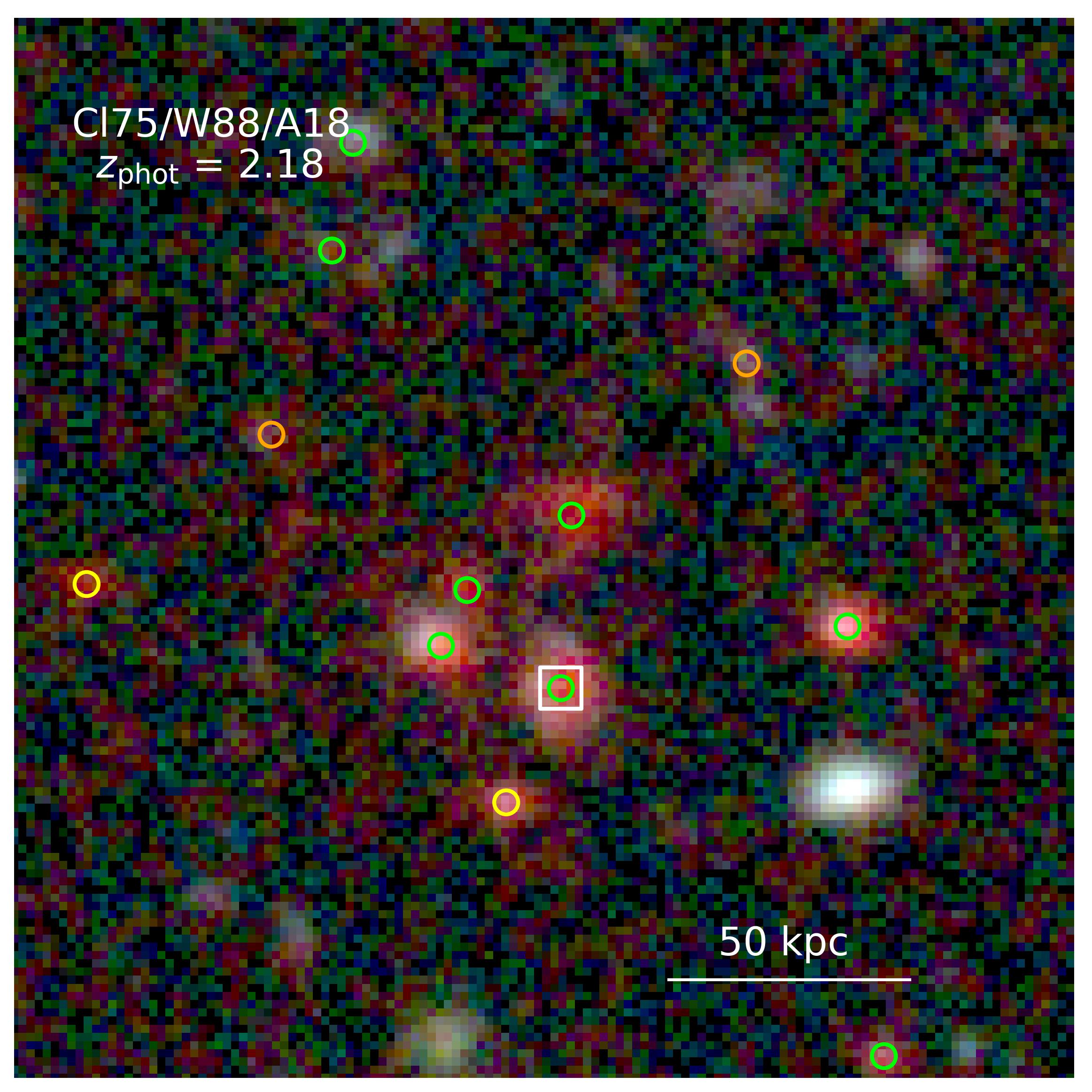} 
    \includegraphics[width=4.25cm]{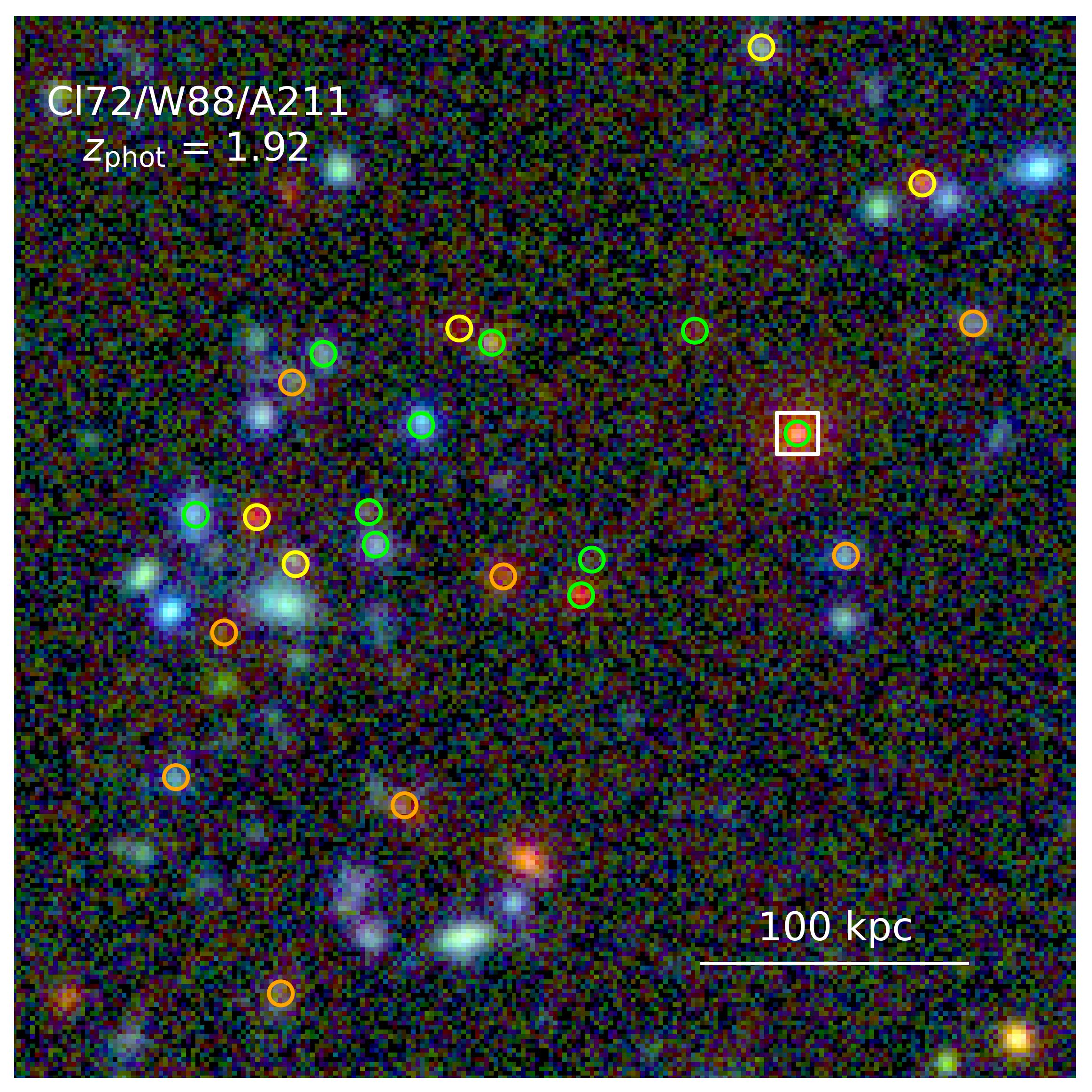}    
    \includegraphics[width=4.25cm]{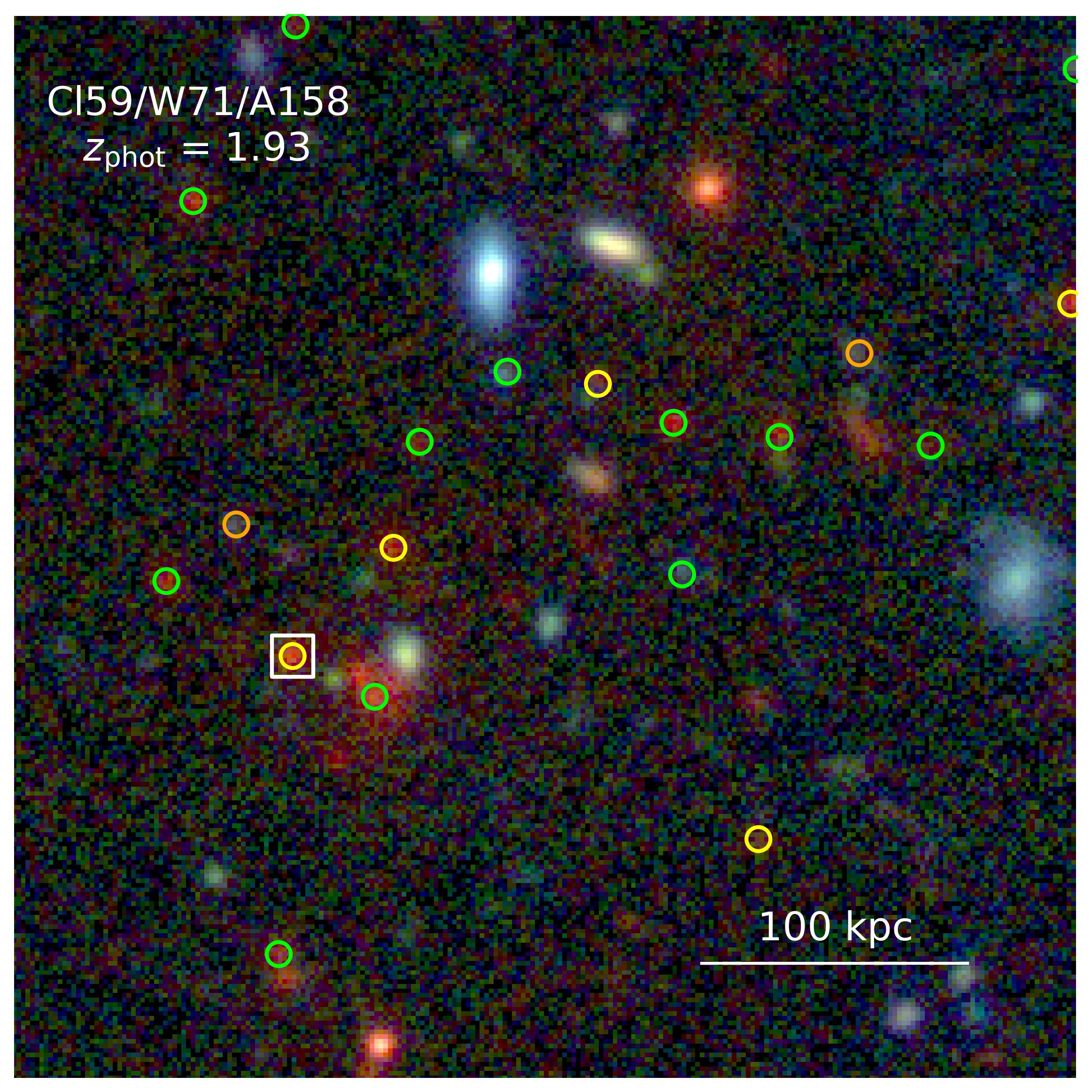}    
    
    \caption{Same as Fig.\ref{figure:RGB_medium_z} but for two XMM-LSS (left) and two CDFS (right) cluster candidates from the joint catalogue at $z_\text{phot, joint} \sim 2$. The FoV was adapted for each candidate to avoid perturbation of the image dynamic by bright foreground sources. The physical scale is indicated by the white line on the bottom of each image. }
    \label{figure:RGB_high_z}   
\end{figure*}

\section{Building of a AMICO-WaZP joint cluster sample } 
\label{section:GS_construction}
    
\noindent From the raw lists of AMICO and WaZP detections, we aimed to construct a sample of robust cluster candidates, identified by both cluster finders. 
Although they both rely on the same photometric redshifts, as they adopt very different assumptions, galaxies are combined in different ways, based on different selection criteria and weights. Thus, the overdensities identified by both algorithms are less likely to be spurious detections.

To cross-match the two samples, we considered a subset of highest SNR detections, ensuring an equal cluster density between the two cluster finders. While high SNR cuts lead to higher purities, lowering the SNR threshold improves completeness. 
To maximise purity while achieving sufficient completeness at high redshift, a good compromise is reached for an initial detection density of $120\text{ deg}^{-2}$ in both fields. This choice was justified by the comparison to galaxy mock catalogues performed in the context of the Euclid survey with H-band depth and photometric redshift uncertainties similar to the VIDEO fields studied here. From the first panel of fig.~8 of \citealt{Adam_2019}, a density of $120\text{ deg}^{-2}$ cluster detections between redshifts $0$ and $2$ leads to purity levels between $85~\%$ and $90~\%$ for WaZP and AMICO. Therefore, we expect our joint catalogue to reach a high level of purity. In addition to this statistical argument, a careful visual inspection of the entire joint sample is performed in the following. A cluster density of $120\text{ deg}^{-2}$ is achieved with the following SNR thresholds:

\begin{itemize}
     \item[-] In the XMM-LSS:
     \begin{itemize}
         \item[] SNR $\geq 5.4$ for WaZP
         \item[] SNR $\geq 11.5$ for AMICO
     \end{itemize}
     \item[-]  In the CDFS: 
        \begin{itemize}
            \item[] SNR $\geq 5.2$ for WaZP
            \item[] SNR $\geq 10.7$ for AMICO
        \end{itemize}
 \end{itemize}

\noindent which correspond initially to $550$ and $500$ AMICO and WaZP detections in the XMM-LSS and CDFS, respectively. Note that the different absolute values of SNR are due to the different SNR definitions used by the two algorithms.

The WaZP-AMICO cross-match was performed with the publicly available \texttt{Python} library \texttt{ClEvaR}\footnote{https://github.com/LSSTDESC/ClEvaR}. The adopted strategy was a two-way cylindrical matching within a given physical radius and redshift offset length.
The two cluster lists were first ranked by decreasing order of SNR. Matched clusters were removed from the list. If a cluster had more than one possible counterpart, the one with the highest SNR was selected. The two matching parameters were empirically determined in two steps to minimise random associations.

We first matched detections based only on their angular separation, with a very small maximum offset of 150~kpc, and examined in Fig.~\ref{figure:bias_std_matching} the resulting redshift offset distribution, $\varepsilon_\text{cl} = (z_\text{AMICO}-z_\text{WaZP})/(1+\Bar{z})$, where $\bar{z}$ is the mean redshift of each pair. Taking into account the typical photometric redshift errors, more than $98\%$ of the matches can be considered as physical pairs (i.e. with $\Delta z/(1+z) < 0.15$). 
The mean and standard deviation as a function of the redshift were estimated with the biweight estimator (\citealt{Beers_1990}). We found essentially no relative redshift bias between the two cluster finders up to $z \sim 1.7$ in both fields. 
The mean scatter of the redshift offsets is $ \overline{\sigma_\text{cl}} \sim 0.01$. Its redshift dependence, $\sigma_\text{cl}(z)$, is modelled as a third-order polynomial, shown as a grey envelope in Fig.~\ref{figure:bias_std_matching}.

To estimate the optimal matching radius, we performed a second matching, selecting pairs with a redshift offset within $\pm 2\sigma_\text{cl}(z)(1+z)$ and with a maximum angular separation of $R = 1.0\,\text{Mpc}$. The resulting angular separation distributions are shown in Fig.~\ref{figure:ang_sep_matching}, where more than $90\%$ of the pairs are separated by less than $400$\,kpc. Based on this, we defined our optimal matching radius to be $R = 400$\,kpc. For the matching parameters we derived, knowing the initial density of detection, the probability to have random association between the two cluster finders was estimated to be lower than $1\%$.

Using the parameters previously defined, the matching leaded to $277$ and $250$ pairs of detections in the XMM-LSS and CDFS, respectively, which correspond to $\sim 50\%$ of both samples. A large fraction of the unmatched systems would have a counterpart, but at a SNR lower than the adopted thresholds. Figure~\ref{figure:S/N_matches_vs_all} shows the SNR distribution of the matched and unmatched detections for the two cluster finders. In general, the higher the SNR, the higher the matching rate. There are a few exceptions at high SNR but these can be explained by the scatter between the SNR of the two algorithms, or by pairs with centring or redshift offsets just outside our matching window. 

A systematic visual inspection of the entire sample was performed. Only very few cases appeared problematic. We identified respectively $6$ and $2$ spurious detections in the XMM-LSS and CDFS due to insufficient masking around foreground bright sources or observation artefacts. These were removed from the following analysis. In a few cases it was also noticed the presence of nearby masked regions that could potentially impact the centring or other measured characteristics associated to the detections. To take this into account, we computed for each candidate the fraction of area within a $1.0$\,Mpc disk centred at the cluster position that is effectively covered by unmasked data around.

Our final working cluster catalogue is therefore composed of $271$ and $248$ cluster candidates in the XMM-LSS and CDFS, respectively. In particular, $31$ (XMM-LSS) and $43$ (CDFS) detections are found at redshifts beyond $1.5$. The redshift and spatial distributions of the joint AMICO-WaZP detections are shown in Fig.~\ref{figure:GS_zdistri_CF} and Fig.~\ref{figure:GS_spatial_distri}, respectively. These latter show a homogeneous coverage over the two footprints, with no suspicious overdensities in redshift or space. We represent in Fig.~\ref{figure:GS_spatial_distri} the limit of the UDS footprint by a red rectangle. This is the region that will be covered by the MOONRISE survey in the XMM-LSS (see Sect.\ref{section:MOONRISE} for more details). Finally, in Fig.~\ref{figure:RGB_medium_z}, we show six examples of high SNR cluster candidates at $z_\mathrm{phot, joint} \sim 1-1.6$ and four examples of high SNR cluster candidates at $z_\text{phot, joint} \sim 2$ in Fig.~\ref{figure:RGB_high_z}. They all present a significant concentration of galaxies. All of them are in the domain of redshift that will be covered by MOONRISE, showing good examples of the kind of systems that could be identified by MOONS.   

In the following, each detection of the joint sample is assigned a new centring, RA$_\text{joint}$ and Dec$_\text{joint}$, and a new redshift, $z_\text{phot, joint}$, defined as the average of the centrings and redshifts provided by each cluster finder. The joint sample detections are labelled with the prefix 'Cl', while their WaZP and AMICO identifiers are labelled, respectively, with the prefixes 'W' and 'A'. In this study, the membership probabilities that are used are coming by default from the WaZP algorithm, except when stated explicitly. Extracts of the XMM-LSS and CDFS catalogues can be found in Table\,\ref{table:catalogue_xmm} and \ref{table:catalogue_cdfs} in Appendix\,\ref{appendix:extract_catalogues}. For each candidate we provide its ID (Col. 1), its joint RA-Dec coordinates (Col. 2-3), its joint $z_\text{phot}$ (Col. 4), its WaZP and AMICO richnesses (Col. 5-6), and SNR (Col. 7-8). If exists, its spectroscopic redshift estimation, and the number of members used to compute it are shown in Col. 9-10 (see Sect. \ref{subsection:zspec_estimation} for more details). If exists, the spectroscopic redshift derived from its BCG is shown in Col.11 (see Sect.\ref{subsection:BCG} and Sect.\ref{subsection:zspec_estimation} for more details). Finally, the coverage fraction is shown in Col. 12. 

\begin{figure*}[t]
    \centering
    \captionsetup{format=plain}
    \captionsetup{labelfont=bf}
    \includegraphics[width=5.66cm]{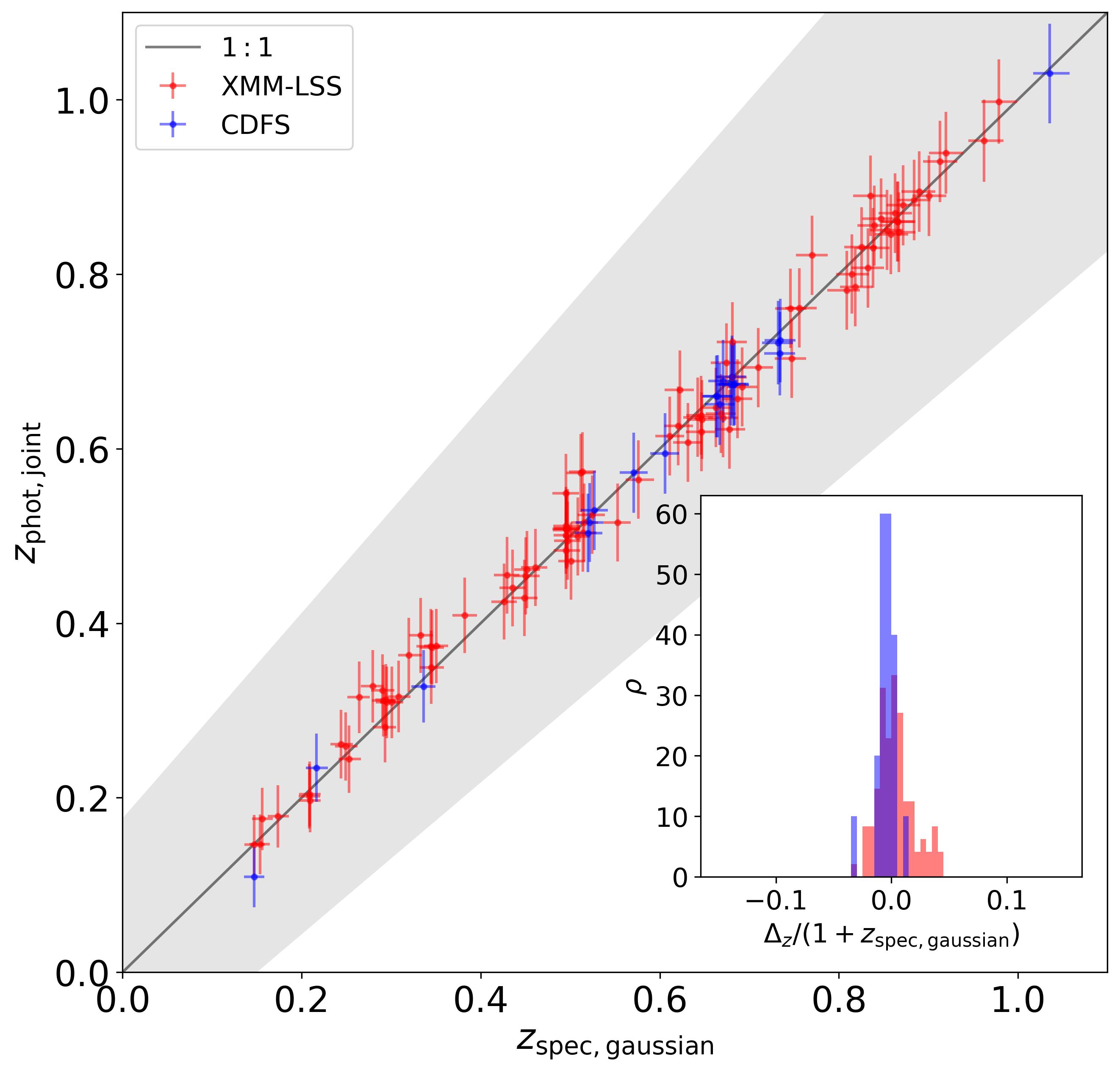}  
    \includegraphics[width=5.66cm]{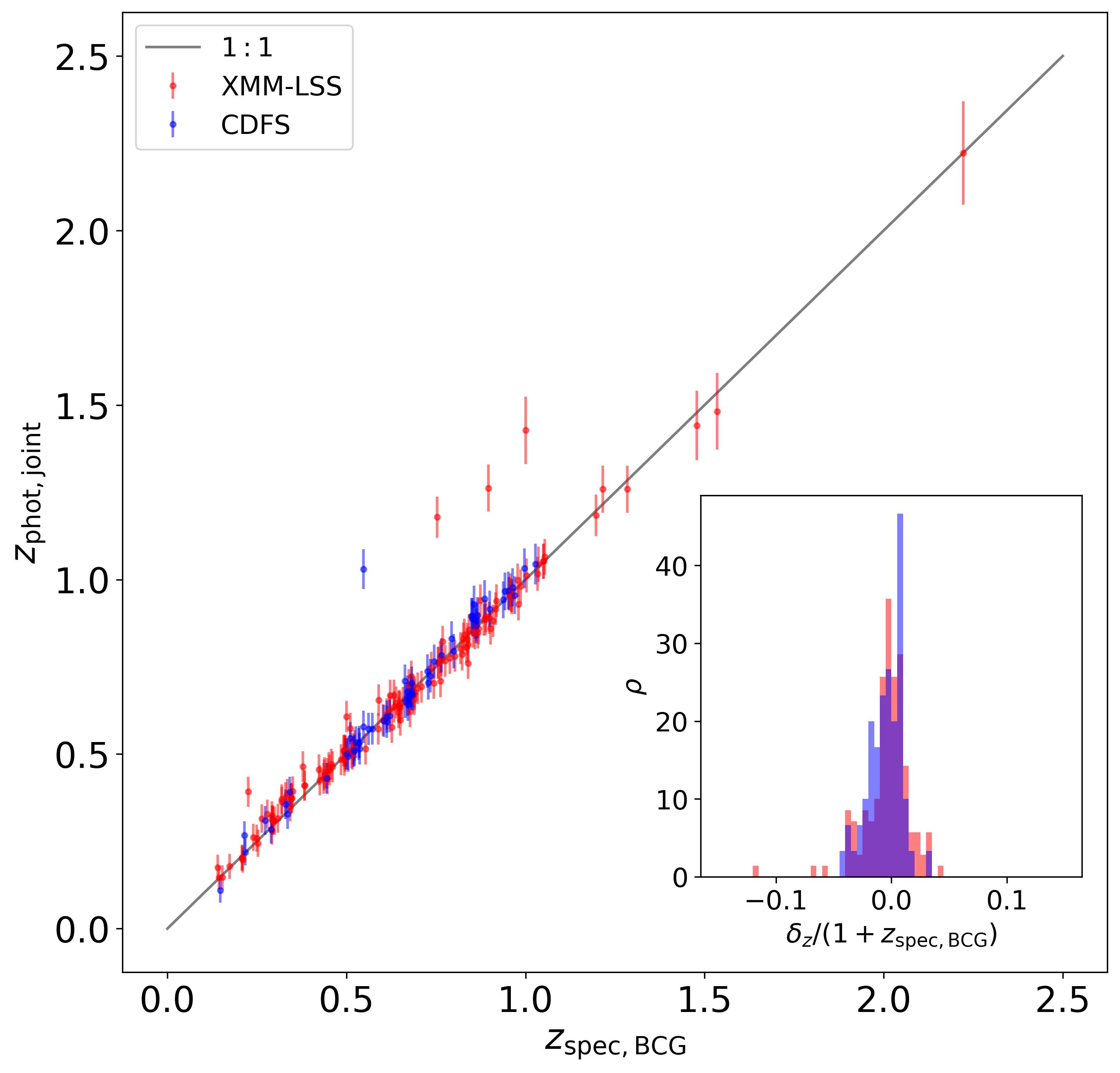} 
    \includegraphics[width=5.66cm]{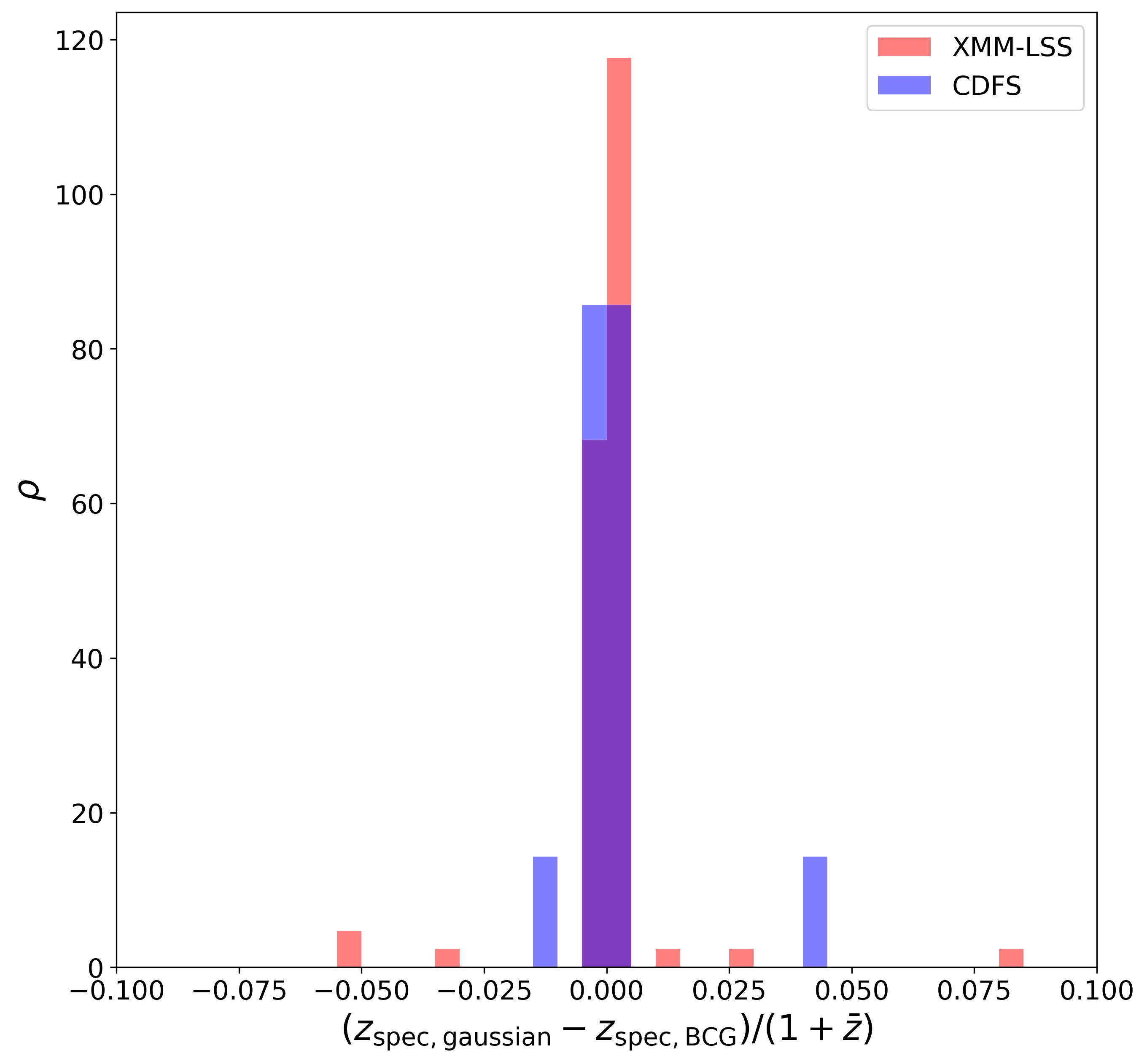}    
    \caption{\textit{Left}: Relation between $z_\text{phot, joint}$ and the estimated $z_\text{spec, Gaussian}$. The x-axis error bars represent the $\pm 3000 \text{ km}\cdot\text{s}^{-1}$ rest-frame selection window, while the y-axis error bars are derived from the galaxies photometric redshift metrics ($1\sigma$ confidence level). The inset panel shows the probability distribution function of the offset $(z_\text{phot, joint}-z_\text{spec, Gaussian})/(1+z_\text{spec, Gaussian})$ between the two redshift measurements. The grey area represents the $\vert(z_\text{spec} - z_\text{phot})/(1+z_\text{phot})\vert < 0.15$ selection window. 
    \textit{Middle}: Relation between $z_\text{phot, joint}$ and  $z_\text{spec, BCG}$. The y-axis error bars are derived from the galaxies photometric redshift metrics ($1\sigma$ confidence level). The inset panel shows the probability distribution function of the offset $(z_\text{phot, joint}-z_\text{spec, BCG})/(1+z_\text{spec, BCG})$ between the two redshift measurements. 
    \textit{Right}: Distribution of the scatter between the $z_\text{spec}$ estimation from statistics over the members and the one derived from the BCG candidates for the $100$ cluster candidates with both measurements available. The x-axis is limited to the range $[-0.1, 0.1]$ for a purpose of readability. One BCG candidate with a large redshift offset ($\sim 0.27$), likely caused by a catastrophic photometric redshift estimate, lies outside the selection window and is therefore not shown in the figure.}
    \label{figure:GS_zphot_zspec}   
\end{figure*}

\begin{figure}[t]
    \centering
    \captionsetup{labelfont=bf}
    \includegraphics[width=0.48\textwidth]{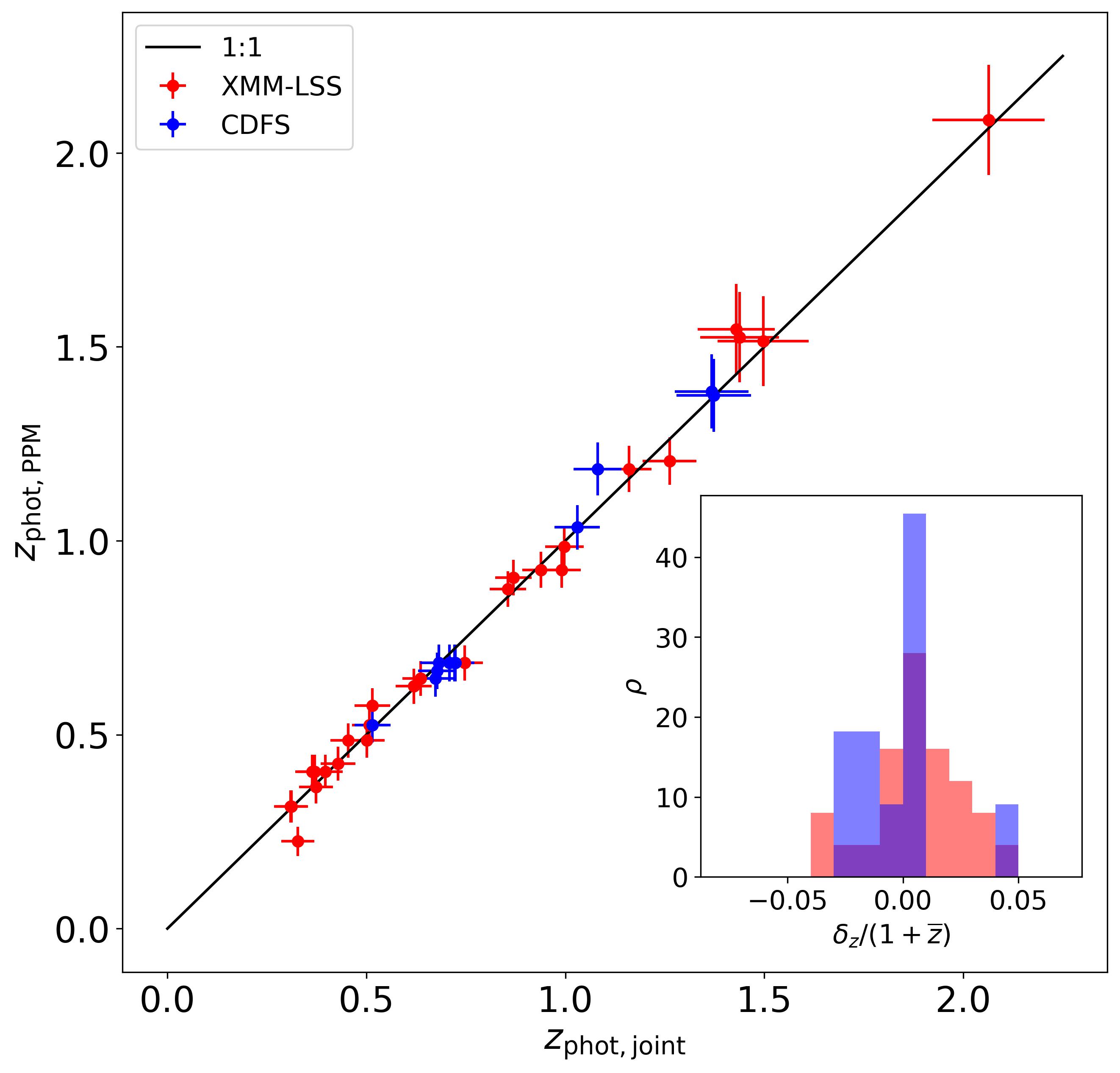}    
    \caption{Relation between the redshift of PPM overdensities and cluster candidates for the associated pairs in the XMM-LSS (red) and CDFS (blue). The inset panel shows the probability distribution function of the offset $(z_\text{phot, PPM}-z_\text{phot, joint})/(1+\bar{z})$. We observe an overall good agreement between the redshifts of the matched candidates.}
    \label{figure:GS-PPM_z}   
\end{figure}

\begin{figure}[t]
    \centering
    \captionsetup{labelfont=bf}
    \includegraphics[width=0.48\textwidth]{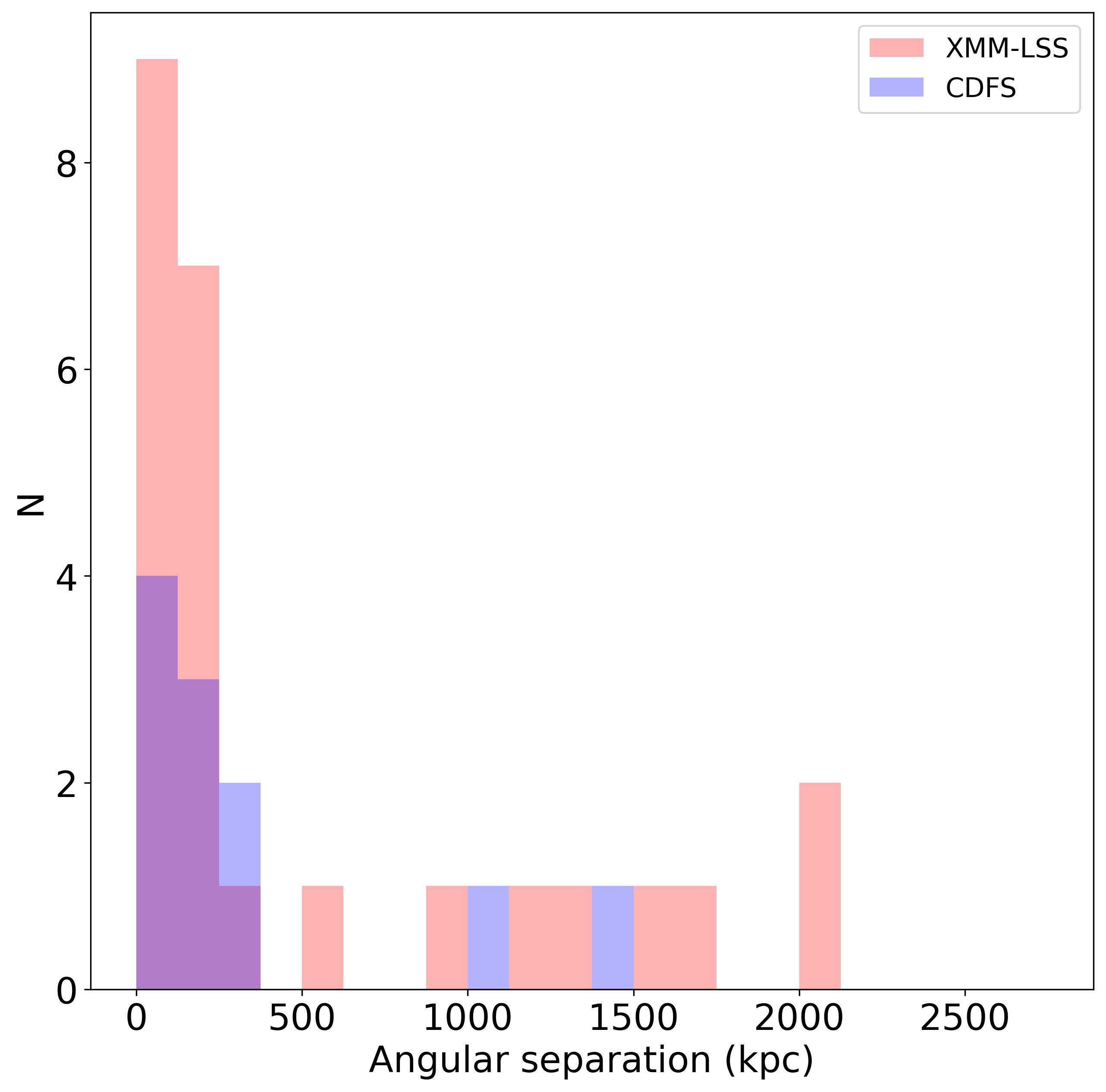}    
    \caption{Distribution of the angular separation between the cluster candidates and their associated PPM overdensities. We used the WaZP cluster centre definition here, as the matching is based on the WaZP membership.}
    \label{figure:GS-PPM_angsep}   
\end{figure}

\begin{figure}[t]
    \centering
    \captionsetup{labelfont=bf}
    \includegraphics[width=0.48\textwidth]{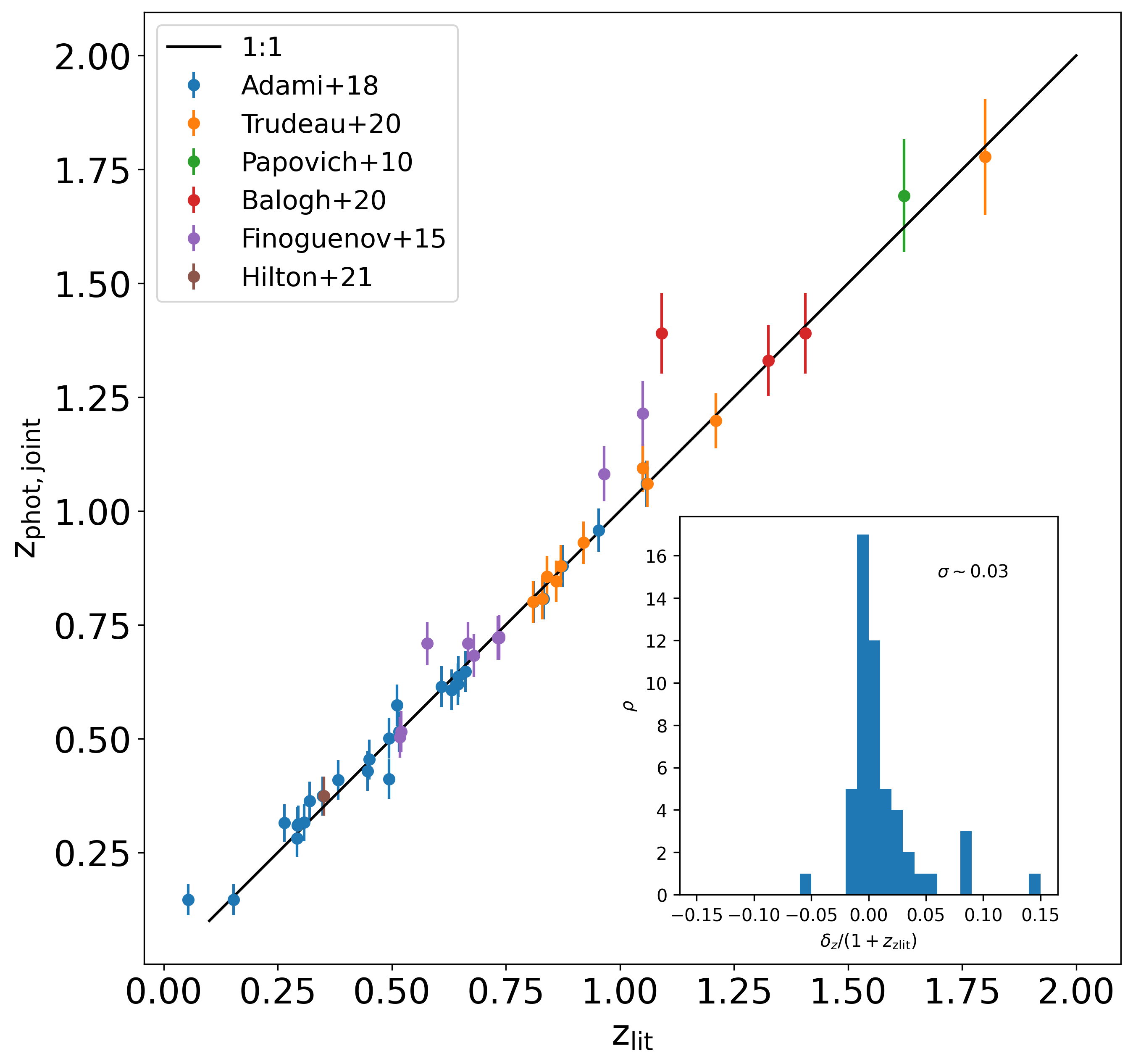}
    \caption{Relation between the spectroscopic redshift from the literature, $z_\text{lit}$, and the mean AMICO-WaZP redshift, $z_\text{phot, joint}$, for candidates that match with the ancillary cluster catalogues. The inset panel shows the distribution of the offset $(z_\text{phot, joint}-z_\text{lit})/(1+z_\text{lit})$. The error bars are derived from the galaxies photometric redshift metrics, at $1\sigma$ level. The standard deviation of the relation is approximately $\sigma \sim 0.03$, highlighting good agreements between our results and those from the literatures.}
    \label{figure:GS_anc}   
\end{figure}

\begin{figure*}[t]
    \centering
    \captionsetup{format=plain}
    \captionsetup{labelfont=bf}
    \includegraphics[width=5.66cm]{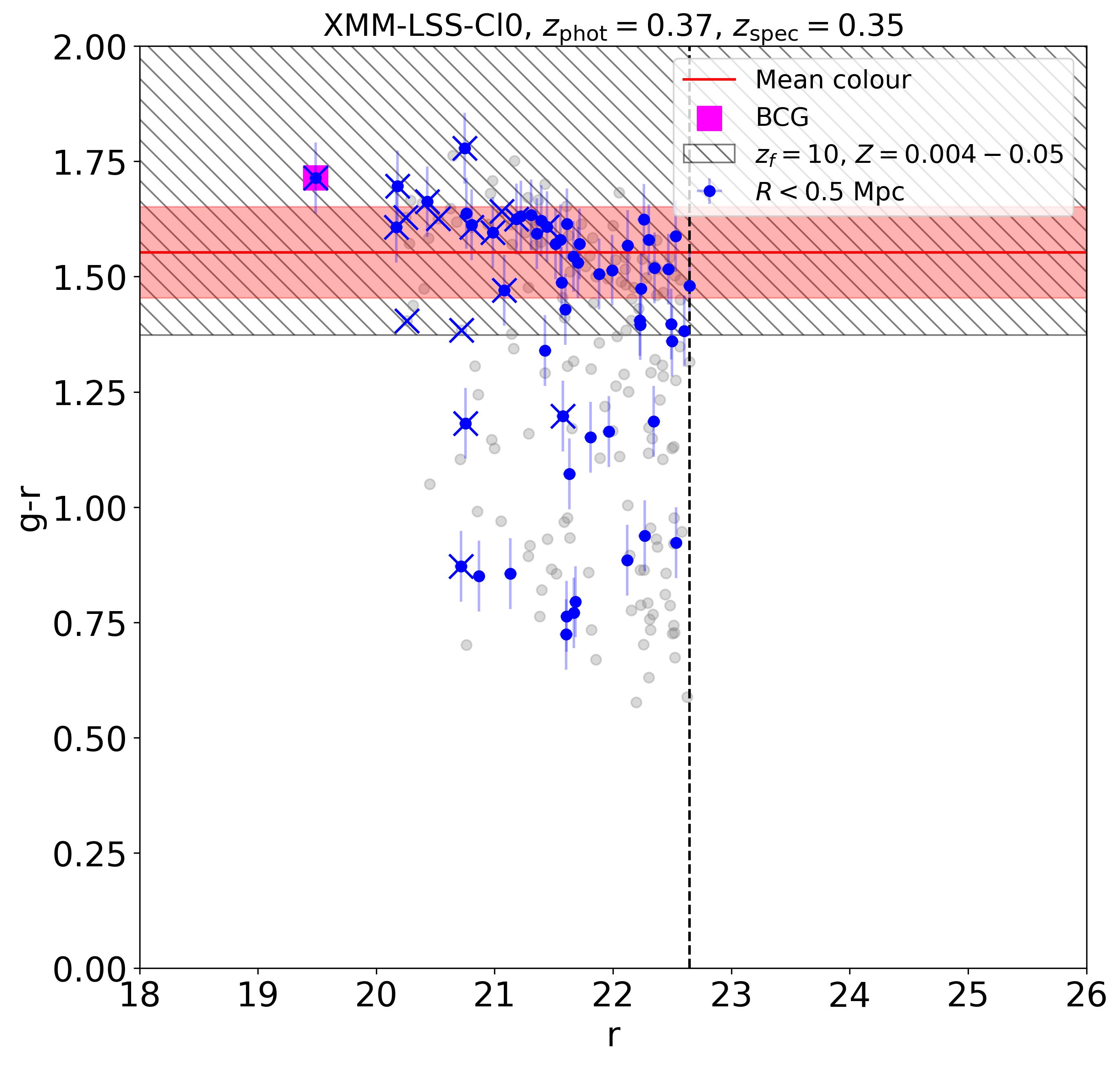}
    \includegraphics[width=5.66cm]{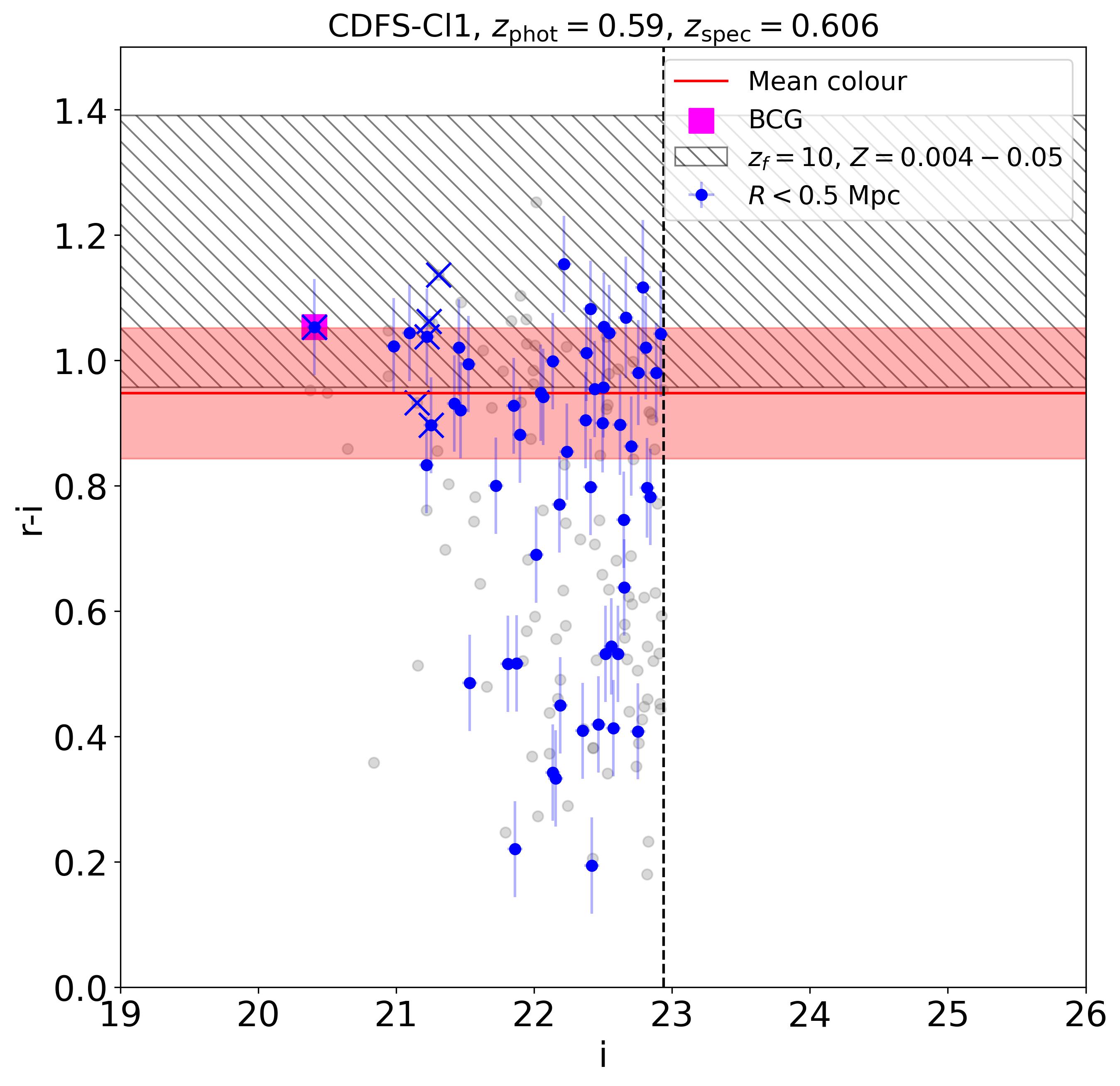}    
    \includegraphics[width=5.66cm]{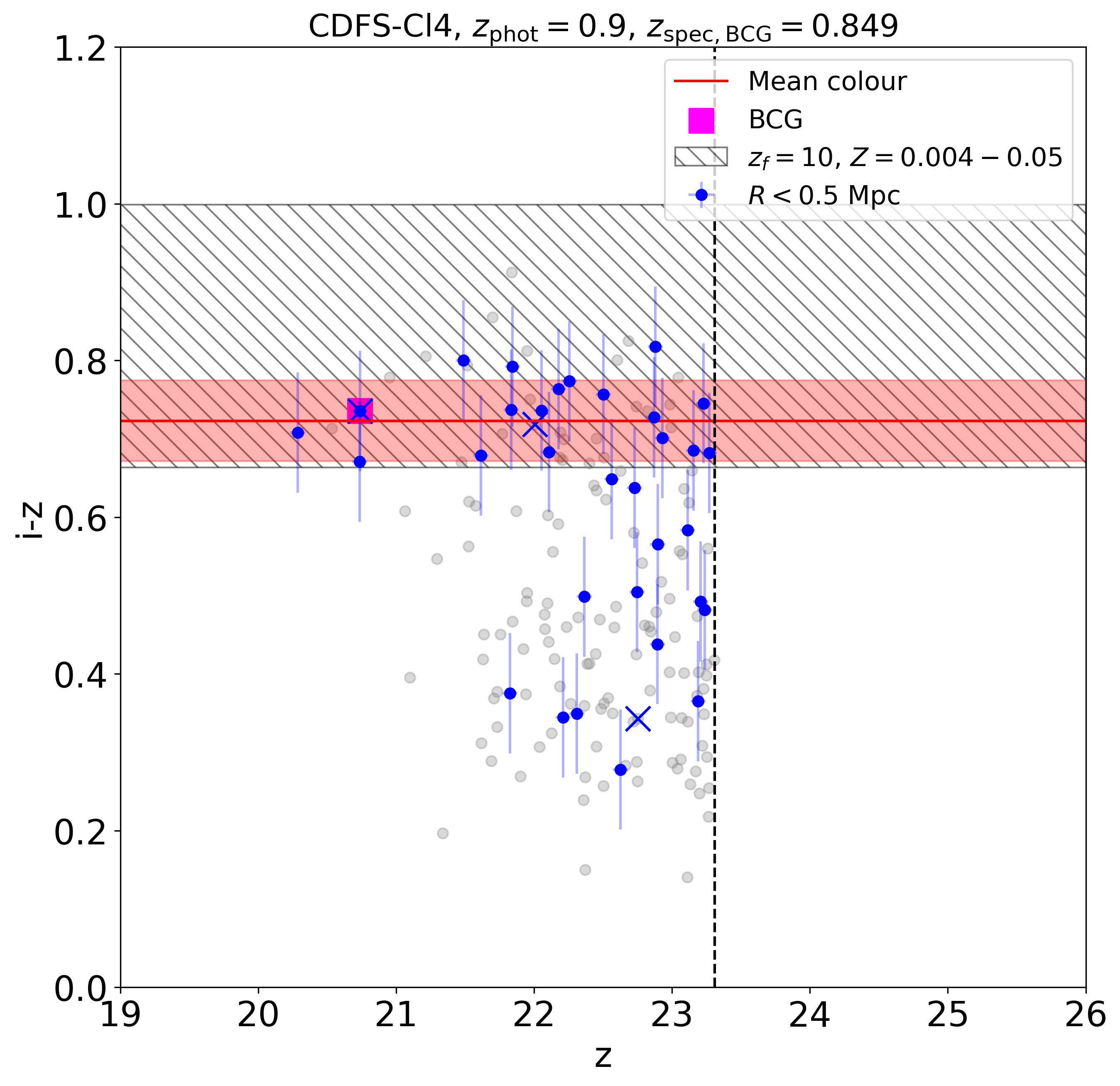} 

    \includegraphics[width=5.66cm]{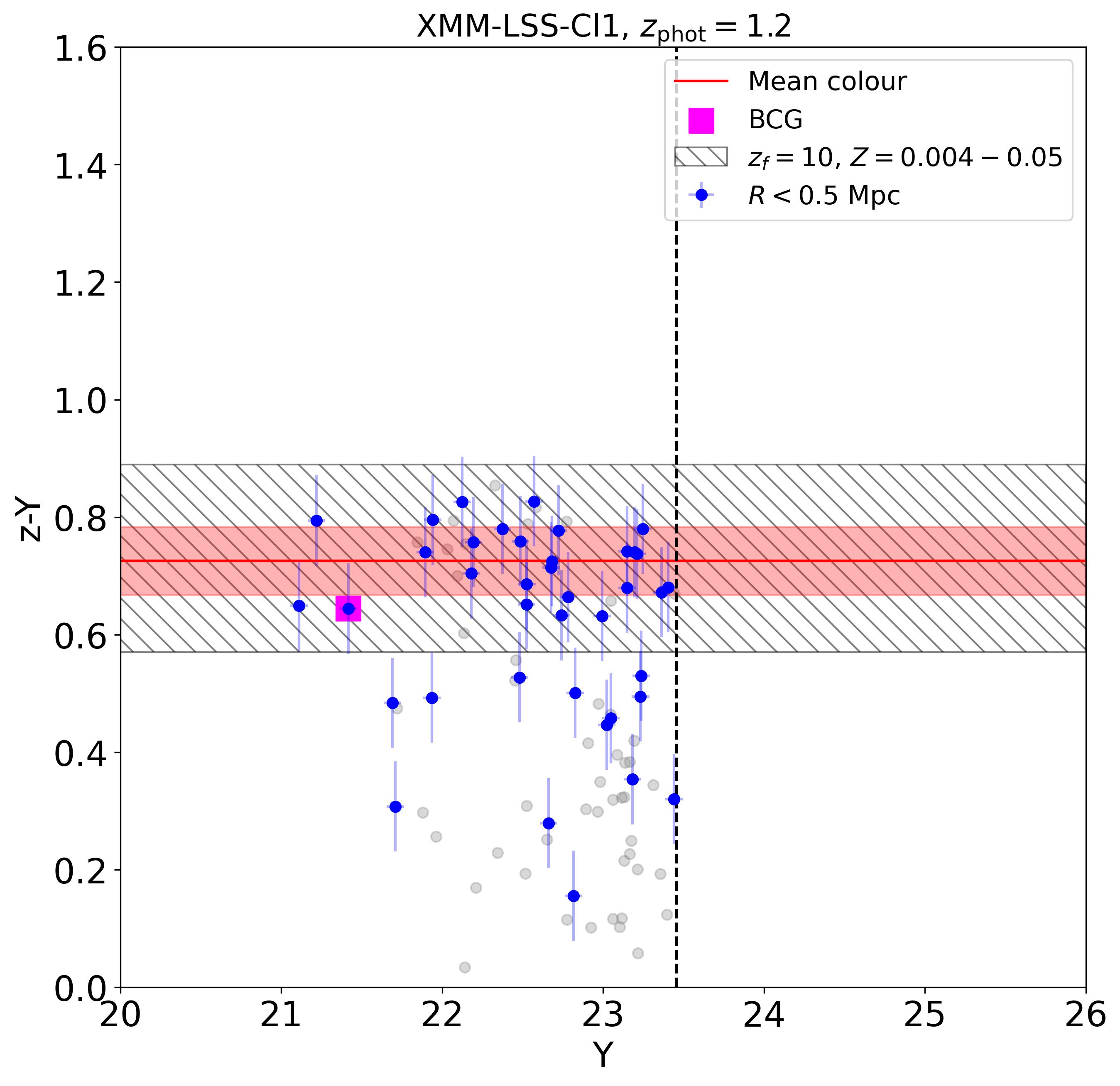}
    \includegraphics[width=5.66cm]{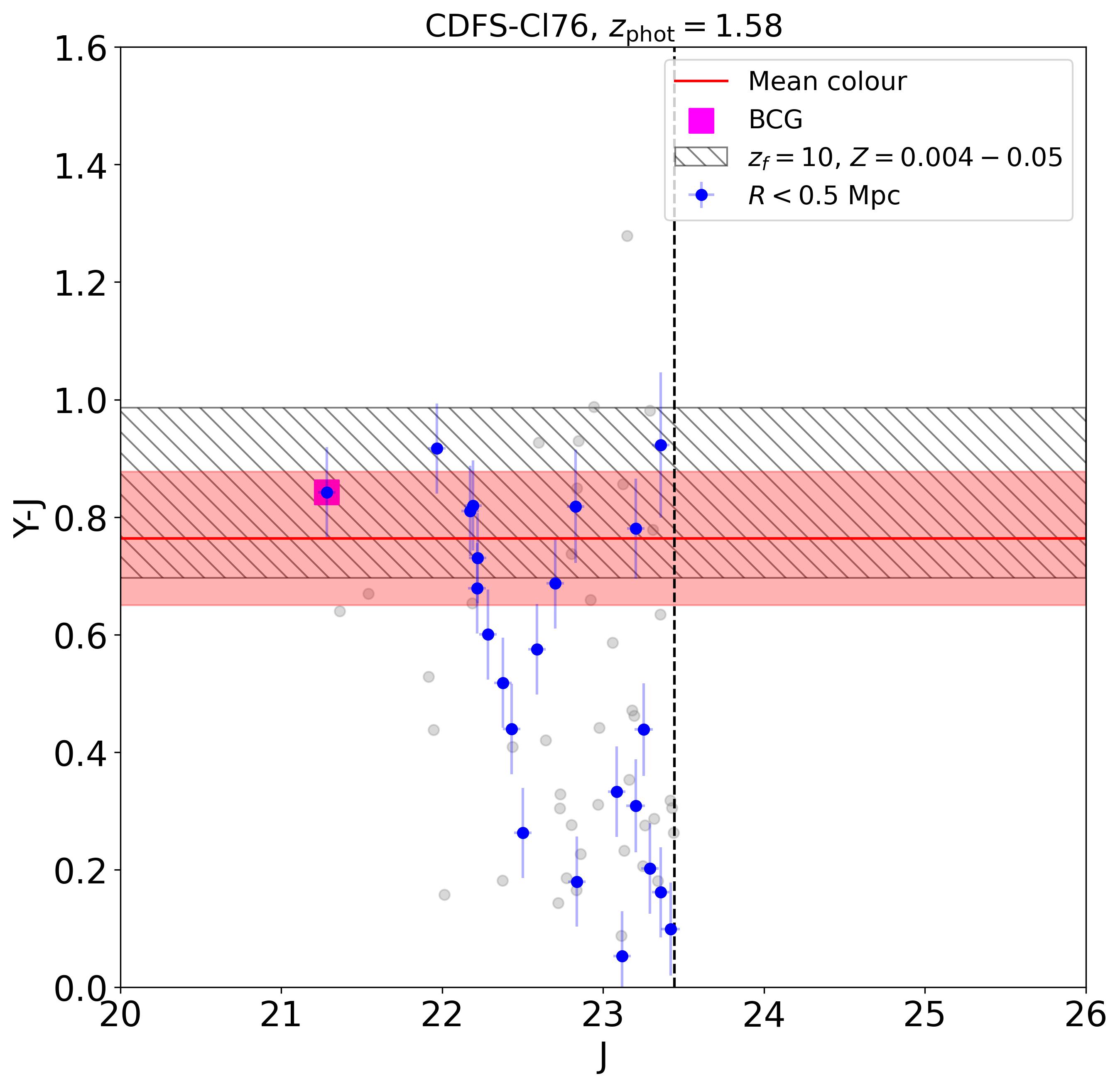} 
    \includegraphics[width=5.66cm]{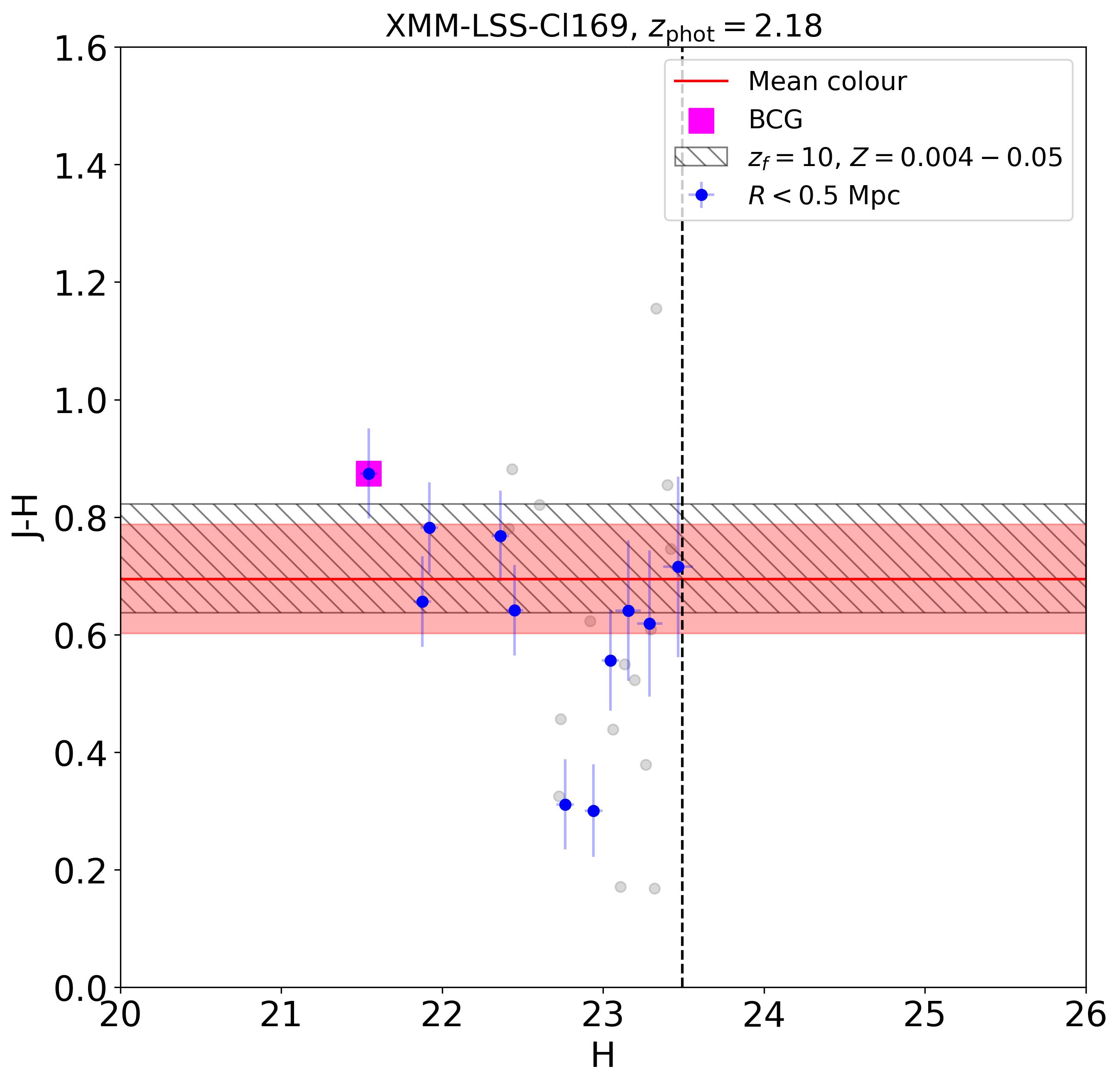}  
    
    \caption{Examples of Colour-magnitude diagrams (CMDs) for candidates at different redshifts. Each example corresponds to one combination of filters, as described in Table~\ref{table:CMD_color_selection}. The grey points represent all members, while the blue ones represent the members selected within $0.5$ Mpc. The red line and the envelope represent the mean colour of the red ridge and its associated  standard deviation derived in Sect. \ref{section:red-sequence}. The vertical dashed black line represents the applied magnitude cut. We compare the mean colour to a model of galaxy evolution (grey dashed region) with a redshift of formation at $z_f = 10$, for the range of metallicity $Z = 0.004-0.05$. We refer to Sect.~\ref{section:red-sequence} for more details. For candidates with spectroscopic redshift estimation computed in Sect.~\ref{subsection:zspec_estimation}, we indicate by a blue cross members within $R = 1.0$\,Mpc and within a window of $\Delta v = \pm 3000 \text{ km}\cdot \text{s}^{-1}$ around the cluster's $z_\text{spec}$. If the spectroscopic redshift estimation is derived from the BCG, we specify it in the title of the panel.}
    \label{figure:CMD_examples}   
\end{figure*}

\begin{figure}[t]
    \centering
    \captionsetup{labelfont=bf}
    \includegraphics[width=0.48\textwidth]{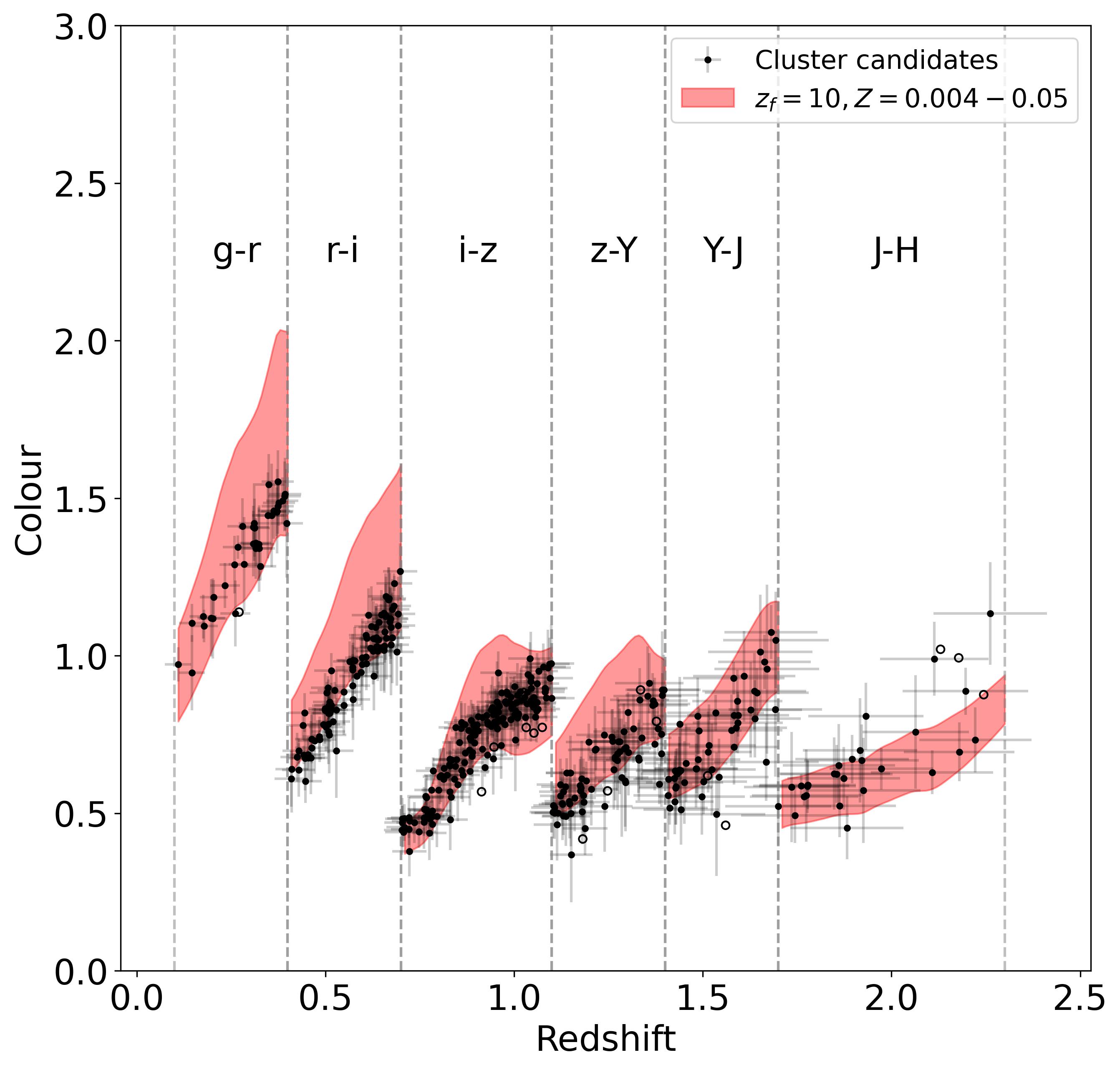}   
    \caption{Evolution of the apparent mean colour of the red-sequence for the AMICO-WaZP joint candidates as a function of redshift. The coloured envelopes represent the expected colours derived from models within the metallicity range $Z \in [0.004, 0.05]$ for a formation redshift at $ z_f = 10$, for a folding time of $\tau_f = 0.75$ Gyr, without any subsequent burst of star formation. The error bars for the x-axis represent the error on the photometric redshift estimated in Sect.~\ref{subsection:zphot_quality}. Open circles represent the $16$ cases that were rejected after visual inspection, for which the red-sequence was too poor to be correctly identified. The limits of the different redshift ranges associated with each set of filters are represented as dashed lines.}
    \label{figure:RS_mean_colour}   
\end{figure}

\begin{figure}[t]
    \centering
    \captionsetup{labelfont=bf}
    \includegraphics[width=0.48\textwidth]{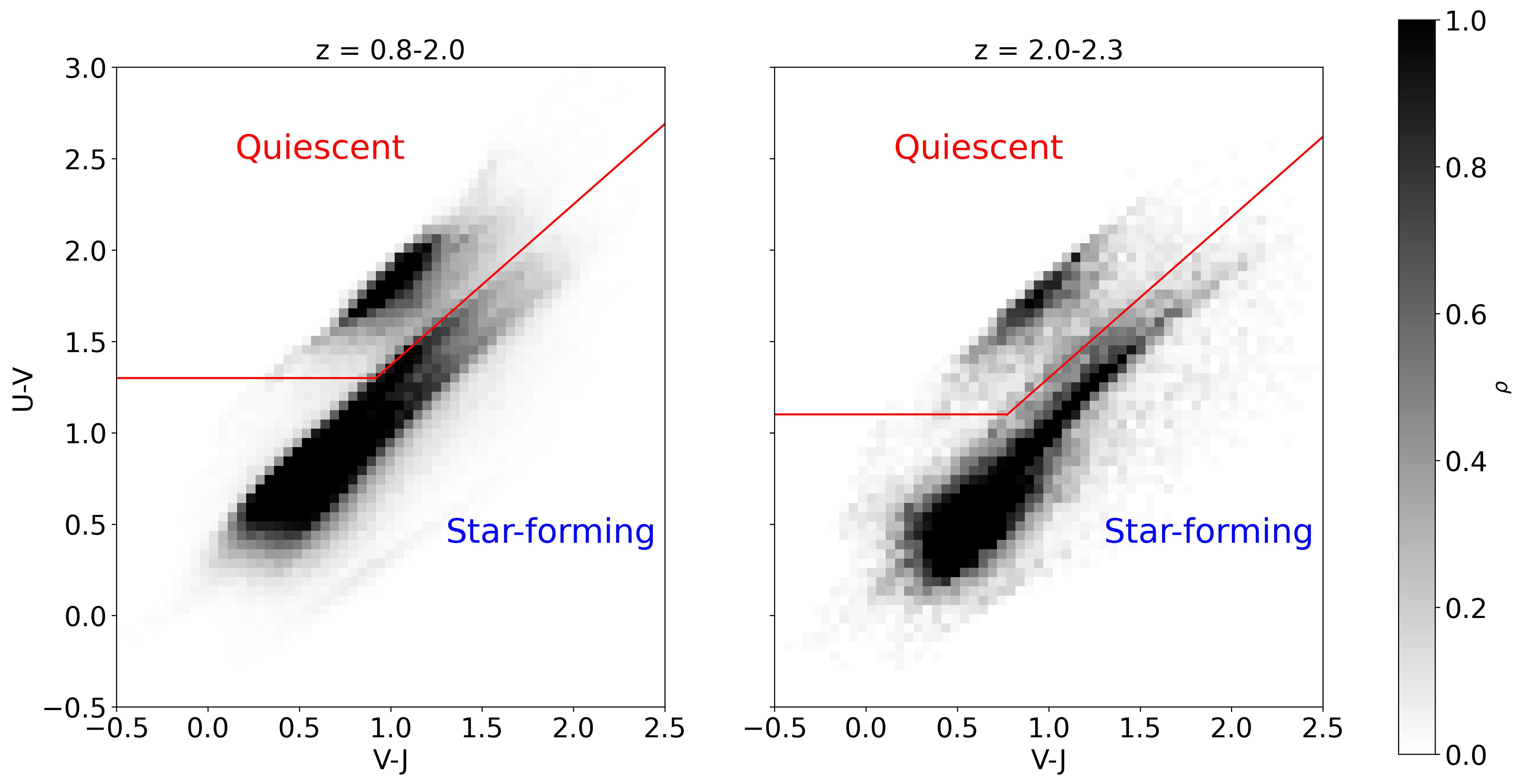}
    \caption{Stacked UVJ diagram of galaxies with $z_\text{phot} \in [0.8-2.0]$ (left) and $z_\text{phot} \in [2.0-2.3]$. The red line represents the limit between quiescent and star-forming galaxies empirically derived, as defined by Eqs. \ref{equation:active_passive_crit_lowz} and \ref{equation:active_passive_crit_highz}.}
    \label{figure:UVJ_stacked}   
\end{figure}

\section{Analysis of the joint WaZP-AMICO sample}
\label{section:GS_analysis}

\subsection{Identification of brightest cluster galaxies}
\label{subsection:BCG}
For all cluster candidates, we aimed to identify their likely brightest cluster galaxies (BCGs). BCGs are generally found very close to the centre of their parent clusters, and even in case of major cluster merger, they are not expected to lie further than a few hundred of kpc away (see, for example, \citealt{Mann_2012}). We identified, for each cluster, the brightest member (in $m_\text{H}$ ) within $400$ kpc from the centre as a likely BCG candidate, here referred to as the brightest central member (BCM). This threshold was derived from the global distribution of the offset between the cluster centre and the brightest members, shown in Fig.~\ref{figure:bcg_ang_sep}. In most cases ($\sim 65\%$), the overall brightest cluster member is located within 400 kpc from the cluster centre. For those beyond $400$~kpc, they are likely to be outliers, with a median membership probability of $\sim 0.3$. In addition, the median difference between the $m_\text{H}$ of the BCM and of the brightest cluster member is $\sim 0.33$, with a BCM among the top three brightest members in $\sim 91 \%$ of the cases. Thus, in order to avoid contamination from potential outliers and as BCGs are expected to be close to the centre of their parent clusters, except in cases of massive merging events, it appeared reasonable to choose the BCMs as the most likely BCG candidates. Comparison with memberships from AMICO showed similar results, where the BCG candidates we identified were mostly in common with the WaZP one, except for a few cases caused by differences in terms of membership computation or angular separation between the different cluster centroids.

\subsection{Spectroscopic redshift estimation.} 
\label{subsection:zspec_estimation}

In this section, we use our compilation of spectroscopic redshifts, described in Sect.~\ref{subsection:zspec_cat}, to estimate the spectroscopic redshifts of the cluster detections when such data are available. Spectroscopic redshifts from the 3D-HST and FIGS surveys were excluded from the analysis, as their corresponding measurement errors are of the order of typical cluster velocity dispersion ($\sigma_v \sim 1000\text{ km}\cdot\text{s}^{-1}$).

We first used the spectroscopic redshift distribution of the identified member galaxies to infer the cluster redshift. 
The main difficulty, particularly at high redshifts, where several concentrations may appear in the spectroscopic redshift distribution, is to identify which one corresponds to the detection made by the cluster finder in photometric redshift space. To mitigate this issue, we restricted the analysis to galaxies identified as members and limited it to a cylindrical region centred on the coordinates of the cluster detection with a radius of 1\,Mpc, and with an initial spectroscopic window $|(z_\text{spec}-z_\text{phot, joint})/(1+z_\text{phot, joint})| \leq 0.15$, where $z_\text{phot, joint}$ is the photometric redshift of the cluster detection. Focussing on the inner region of the cluster helps minimise outlier contamination in the outskirts of the cluster. The use of a large initial redshift-space window is intended to account for the bias and scatter previously observed between photometric and spectroscopic redshifts. 
We convolved the $z_\text{spec}$ distribution based on a Gaussian kernel density estimation of bandwidth $0.005(1+z_\text{phot, joint})$ and identified the main concentration in the smoothed $z_\text{spec}$ distribution as the location of the highest peak. The large value for the bandwidth (corresponding to the velocity dispersion of a rich cluster) has been chosen in order to erase peaks in the distribution caused by small clumps along the line of sight. We then selected a rest-frame window of $\pm 3000 \text{ km}\cdot \text{s}^{-1}$ around the main concentration and estimated the cluster redshift $z_\text{spec, gaussian}$ using the location bi-weight estimator from \citealt{Beers_1990}, shown to be robust to non-Gaussian  underlying populations (thus to interloper contamination). Only measurements with more than five members were kept in the following analysis.

Following this process, we estimated spectroscopic redshift measurements for $96$ and $20$ cluster candidates in the XMM-LSS and CDFS, respectively. The left panel of Fig.~\ref{figure:GS_zphot_zspec} shows the $z_\text{spec, gaussian}$ - 
$z_\text{phot, joint}$ relation for these $116$ cluster candidates, together with the 1:1 relation. 
The mean scatter of the points throughout the redshift range with respect to the 1:1 relation is estimated to be $\sim 0.01$–$0.02$, indicating a good overall agreement between the spectroscopic and photometric redshift estimates.

As a second approach to estimate the spectroscopic redshift of our cluster candidates, we used the spectroscopic redshift of the BCGs candidate when available. 
Among the BCG candidates we identified previously, $204$ have $z_\text{spec}$ measurements. The central panel of Fig.~\ref{figure:GS_zphot_zspec} shows the relation between the BCGs $z_\text{spec}$ and the $z_\text{phot, joint}$ of their parent clusters. First, the four outliers above the 1:1 relation correspond to galaxies with catastrophic photometric redshifts. The inset histogram shows the distribution of the offset $(z_\text{phot, joint}-z_\text{spec, Gaussian})/(1+z_\text{spec, Gaussian})$ between the joint photometric redshift and the spectroscopic ones of the BCGs, without the four catastrophic cases. 
The standard deviation of the $(z_\text{phot, joint}-z_\text{spec, BCG})/(1+z_\text{spec, BCG})$ distribution is comparable to the one for cluster spectroscopic redshift estimation with a Gaussian kernel, with $\sigma \sim 0.02$ in both fields. There are $85$ and $15$ cluster candidates with $z_\text{spec}$ estimation from both BCGs and statistics over memberships in the XMM-LSS and CDFS, respectively. The right panel of Fig.~\ref{figure:GS_zphot_zspec} shows the distribution of $(z_\text{spec, Gaussian}-z_\text{spec, BCG})/(1+\bar{z})$, with $\bar{z} = (z_\text{spec, Gaussian}+z_\text{spec, BCG})/2$. The standard deviation of the relationship between these two estimators of clusters $z_\text{spec}$ is $\sim 0.001$, showing the general very good agreement that we have between these two estimators. We have nine cluster candidates where the offset between the $z_\text{spec, BCG}$ and the $z_\text{spec, Gaussian}$ is larger than $0.01$. One is due to the catastrophic photometric redshift of its BCG, with $z_\text{phot} = 1.03$ but $z_\text{spec} = 0.547$, making it a foreground galaxy. For the remaining ones, their photometric redshift measurements are all compatible with their spectroscopic ones. However, compared to the photometric redshift of their parent clusters, their $z_\text{phot}$ is slightly shifted, with offsets between $0.01$ and $0.06$, in absolute values.
We conclude that these galaxies may possibly be foreground or background objects, and so could be less likely BCGs candidates.

\subsection{Search for overdensities around radio sources}

\noindent Instead of systematic, blind, searches for cluster candidates, another approach is to search for overdensities around some subclasses of galaxies which are often considered as signposts of galaxy clusters and galaxy overdensities in general, such as radio emitting galaxies (e.g. \citealt{Wylezalek_2013}, \citealt{castignani_2014b}), or sub-millimetre galaxies (e.g, \citealt{calvi23}). In this paper, we applied the well established Poisson probability method (PPM) around cluster candidates members identified as radio emitting sources. The PPM searches for megaparsec-scale overdensities of galaxies around a given target.
Through the use of a solid positional prior and an accurate photometric redshift sampling, PPM partially overcomes the limitations deriving from low number-count statistics and shot-noise fluctuations, which are particularly relevant in the high-$z$ Universe, such as in the case of proto-clusters. More specifically, the PPM method uses photometric redshifts of galaxies to search for overdensities around each target along the line of sight. 
To search for overdensities around a sample of radio sources, the PPM adopts an accurate sampling of the photometric redshift information to the detriment of a less sophisticated tessellation of the projected space, which is performed in terms of concentric annuli centred around each target. We refer to previous studies for a detailed description of the method (\citealt{castignani_2014a, castignani_2014b}), its wavelet-based extension (\citealt{castignani2019}), and the applications (\citealt{castignani_2014b,castignani2019}, \citealt{calvi23}).
\\
Here, we used VLA radio sources from \citealt{Kellermann_2008} for the CDFS, and from the VLA FIRST survey (\citealt{Becker1995}) for the XMM-LSS. Each radio source was associated with its optical counterpart within our catalogue with a $2$~arcseconds separation criterion and using its photometric redshift, we computed the $1.4$ GHz rest-frame luminosity using:  

\begin{equation}
    \text{L}_{1.4\text{GHz}} = 4\pi\text{F}_{1.4\text{GHz}}\cdot \text{D}_\text{L}(z_\text{phot})^2 \cdot (1+z_\text{phot})^{\alpha-1}
\end{equation}

\noindent with D$_\text{L}$ the luminosity distance and $\alpha$ the radio spectral index. Here, we used $\alpha = 0.8$, according to \citealt{Chiaberge_2009}. 
To homogenise both catalogues of radio sources, we considered only sources with $\text{L}_{1.4\text{GHz}} > 10^{23.2}$ W$\cdot$Hz$^{-1}$. This selection allowed us to safely exclude star forming galaxies, which typically have lower radio luminosity densities. Higher values such as those considered in this work are instead typically associated with radio loud active galactic nuclei, including radio galaxies and radio-loud quasars.

Following this criterion, we identified $213$ and $158$ radio sources in the XMM-LSS and  CDFS, respectively. Since the footprint of the radio survey only covers a fraction of the CDFS, we only considered cluster candidates that are at least partially covered by the footprint of the radio catalogue. We identified $27$ and $11$ cluster candidates containing, at least, one radio source in the XMM-LSS and CDFS, respectively. Using PPM, we searched for galaxy overdensities around these sources. For clusters with several radio-loud members, we identified the main PPM counterpart as the closest to the cluster centre. We identified a PPM counterpart for $26$ XMM-LSS and $11$ CDFS cluster candidates. Among them, there are two XMM-LSS and one CDFS pairs of cluster candidates that share common PPM detections, as their associated radio sources are common members. Figure~\ref{figure:GS-PPM_z} shows the relation between the redshift of the cluster candidates and the ones of their associated PPM overdensities. We have overall good agreement in terms of redshift between them, with an offset lower than $0.05$ in absolute value. Figure~\ref{figure:GS-PPM_angsep} shows the distribution of the angular separation between the cluster candidates and their associated PPM overdensities. Cases with an angular separation larger than $1.0$\,Mpc represent $\sim 27\%$ and $\sim 18\%$ of the entire set of pairs in the XMM-LSS and CDFS, respectively, and could be seen as less likely associations. On the other hand, as PPM is designed to search for overdensities of galaxies on Mpc scales, these cases could also correspond to clusters embedded within larger structures. Spectroscopic follow-ups at the Mpc scale are required to better understand these cases (which will be one of the purposes of MOONRISE).

Altogether, 37 of the 38 AMICO-WaZP cluster candidates containing a radio source are also detected by PPM.
Except for a few cases, we observe very good redshift and angular separation agreement. Overall, as these candidates are also well identified by a third, independent cluster finder, the PPM analysis strengthens the robustness of these joint AMICO-WaZP detections.

\subsection{Comparison with ancillary cluster catalogues}
\label{subsection:ancillary}

\noindent Following the work described above, we compared our sample of cluster candidates to spectroscopically confirmed, multi-$\lambda$, clusters available in the literature. For this purpose, we used $49$ XXL clusters from \citealt{Adami_2018}, ten from \citealt{Trudeau_2020}, six GOGREEN/GCLASS clusters from \citealt{Balogh_2020}, one Spitzer-selected high redshift cluster from \citealt{Papovich_2010}, one ACT-DR5 cluster from \citealt{Hilton_2021} and $29$ ECDFS groups from \citealt{Finoguenov_2015}. 

We cross-matched our cluster candidates with the ancillary data using a maximal radius criterion of $R = 1.0$~Mpc and $\delta z/(1+z) < 0.15$. If we had several matches with our list of candidates, we considered only the one with the highest WaZP SNR. Following this process, we found a counterpart for $27$ clusters from \citealt{Adami_2018}, for all clusters from \citealt{Trudeau_2020}, for three GOGREEN/GCLASS clusters, ten ECDFS clusters, one for the cluster from \citealt{Papovich_2010}, and one for the ACT cluster (the candidate with the highest WAZP/AMICO SNR). Excepted for a few cases, clusters from the literature were always recovered by at least one of the cluster finders. However, due to the matching criteria used to build the AMICO-WaZP joint sample, some of them were excluded during the process and appeared as unmatched when compared to our final list of cluster candidates. Figure~\ref{figure:GS_anc} shows the comparison between the redshifts of the cluster candidates and those of their counterparts in the literature. Overall, a good agreement is found, except for a few outliers. The standard deviation of the offset $(z_\text{phot, joint}-z_\text{lit})/(1+z_\text{lit})$ is approximately $0.03$. We have four cluster candidates for which we measure an offset of $0.06$ or more, representing a deviation larger than $\pm 2\bar{\sigma_\varepsilon}$. One of them matches with a GOGREEN/GCLASS cluster, one with \citealt{Adami_2018} clusters, and two with ECDFS clusters. All of these cases could possibly be due to random associations with foreground low-mass groups of galaxies, induced by the large redshift criterion we use here. We refer to Appendix \ref{appendix:lit_bad_match} for a detailed description of them.

\setlength{\tabcolsep}{10pt} 
\renewcommand{\arraystretch}{1.25} 
\begin{table}[h!]
\captionsetup{labelfont=bf}
\adjustboxset{width=\textwidth}
\centering
\caption{\label{table:CMD_color_selection} Filters selection as a function of cluster redshifts $z_\text{phot, joint}$ for the construction of the CMD described in Sect. \ref{section:red-sequence}. }
\begin{tabular}{ccc} 
 \hline\hline
 Redshift range & colour & magnitude \\
 \hline 
     $0.1-0.4$ & \textit{g}-\textit{r} & \textit{r} \\ 
     $0.4-0.7$ & \textit{r}-\textit{i} & \textit{i} \\
     $0.7-1.1$ & \textit{i} - \textit{z} &  \textit{z} \\
     $1.1-1.4$ & \textit{z} - Y & Y \\
     $1.4-1.7$ & Y - J & J \\
     $1.7-2.3$ & J - H & H \\
  \hline
  
\end{tabular}
\tablefoot{Lowercase italic letters are for the HSC filters, while capital letters are for the VISTA ones.}
\end{table}

\section{Colour-magnitude relation.}
\label{section:red-sequence}

In this section, we analyse the Colour-Magnitude diagrams (CMDs) in apparent magnitudes of the cluster candidates previously selected. We aim to test for the presence of a passive, red population of galaxies in the inner parts of these clusters in the different redshift bins probed. Indeed, many studies have shown the presence of passive galaxies at the centre of rich clusters organised as a ridge-line in the CMD, the so-called red-sequence (RS), widely observed at low and intermediate redshifts (e.g. \citealt{Lopez-Cruz_2004}). However, despite some evidence of the existence of a RS up to $z \sim 2$, a consensus is not yet achieved for $z > 1$.  

For each cluster, the CMD is built considering only members within $0.5$\,Mpc from the cluster centre, and brighter than $m_\text{bright} + 1.5$. Here, $m_\text{bright}$ is defined as the brightest galaxy (in the reddest band of the CMD) as a function of redshift in the entire galaxy sample. In practice, to avoid outliers, we computed the mean magnitude of the $95\%$ brightest galaxies per redshift bin. In addition, we excluded members fainter than $m_\text{H} = 24$ which fell beyond the completeness limit of the sample.

In order to optimally frame the 4000\,\AA\, break, we adopted the redshift-dependent filter selections described in Table\,\ref{table:CMD_color_selection}. The analysis was interrupted at $z_\text{phot, joint} = 2.3$ as the Balmer break cannot be bracketed any more with the available filters. As some regions are not covered by the four HSC filters, in the first four redshift bins, the analysis was limited to clusters that are covered by the relevant filters at least at a $50\%$ level within a radius of $1.0$\,Mpc. In the end, we were left with $461$ candidates.

In Fig.~\ref{figure:CMD_examples} are shown examples of CMDs for clusters at different redshifts. They display a bimodal distribution, with a blue component and a more or less pronounced redder ridge-line. To measure the mean colour of the red ridge-line, the distribution of colours is modelled as a two-components Gaussian mixture. 
We estimated the mean colour $\mu_r$ of the ridge and its scatter $\sigma_r$ by identifying the red Gaussian component. We considered the estimation of the mean colour as valid if we had more than three members in the colour range $\mu_r \pm \sigma_r$. Applying this criterion, $11$ cases were rejected. After visual inspection of each CMD, we identified $16$ additional cases where the red ridge was either too poorly populated or not sufficiently prominent to be fitted reliably. The excluded cases typically correspond to candidates that are very poor or systems at high-redshift, where a well-established RS is not expected to be systematically present. For the other candidates, the bi-Gaussian successfully isolated the red ridge-line, yielding a mean normalised scatter $\overline{\sigma_r/(1+z)} \sim 0.05$\,mag, consistent with the results of similar studies at lower redshifts (e.g. \citealt{Andreon2003}, \citealt{Jaffe2011}).

We compared the mean colour of the red component with predictions from models of galaxy evolution. To model colours of a typical quiescent galaxy population, we followed the prescription from \citealt{Andreon_2006}, assuming a high-redshift formation over a relatively short timescale, followed by purely passive evolution. For this purpose, we used the \verb|CIGALE| SED fitting code (\citealt{CIGALE}) and compared the resulting mean colours with those measured for our cluster candidates. We modelled an exponentially declining star formation history (\verb|sfh2exp| module) with a formation redshift at $z_f = 10$ and assumed a folding timescale of the main stellar population of $\tau_\text{main} = 0.75$\,Gyr. We used a single stellar population derived from the \citealt{BC03} model (\verb|bc03| module), computed for different metallicities: $Z = 0.004, 0.008, 0.02, \text{ and } 0.05$. Here, a \citealt{Chabrier_2003} initial mass function (IMF) was used.

Figure \ref{figure:RS_mean_colour} shows the evolution of the mean colour of the red component as a function of the redshift of the AMICO-WaZP joint cluster candidates. We represent by open circles the $16$ cases we rejected following the visual inspection. For the remaining $434$ candidates, we observe an overall good agreement between the model predictions and the mean colour of the red component measured for the candidates. This suggests the presence of a population of quiescent galaxies already in place at $z \sim 2$, as reported by previous studies (e.g,  \citealt{Kodama2007}, \citealt{Andreon_2014}, \citealt{Cerulo16}, \citealt{Strazzulo_2019}, \citealt{Willis2020}, \citealt{Toni2025b}). We also observe a few systems in the last bin of redshifts that are redder than the predictions of the model used here. Nevertheless, we find that by slightly reducing the folding timescale to $0.5$\,Gyr, we can reproduce the mean colours of these candidates. This may indicate the presence of progenitors of massive elliptical galaxies, as suggested by \citealt{De_Lucia_2006}. In any case, spectroscopic follow-ups are required to better understand the galaxy population within our candidates, since they might have different evolutionary histories depending on the mass or evolutionary stage of their parent clusters. Comparisons with AMICO memberships return similar results, with the presence of a population of red galaxies already in place at $z \sim 2$, compatible with a scenario of high-redshift formation and passive evolution.

 \begin{figure*}[t]
    \centering
    \captionsetup{format=plain}
    \captionsetup{labelfont=bf}
    \includegraphics[width=5.66cm]{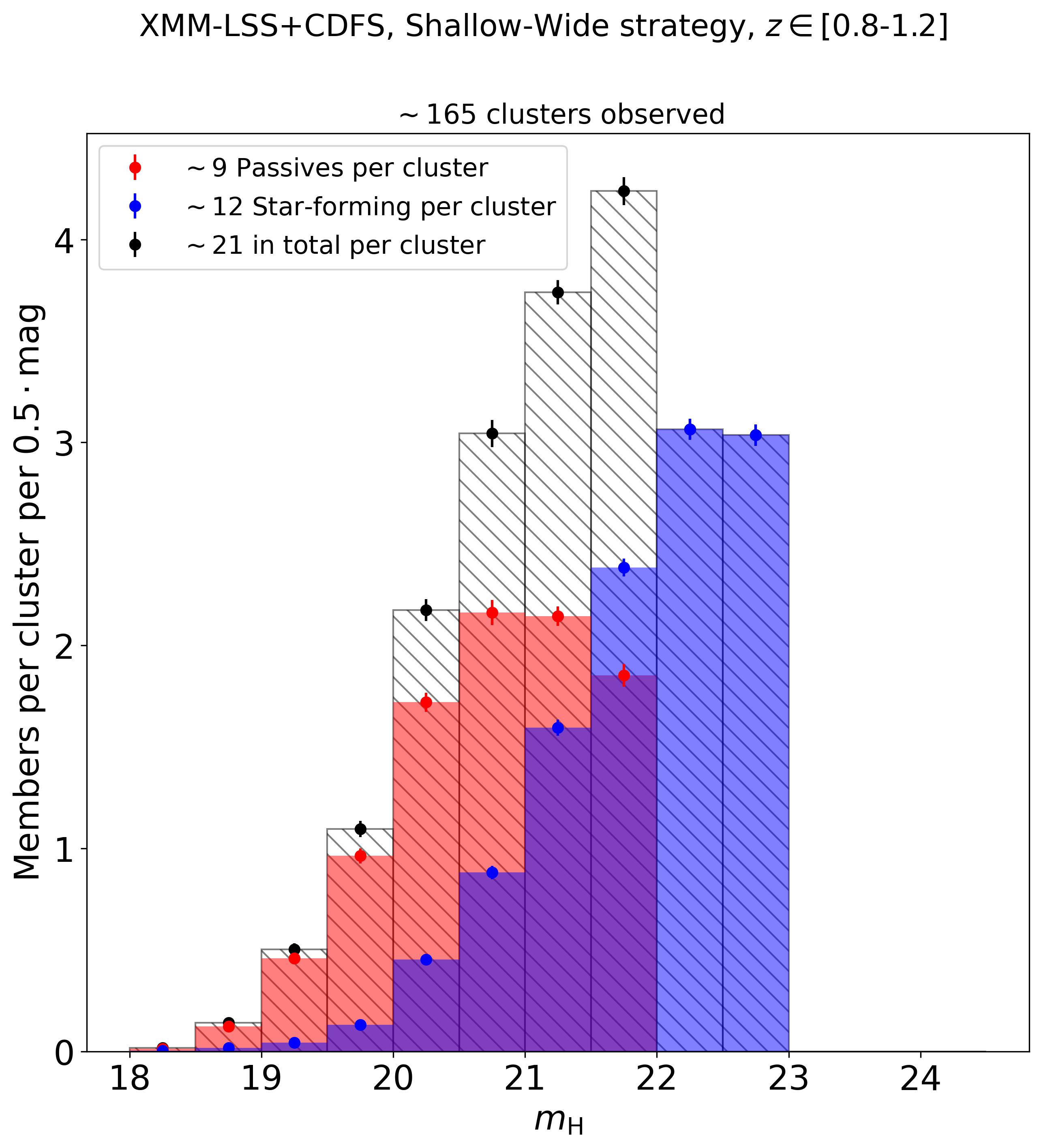}    
    \includegraphics[width=5.66cm]{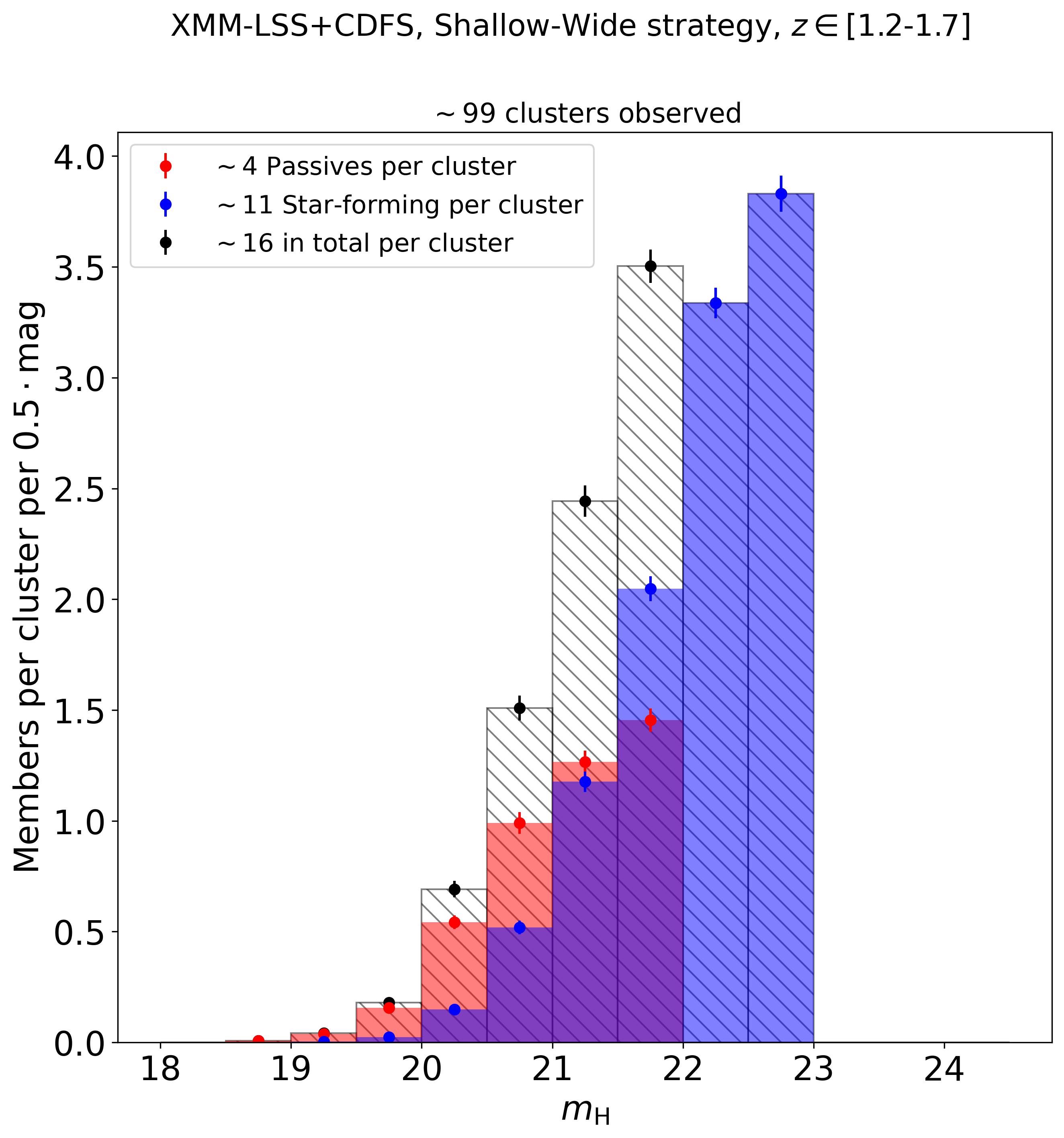}
    \includegraphics[width=5.66cm]{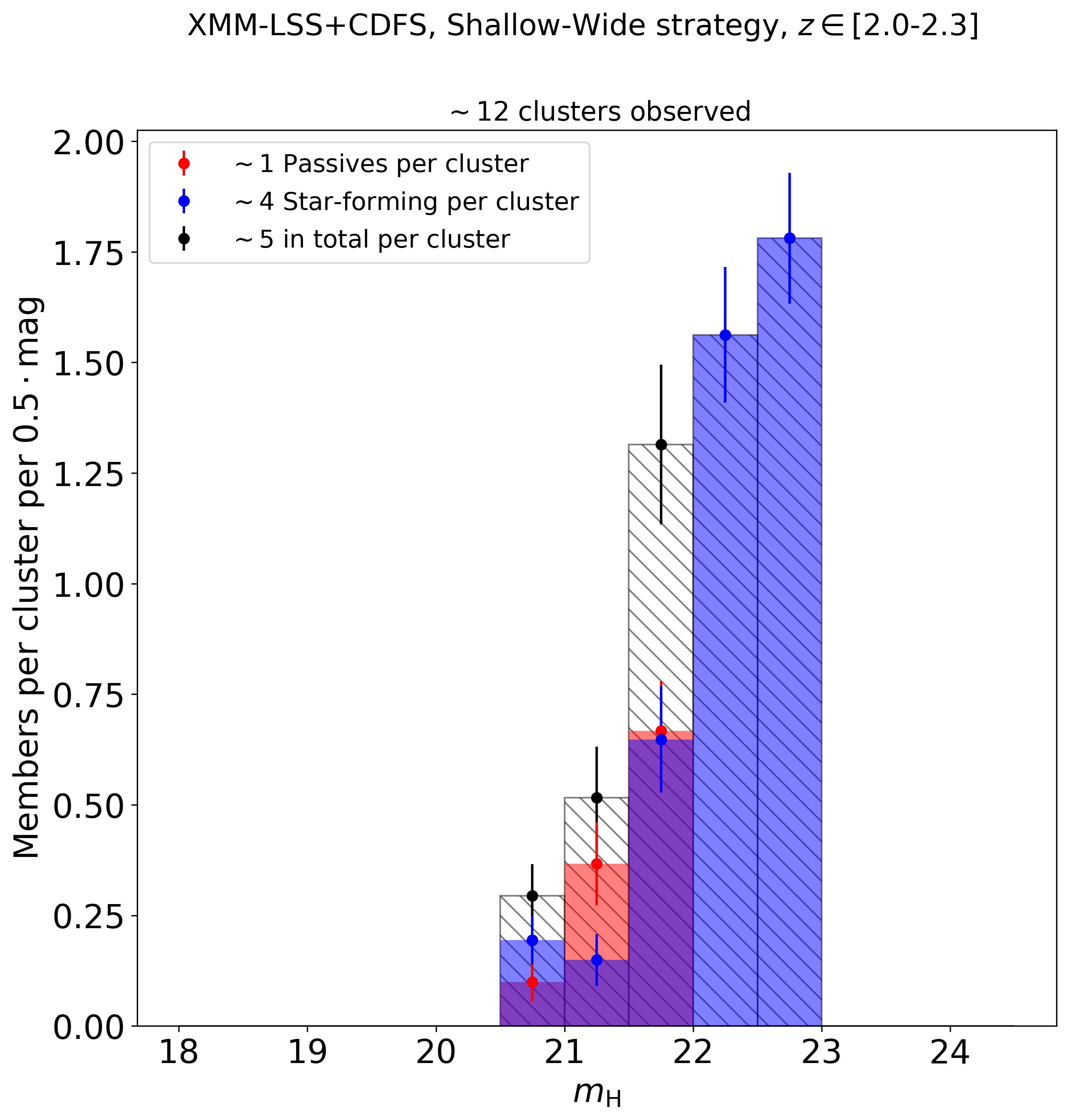}    
    \caption{Mean weighted distribution of the H-band magnitude of the detected members (in red for passive galaxies, in blue for star-forming galaxies. The dashed black bars are for all members) per cluster candidate when applying the Shallow-Wide MOONRISE strategy to the CDFS + XMM-LSS. From left to right, redshift ranges $[0.8, 1.2]$, $[1.2, 1.7]$ and $[2.0, 2.3]$ are shown. In each box, a legend describing the mean number of galaxies of different types expected to be observed by MOONRISE is added.}
    \label{figure:CDFS_Shallow_wide}   
\end{figure*}

\begin{figure*}[t]
    \centering
    \captionsetup{format=plain}
    \captionsetup{labelfont=bf}
    \includegraphics[width=8cm]{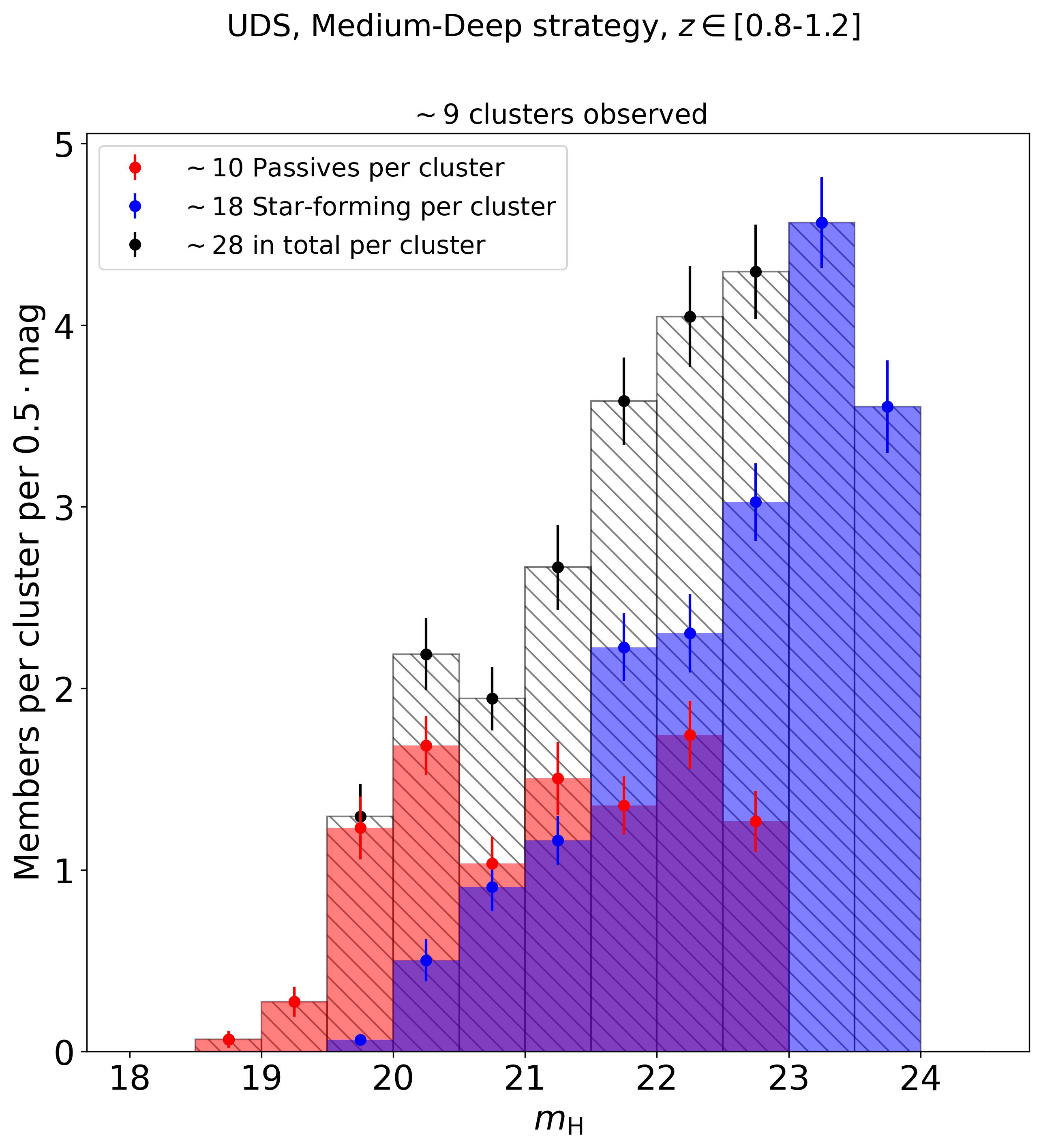}    
    \includegraphics[width=8cm]{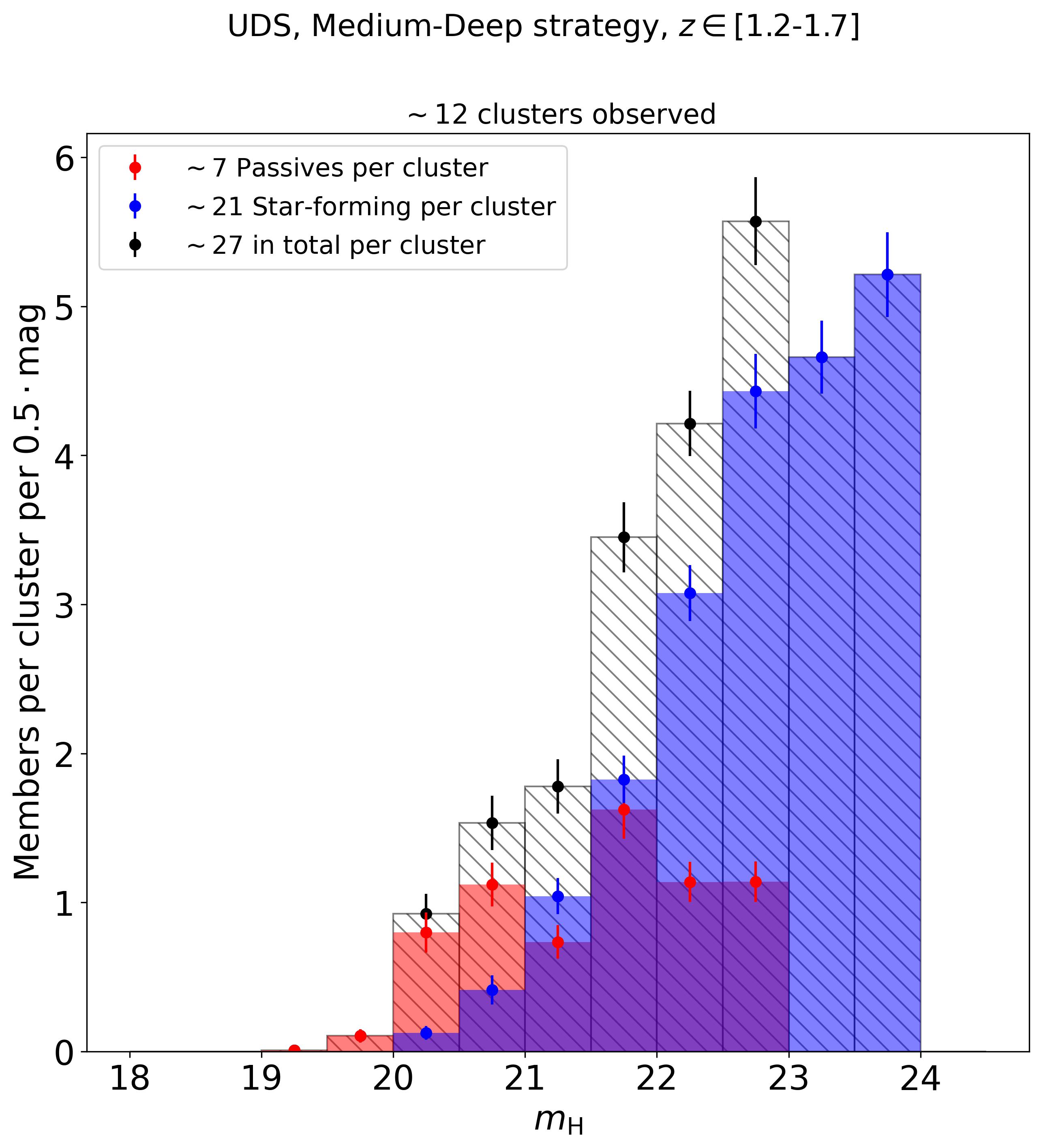}
    \caption{Same as Fig.\ref{figure:CDFS_Shallow_wide} but for the UDS when applying the Medium-Deep strategy. Only the redshift ranges $[0.8, 1.2]$ and $[1.2, 1.7]$ are considered here, as no cluster candidate was found in the UDS in the redshift range $[2.0, 2.3]$.}
    \label{figure:UDS_Medium_Deep}   
\end{figure*}

\section{Probing the galaxy population of clusters with MOONRISE}
\label{section:MOONRISE}

\noindent 

In the previous sections, we have demonstrated the potential of applying state-of-the-art cluster-finding algorithms to deep, multi-band photometric VIDEO fields for which high-quality photometric redshifts have been derived. By cross-matching the detections obtained with the two cluster finders, using an optimised SNR threshold for both codes, we constructed a joint cluster-candidate catalogue that reaches redshift $2$ with good purity ($> 80\%$ based on numerical simulations). 

Parts of the analysed VIDEO fields (XMM-LSS and CDFS) are expected to be covered in spectroscopy in the near future by MOONRISE. We hereafter investigate how many spectroscopic redshift members would be confirmed by MOONRISE in cluster candidates previously identified in those fields using photometric redshifts. MOONRISE observational strategies are currently being defined within the consortium
, in order to meet the broad scientific objectives. In this section, we illustrate, using real data, how the group and cluster populations detected in the VIDEO fields are expected to be sampled under these different strategies.

We focus here on two observational strategies designed to optimally trace the effects of environment on galaxy evolution (the Medium-Deep strategy) and the evolution of large-scale structure (the Shallow-Wide strategy). To meet scientific objectives, observational requirements in terms of redshift coverage, magnitude limits, and sampling rate were derived from mock catalogues simulating the properties of the VIDEO fields (\citealt{deLucia2007}).
The resulting goals are as follows: a survey completeness of $70\%$, redshift coverage of $z = 0.8-1.7$ and $z = 2.0 - 2.3$, with different limiting magnitudes for each observational strategy:
\begin{itemize}
    \item[-] For the Medium-Deep: $m_\text{H} < 24$ for star-forming galaxies, $m_\text{H} < 23$ for passive galaxies
    \item[-] For the Shallow-Wide: $m_\text{H} < 23$ for star-forming galaxies, $m_\text{H} < 22$ for passive galaxies
\end{itemize}
To mimic these observing strategies and assess their impact on the sampling of the cluster population, we applied the corresponding galaxy selections to the members of the joint cluster candidates and estimated the expected recovery rate as a function of redshift.
We used the U-V-J rest frame colour-colour diagram shown in Fig.~\ref{figure:UVJ_stacked} to separate passive and star-forming galaxies.  
They are considered quiescent if their rest frame  U, V, and J rest-frame magnitudes follow the relation:     

\begin{equation}
\left\{ 
    \begin{array}{ll}
         U-V &> 0.88 \times(V-J)+0.49  \\
         U-V &> 1.3 
    \end{array}
\right.
\label{equation:active_passive_crit_lowz}
\end{equation}
\noindent for $z \in [0.8, 2.0]$, and 

\begin{equation}
\left\{ 
    \begin{array}{ll}
         U-V &> 0.88 \times(V-J)+0.42  \\
         U-V &> 1.1
    \end{array}
\right.
\label{equation:active_passive_crit_highz}
\end{equation}
\noindent for $z \in [2.0, 2.3]$. 

We made use of rest-frame magnitudes and physical parameters  provided by \verb|BAGPIPES| spectral modelling and fitting tool (\citealt{Carnall_2018}, \citealt{Carnall_2019}), with models redshift fixed at the $z_\text{phot}$ of each galaxy (see Sect. \ref{subsection:zphot_cat}). The code uses the $2016$ updated version of \citealt{BC03} as a stellar population model, based on a \citealt{Kroupa_2021} IMF. A \citealt{Calzetti2000} dust attenuation law was used, with allowed reddening $A_{V}$ between $0$ and $4$. 
\noindent

In order to maximise statistics, we used the full XMM-LSS and CDFS fields to test the Shallow-Wide strategy. For the Medium-Deep strategy, we focused, as an example, on the deepest region covered by the UDS to meet the depth requirements. In this case, because of the limited field size, our results may be affected by cosmic variance. We divided the redshift range into three bins: $[0.8, 1.2], [1.2, 1.7]$, and $[2.0, 2.3]$. For each strategy, we adopted the assumption of a uniform random selection of galaxies with an overall completeness of $70\%$ across the entire field for the given strategy, without accounting for fibre positions. We first applied the observational strategy criterion to the full catalogue and then randomly selected $70\%$ of the galaxies, repeating this procedure $100$ times. 
Within each redshift bin, we then computed the total number of member galaxies recovered per cluster, as well as their distribution in magnitude bins. These quantities were weighted by the cluster membership probabilities. Finally, we averaged over the one hundred realisations to estimate the mean number of members recovered per cluster for a given strategy.

Figures~\ref{figure:CDFS_Shallow_wide} and \ref{figure:UDS_Medium_Deep} show the mean weighted number of recovered members per bins of H magnitude per cluster for the XMM-LSS+CDFS under the Shallow-Wide strategy, and for the UDS under the Medium-Deep strategy, respectively. 
These figures clearly indicate that the Shallow-Wide Strategy will be key for massive redshift confirmation of clusters candidates ($264$ on the whole region, corresponding to $\sim 30 \text{ deg}^{-2}$) over the redshift range [$0.8$-$1.7$] with more than $15$ members per cluster. 
The Medium-Deep strategy yields a higher number of members per cluster (on average $\sim 30$) thereby enabling detailed kinematic studies and cluster mass measurements through velocity dispersions for a comparable density of detections ($\sim 24 \text{ deg}^{-2}$). It is worth noting also the good balance achieved between passive and star-forming sampling, thanks to the use of different magnitude selection criteria for each population.  
At higher redshifts ($z > 2$), we enter the regime of rare systems. In this case, spectroscopic confirmation appears barely achievable for only $12$ candidates in the XMM-LSS+CDFS fields with the Shallow-Wide strategy, corresponding to a projected density of $\sim 1\,\mathrm{deg}^{-2}$. The Medium-Deep strategy could not be tested in this redshift range but may be investigated in the future using other existing deep surveys (i.e. COSMOS). 

We would like to emphasise the caveats of this approach. First, to better control purity and ensure a robust analysis, we decided to consider only high SNR detections in common for both algorithms, which strongly limits the number of detections. Second, we cannot exclude the possibility that a small fraction of objects suffers from line-of-sight contamination. Estimating this fraction accurately would require a detailed characterisation of the selection function on the data, which is beyond the scope of this paper. 
However, given the use of the joint approach and a conservative SNR threshold, this fraction is expected to remain well below $20\%$. 
Despite these limitations, the analysis presented in this section demonstrates that the Shallow-Wide and Medium-Deep strategies, defined using mock catalogues, appear very promising when applied on real data and complement each other very effectively in the characterisation of galaxy populations in dense environments. MOONRISE will be instrumental in validating these systems and thus unambiguously characterising the selection function of groups and clusters up to redshift $2$.

\section{Discussions and conclusion}
\label{section:conclusion}
 
We presented a catalogue of $519$ cluster candidates in the redshift range $ [0.1, 3]$, jointly detected by the WaZP and AMICO cluster finders in the XMM-LSS and CDFS VIDEO fields. Selecting only the common detections from the two cluster finders resulted in a catalogue of cluster candidates with enhanced purity, but it may exclude potentially valid detections identified by only one of the two methods, making this a very conservative choice.

Comparisons with spectroscopically confirmed clusters from the literature and the various estimations of spectroscopic redshift measurement for eligible cluster candidates showed, overall, good agreement with the estimated photometric redshift. We
showed that cluster candidates containing radio-loud galaxies can be efficiently identified by the PPM algorithm, a third, different cluster finder.

For each cluster candidate, we provided a list of likely members, which were used to
address the evolution of galaxy properties through redshift. The study of the apparent Colour-Magnitude diagram of each individual cluster candidate allowed the identification of a red-sequence-like component up to $z\sim 2$. Comparison with models of galaxy evolution highlighted the presence of a population of red, ancient galaxies already in place at $z \sim 2$. These results are consistent with previous studies, reinforcing the hypothesis that the galaxy populations within clusters has undergone a peak in their transformation process at $z \sim 2-3$, leading to the population we observe at lower redshift.

All these results tended to reinforce our confidence about the true nature of these candidates and leaded to preliminary insights into the properties of galaxies within clusters at $z > 1$. Due to our method of selection and the limited surface of both fields, our conclusions are limited to this specific subpopulation, and further work is needed to determine the selection function and assess the representativeness of the cluster catalogue. We also plan to investigate the properties of these candidates in greater detail in future studies. They will include their stellar mass function and the characteristics of their galaxy populations.

Spectroscopic follow-ups of these candidates will be essential to confirm their redshifts and physical properties. While targeted follow-ups can be performed for a limited number of clusters, a comprehensive three-dimensional mapping from a large-scale spectroscopic survey is necessary to enable systematic redshift confirmation and to assess line-of-sight contamination. Indeed, massive spectroscopy have proven to be essential for understanding the selection function of optical cluster catalogues, as demonstrated by the use of DESI spectroscopy to characterise RedMapper cluster catalogues \citep{Myles2025}. MOONRISE will allow us to go a step further in redshift and to characterise the selection function up to redshift of $2$. 
It will open a new window on cluster formation in the redshift range $[0.8, 2.3]$. Its spectral coverage is designed for this regime, and its combination of high multiplexing and a wide field of view will allow full coverage of the cluster virial regions-essential for conducting robust dynamical analyses. These observations will provide a unique dataset to investigate the assembly of mass and stars in massive clusters within their large-scale environments and to study the physical mechanisms that regulate star formation as a function of the environment. 

\section*{Data availability}

Tables \ref{table:catalogue_xmm} and \ref{table:catalogue_cdfs} are only available in electronic form at the CDS via anonymous ftp to \url{cdsarc.u-strasbg.fr} (130.79.128.5) or via \url{http://cdsweb.u-strasbg.fr/cgi-bin/qcat?J/A+A/.}

\begin{acknowledgements}

Funding for the Sloan Digital Sky Survey V has been provided by the Alfred P. Sloan Foundation, the Heising-Simons Foundation, the National Science Foundation, and the Participating Institutions. SDSS acknowledges support and resources from the Center for High-Performance Computing at the University of Utah. SDSS telescopes are located at Apache Point Observatory, funded by the Astrophysical Research Consortium and operated by New Mexico State University, and at Las Campanas Observatory, operated by the Carnegie Institution for Science. The SDSS web site is \url{www.sdss.org}.
\\
SDSS is managed by the Astrophysical Research Consortium for the Participating Institutions of the SDSS Collaboration, including Caltech, The Carnegie Institution for Science, Chilean National Time Allocation Committee (CNTAC) ratified researchers, The Flatiron Institute, the Gotham Participation Group, Harvard University, Heidelberg University, The Johns Hopkins University, L’Ecole polytechnique fédérale de Lausanne (EPFL), Leibniz-Institut für Astrophysik Potsdam (AIP), Max-Planck-Institut für Astronomie (MPIA Heidelberg), Max-Planck-Institut für Extraterrestrische Physik (MPE), Nanjing University, National Astronomical Observatories of China (NAOC), New Mexico State University, The Ohio State University, Pennsylvania State University, Smithsonian Astrophysical Observatory, Space Telescope Science Institute (STScI), the Stellar Astrophysics Participation Group, Universidad Nacional Aut\'{o}noma de México, University of Arizona, University of Colorado Boulder, University of Illinois at Urbana-Champaign, University of Toronto, University of Utah, University of Virginia, Yale University, and Yunnan University.
\\
GAMA is a joint European-Australasian project based around a spectroscopic campaign using the Anglo-Australian Telescope. The GAMA input catalogue is based on data taken from the Sloan Digital Sky Survey and the UKIRT Infrared Deep Sky Survey. Complementary imaging of the GAMA regions is being obtained by a number of independent survey programmes including GALEX MIS, VST KiDS, VISTA VIKING, WISE, Herschel-ATLAS, GMRT and ASKAP providing UV to radio coverage. GAMA is funded by the STFC (UK), the ARC (Australia), the AAO, and the participating institutions. The GAMA website is \url{http://www.gama-survey.org/}.
\\
This paper uses data from the VIMOS Public Extragalactic Redshift Survey (VIPERS). VIPERS has been performed using the ESO Very Large Telescope, under the 'Large Programme' 182.A-0886. The participating institutions and funding agencies are listed at \url{http://vipers.inaf.it}.
\\
This work is based [in part] on observations made with the Spitzer Space Telescope, which was operated by the Jet Propulsion Laboratory, California Institute of Technology under a contract with NASA.
\\ 
This work is based [in part] on data from the ESO Programme 180.A-0776. More details are available here \url{https://www.nottingham.ac.uk/astronomy/UDS/UDSz/}
\\
This research used data obtained with the Dark Energy Spectroscopic Instrument (DESI). DESI construction and operations is managed by the Lawrence Berkeley National Laboratory. This material is based upon work supported by the U.S. Department of Energy, Office of Science, Office of High-Energy Physics, under Contract No. DE–AC02–05CH11231, and by the National Energy Research Scientific Computing Center, a DOE Office of Science User Facility under the same contract. Additional support for DESI was provided by the U.S. National Science Foundation (NSF), Division of Astronomical Sciences under Contract No. AST-0950945 to the NSF’s National Optical-Infrared Astronomy Research Laboratory; the Science and Technology Facilities Council of the United Kingdom; the Gordon and Betty Moore Foundation; the Heising-Simons Foundation; the French Alternative Energies and Atomic Energy Commission (CEA); the National Council of Humanities, Science and Technology of Mexico (CONAHCYT); the Ministry of Science and Innovation of Spain (MICINN), and by the DESI Member Institutions: www.desi.lbl.gov/collaborating-institutions. The DESI collaboration is honored to be permitted to conduct scientific research on I’oligam Du’ag (Kitt Peak), a mountain with particular significance to the Tohono O’odham Nation. Any opinions, findings, and conclusions or recommendations expressed in this material are those of the author(s) and do not necessarily reflect the views of the U.S. National Science Foundation, the U.S. Department of Energy, or any of the listed funding agencies.
\\
This work is based on observations taken by the 3D-HST Treasury Program (GO 12177 and 12328) with the NASA/ESA HST, which is operated by the Association of Universities for Research in Astronomy, Inc., under NASA contract NAS5-26555.
\\
GC acknowledges the support from the Next Generation EU funds within the National Recovery and Resilience Plan (PNRR), Mission 4 - Education and Research, Component 2 - From Research to Business (M4C2), Investment Line 3.1 - Strengthening and creation of Research Infrastructures, Project IR0000012 – “CTA+ - Cherenkov Telescope Array Plus”.
\\
LM acknowledges the financial contribution from the 
PRIN-MUR 2022 20227RNLY3 grant “The concordance cosmological model: stress-tests with galaxy clusters” supported by Next Generation EU and from the grant ASI n. 2024-10-HH.0 “Attività scientifiche per la missione Euclid – fase E”
\\
MV acknowledges financial support from the Inter-University Institute for Data Intensive Astronomy (IDIA), a partnership of the University of Cape Town, the University of Pretoria and the University of the Western Cape, and from the South African Department of Science and Innovation's National Research Foundation under the ISARP RADIOMAP Joint Research Scheme (DSI-NRF Grant Number 150551) and the CPRR HIPPO Project (DSI-NRF Grant Number SRUG22031677).
\\
CB, PG and SM acknowledge financial support from the 'Bonus Qualit\'e Recherche' of Laboratoire J.L. Lagrange. 
\\
DJM acknowledges the support of the Royal Society through the award of a Royal Society University Research Professorship to Prof. James Dunlop.
\\
This work was supported by the Swiss National Science Foundation (SNSF) under funding reference 200021\_213076 'Galaxy evolution in the cosmic web'.
\end{acknowledgements}

\bibliographystyle{aa}
\bibliography{biblio}

\hrule

\begin{itemize}
    \item[$^{1}$] Université Côte d’Azur, Observatoire de la Côte d’Azur, CNRS, Laboratoire Lagrange, Bd de l’Observatoire, CS 34229, 06304 Nice cedex 4, France\label{1} 
    
    \item[$^{2}$] INAF - Osservatorio  di  Astrofisica  e  Scienza  dello  Spazio di  Bologna,  via  Gobetti  93/3,  I-40129,  Bologna,  Italy\label{2}
    
    \item[$^{3}$] Center for Astronomy - University of Heidelberg, Albert-Ueberle-Stra{\ss}e 2, 69120 Heidelberg, Germany\label{3}
    
    \item[$^{4}$] Institute of Theoretical Physics - University of Heidelberg, Albert-Ueberle-Stra{\ss}e 2, 69120 Heidelberg, Germany\label{4}
    
    \item[$^{5}$] Laboratoire d’astrophysique, \'{E}cole Polytechnique F\'{e}d\'{e}rale de Lausanne (EPFL), Chemin Pegasi 51, 1290 Versoix, Switzerland\label{5}
    
    \item[$^{6}$] LUX, Observatoire de Paris, Université PSL, CNRS, Place Jules Janssen, F-92195 Meudon, France\label{6}

    \item[$^{7}$] School of Physics and Astronomy, University of Nottingham, University Park, Nottingham NG7 2RD, UK\label{7}

    \item[$^{8}$] Institute for Astronomy, University of Edinburgh, Royal Observatory, Edinburgh, EH9 3HJ, UK\label{8}

    \item[$^{9}$] Jodrell Bank Centre for Astrophysics, Department of Physics and Astronomy, School of Natural Sciences, The University of Manchester, Manchester, M13 9PL, UK\label{9}

    \item[$^{10}$] Dipartimento di Fisica e Astronomia ”Augusto Righi”, Alma Mater Studiorum Università di Bologna, Via Gobetti 93/2, I-40129 Bologna, Italy\label{10}

    \item[$^{11}$] INFN - Sezione di Bologna, Viale Berti Pichat 6/2, 40127 Bologna, Italy\label{11}

    \item[$^{12}$] INAF-IAPS, Via Fosso del Cavaliere 100, I-00133 Roma, Italy\label{12}

    \item[$^{13}$] ESO\label{13}

    \item[$^{14}$] Department of Physics “E. Pancini”, University of Naples Federico II, Naples, Italy\label{14}

    \item[$^{15}$] INAF – Osservatorio Astronomico di Capodimonte, Salita Moiariello 16, I-80131, Napoli, Italy\label{15}

    \item[$^{16}$] INFN, Sezione di Napoli, Via Cintia, I-80126 Napoli, Italy\label{16}

    \item[$^{17}$] Department of Astronomy, University of Geneva, ch. d’Ecogia 16, CH-1290 Versoix, Switzerland\label{17}

    \item[$^{18}$] Kavli Institute for Cosmology, University of Cambridge, Madingley Road, Cambridge CB3 OHA, UK\label{18}

    \item[$^{19}$] Cavendish Laboratory – Astrophysics Group, University of Cambridge, 19 JJ Thomson Avenue, Cambridge CB3 OHE, UK\label{19}

    \item[$^{20}$] Department of Physics and Astronomy, University College London, Gower Street, London WC1E 6BT, UK\label{20}

    \item[$^{21}$] CSST Science Center for Guangdong-Hong Kong-Macau Great Bay Area, 519082 Zhuhai, PR China\label{21}

    \item[$^{22}$] School of Physics and Astronomy, Sun Yat-sen University Zhuhai Campus, 2 Daxue Road, Tangjia, Zhuhai, Guangdong 519082, PR China\label{22}

    \item[$^{23}$] INAF - Istituto di Radioastronomia, via Gobetti 101, 40129 Bologna, Italy\label{23}

    \item[$^{24}$] Inter-University Institute for Data Intensive Astronomy, Department of Astronomy, University of Cape Town, 7701 Rondebosch, Cape Town, South Africa\label{24}

    \item[$^{25}$] Inter-University Institute for Data Intensive Astronomy, Department of Physics and Astronomy, University of the Western Cape, 7535 Bellville, Cape Town, South Africa\label{25}
    \label{app:affiliations}
\end{itemize}

\clearpage
\begin{appendices} 
 
\renewcommand{\thesection}{\Alph{section}}
\setcounter{section}{0}
\setcounter{table}{0}
\renewcommand{\thetable}{C.\arabic{table}}

\counterwithin{figure}{section}

\section{Spectroscopic properties}
\label{appendix:zspec_mag}

\vspace{-1.25cm}

\begin{figure}[h]
    \centering
    \captionsetup{labelfont=bf}
    \includegraphics[width=0.48\textwidth]{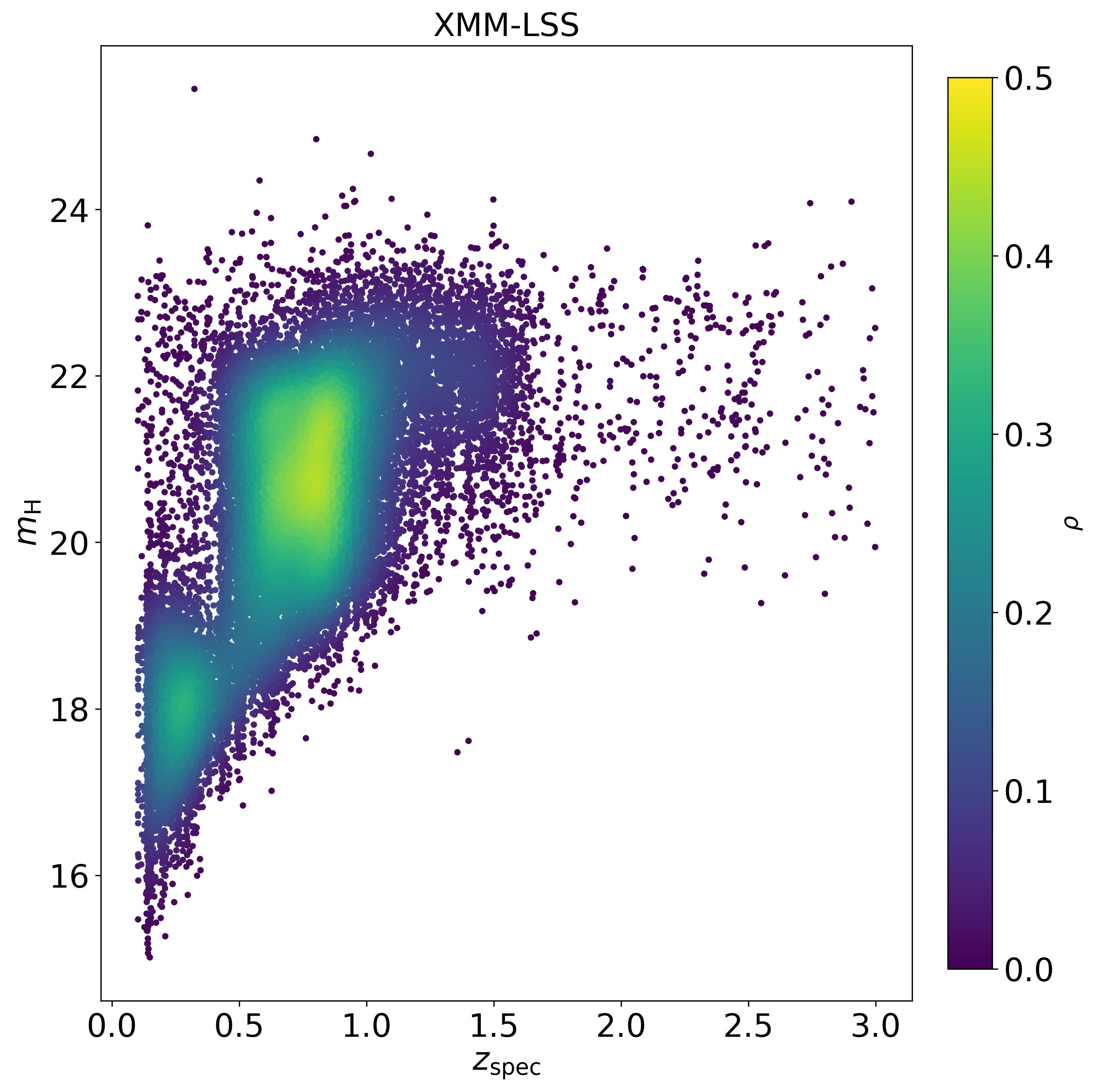}
    \caption{Redshift-$m_\text{H}$ relation for galaxies with spectroscopic redshift in the XMM-LSS. The colour maps represent the amplitude of the probability density function of the points distribution.}
    \label{figure:zspec_mag_xmm}   
\end{figure}


\begin{figure}[h]
    \centering
    \captionsetup{labelfont=bf}
    \includegraphics[width=0.48\textwidth]{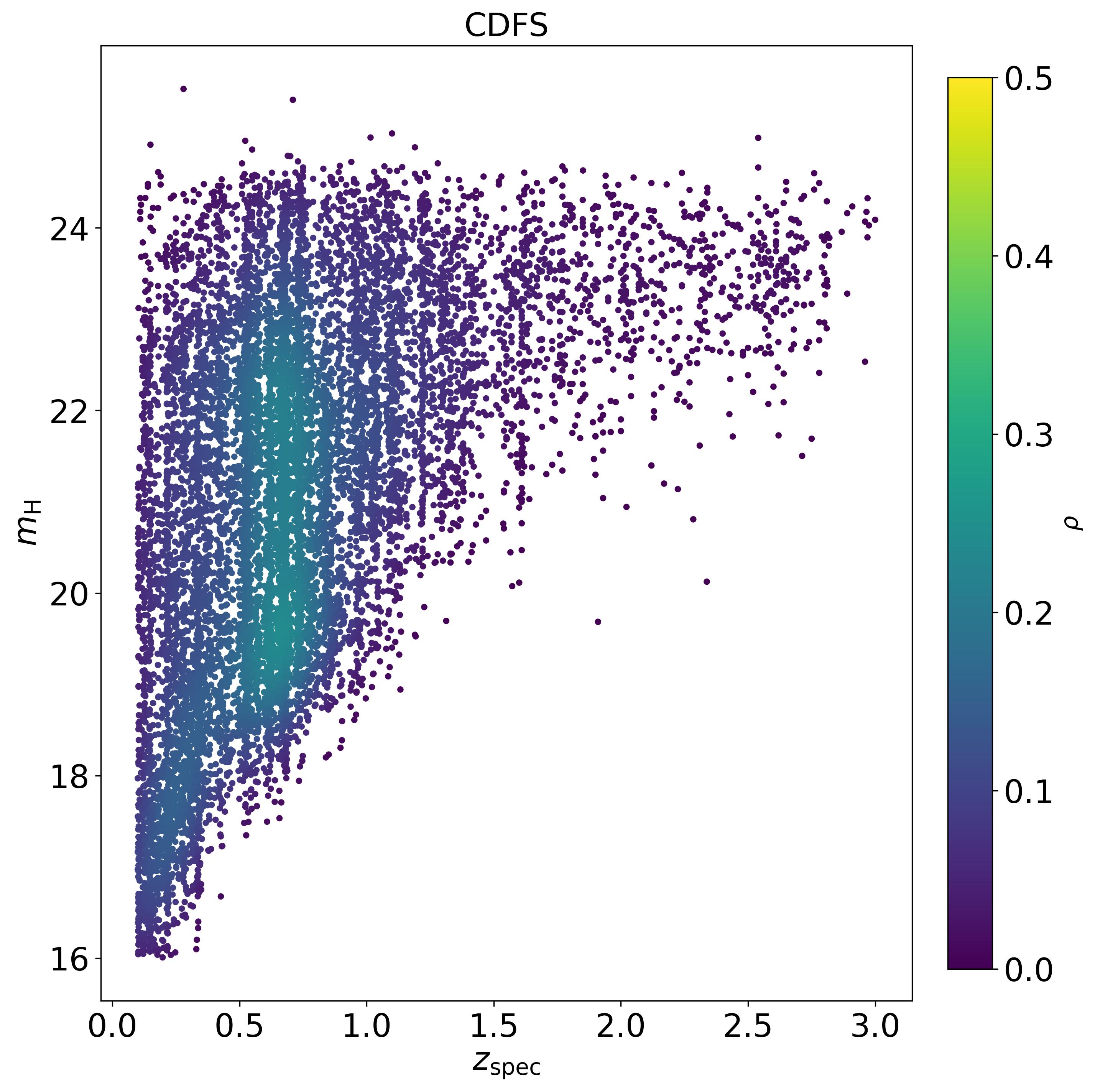}
    \caption{Same as Fig. \ref{figure:zspec_mag_xmm} but for the CDFS}
    \label{figure:zspec_mag_cdfs}   
\end{figure}

\section{Angular separation of the brightest cluster members}
\label{appendix:bcm_offset}

\begin{figure}[H]
    \centering
    \captionsetup{labelfont=bf}
    \includegraphics[width=0.48\textwidth]{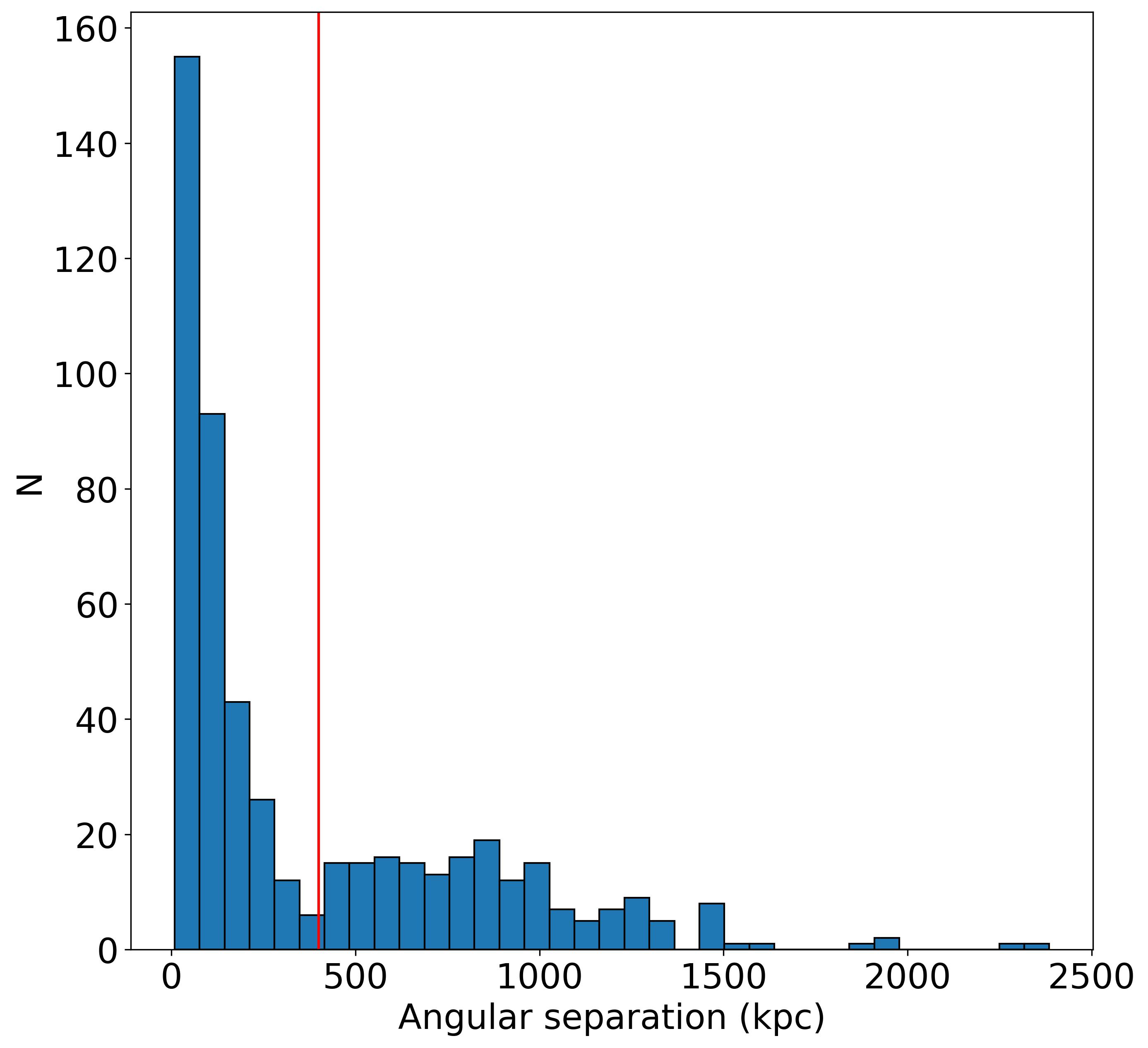}     
    \caption{Angular separation of the brightest cluster members with the cluster centre. Based on this distribution, we kept as fiducial BCG candidates only the brightest members within $400$ kpc. The $400$ kpc limit is represented as a vertical red line.}
    \label{figure:bcg_ang_sep}   
\end{figure}

\onecolumn
\section{Cluster catalogues}
\label{appendix:extract_catalogues}

\setlength{\tabcolsep}{6pt}
\renewcommand{\arraystretch}{1}

\hspace{-0.8cm}
\begin{minipage}{1.12\textwidth}

\adjustboxset{width=\textwidth}
\centering
\captionsetup{labelfont=bf}
\captionof{table}{\label{table:catalogue_xmm} First $20$ AMICO-WaZP joint cluster candidates within the XMM-LSS. }  

\begin{tabular}{cccccccccccc} 
 \hline\hline
 ID & RA & Dec  & 
 $z_\text{phot}$ & 
 $\text{rich}_\text{WaZP}$ & 
 $\text{rich}_\text{AMICO}$ & $\text{SNR}_\text{WaZP}$ & 
 $\text{SNR}_\text{AMICO}$ & $z_\text{spec, mem}$ &  
 $N_\text{zspec}$ & $z_\text{spec, BCG}$ & \makecell{Coverage \\ fraction ($\%$)}  \\
\hline \\
$0$&$33.8711$&$-4.6807$&$0.37$&$78.95$&$48.22$&$21.64$&$35.38$&$0.35$&$18$&$0.348$&$79.88$\\
$1$&$35.7645$&$-4.604$&$1.2$&$39.6$&$46.29$&$20.51$&$41.03$&Nan&Nan&Nan&$90.91$\\
$2$&$37.1303$&$-4.7336$&$0.61$&$68.83$&$40.18$&$17.37$&$28.63$&$0.611$&$19$&$0.612$&$100.0$\\
$3$&$34.9847$&$-5.4697$&$0.33$&$23.76$&$19.81$&$17.09$&$21.25$&$0.28$&$15$&$0.278$&$100.0$\\
$4$&$37.1202$&$-4.9964$&$0.39$&$27.47$&$25.93$&$15.68$&$24.08$&$0.333$&$6$&$0.334$&$100.0$\\
$5$&$36.0156$&$-4.2221$&$1.09$&$67.62$&$43.03$&$15.56$&$25.14$&Nan&Nan&Nan&$92.65$\\
$6$&$35.7909$&$-4.2216$&$0.61$&$38.13$&$25.82$&$15.47$&$18.42$&$0.631$&$17$&$0.501$&$100.0$\\
$7$&$34.634$&$-5.0175$&$0.88$&$42.58$&$30.8$&$15.11$&$28.21$&$0.872$&$5$&Nan&$100.0$\\
$8$&$35.9315$&$-5.0257$&$0.85$&$50.11$&$34.13$&$14.63$&$24.13$&$0.858$&$16$&$0.858$&$100.0$\\
$9$&$36.4189$&$-4.3822$&$0.92$&$49.78$&$38.72$&$14.25$&$25.05$&Nan&Nan&$0.916$&$98.59$\\
$10$&$35.1062$&$-4.8039$&$1.05$&$33.4$&$31.05$&$13.79$&$20.14$&Nan&Nan&Nan&$100.0$\\
$11$&$35.3723$&$-4.0956$&$1.02$&$36.97$&$37.44$&$13.47$&$19.21$&Nan&Nan&$1.033$&$77.14$\\
$12$&$34.1788$&$-4.1819$&$1.26$&$18.65$&$22.23$&$13.45$&$16.41$&Nan&Nan&Nan&$83.33$\\
$13$&$36.2759$&$-4.8045$&$0.83$&$32.66$&$31.77$&$13.38$&$20.24$&Nan&Nan&Nan&$100.0$\\
$14$&$36.119$&$-4.8259$&$0.51$&$59.06$&$40.25$&$13.37$&$31.7$&$0.495$&$16$&$0.495$&$99.15$\\
$15$&$34.0497$&$-4.2399$&$0.15$&$69.26$&$29.85$&$13.33$&$33.64$&$0.153$&$48$&$0.154$&$89.24$\\
$16$&$35.7438$&$-4.1153$&$0.32$&$30.25$&$34.62$&$13.29$&$25.12$&$0.29$&$8$&$0.292$&$83.84$\\
$17$&$34.6761$&$-5.5494$&$0.41$&$38.99$&$27.92$&$13.05$&$25.68$&$0.382$&$6$&$0.383$&$98.64$\\
$18$&$36.6165$&$-4.9974$&$0.41$&$38.41$&$33.86$&$12.95$&$29.98$&Nan&Nan&$0.383$&$100.0$\\
$19$&$35.599$&$-4.6736$&$1.18$&$25.14$&$22.47$&$12.71$&$20.71$&Nan&Nan&$0.754$&$100.0$\\
 \hline

\end{tabular}$a$

\raisebox{0pt}[0pt][0pt]{\makebox[0pt][l]{\scriptsize $a$}}
\bigskip \bigskip 
\centering

\captionof{table}{\label{table:catalogue_cdfs} First $20$ AMICO-WaZP joint cluster candidates within the CDFS.}    

\begin{tabular}{cccccccccccc} 
 \hline\hline
 ID & RA & Dec  & 
 $z_\text{phot}$ & 
 $\text{rich}_\text{WaZP}$ & 
 $\text{rich}_\text{AMICO}$ & $\text{SNR}_\text{WaZP}$ & 
 $\text{SNR}_\text{AMICO}$ & $z_\text{spec, mem}$ &  
 $N_\text{zspec}$ & $z_\text{spec, BCG}$ &  \makecell{Coverage \\ fraction ($\%$)}  \\
\hline \\
$0$&$53.5646$&$-28.4127$&$0.66$&$136.45$&$60.64$&$19.26$&$37.81$&$0.664$&$9$&Nan&$98.88$\\
$1$&$53.4671$&$-27.3556$&$0.59$&$62.51$&$51.53$&$18.7$&$41.38$&$0.606$&$6$&$0.604$&$98.02$\\
$2$&$52.0146$&$-27.4462$&$1.2$&$32.04$&$28.52$&$15.84$&$28.18$&Nan&Nan&Nan&$87.88$\\
$3$&$52.374$&$-28.3276$&$0.67$&$77.86$&$51.53$&$15.62$&$38.51$&$0.682$&$7$&Nan&$85.71$\\
$4$&$52.8823$&$-28.7031$&$0.9$&$51.63$&$30.1$&$15.58$&$18.26$&Nan&Nan&$0.849$&$98.63$\\
$5$&$52.2892$&$-28.8103$&$0.58$&$37.03$&$24.24$&$14.9$&$21.99$&Nan&Nan&$0.548$&$85.15$\\
$6$&$54.043$&$-27.5429$&$0.97$&$30.19$&$32.6$&$14.33$&$21.89$&Nan&Nan&$0.942$&$82.19$\\
$7$&$51.7824$&$-28.7792$&$0.36$&$38.12$&$20.27$&$14.04$&$16.06$&Nan&Nan&$0.33$&$94.29$\\
$8$&$53.9934$&$-27.7097$&$0.94$&$30.13$&$27.31$&$13.52$&$25.18$&Nan&Nan&$0.937$&$88.57$\\
$9$&$53.4959$&$-28.6365$&$0.65$&$85.57$&$42.66$&$13.04$&$25.88$&$0.667$&$6$&$0.664$&$90.32$\\
$10$&$53.5602$&$-28.4522$&$0.66$&$103.84$&$18.08$&$12.91$&$14.92$&$0.663$&$6$&Nan&$92.22$\\
$11$&$52.7366$&$-28.7109$&$1.58$&$65.01$&$42.99$&$12.89$&$24.71$&Nan&Nan&Nan&$93.55$\\
$12$&$54.376$&$-28.4952$&$0.4$&$18.21$&$17.3$&$12.75$&$14.34$&Nan&Nan&Nan&$94.08$\\
$13$&$54.2049$&$-28.3408$&$0.39$&$21.71$&$21.28$&$12.64$&$15.69$&Nan&Nan&Nan&$85.71$\\
$14$&$52.034$&$-28.5896$&$1.46$&$50.48$&$29.53$&$12.45$&$22.03$&Nan&Nan&Nan&$82.54$\\
$15$&$53.9188$&$-27.3443$&$0.54$&$19.15$&$18.89$&$12.16$&$21.14$&Nan&Nan&$0.536$&$97.32$\\
$16$&$52.62$&$-27.934$&$1.11$&$28.14$&$27.13$&$11.61$&$19.78$&Nan&Nan&Nan&$100.0$\\
$17$&$52.9558$&$-27.6622$&$1.03$&$21.89$&$23.86$&$11.5$&$18.75$&$1.035$&$5$&$0.547$&$91.3$\\
$18$&$53.3383$&$-27.2057$&$0.87$&$49.12$&$41.86$&$11.16$&$24.5$&Nan&Nan&$0.862$&$86.84$\\
$19$&$51.7382$&$-28.8182$&$1.86$&$32.41$&$22.18$&$11.0$&$11.53$&Nan&Nan&Nan&$95.24$\\
 \hline

\end{tabular}$a$

\raisebox{0pt}[0pt][0pt]{\makebox[0pt][l]{\scriptsize $a$}}

\footnotetext{$^a$ Tables ordered by decreasing order of WaZP SNR. See Sect.\ref{section:GS_construction} for columns description.}
\end{minipage}

\twocolumn

\section{Spurious associations with clusters from the literature}
\label{appendix:lit_bad_match}

In this Appendix, we describe the $4$ cases of cluster candidates with a redshift offset larger than $0.06$ relative to the matched cluster catalogue taken from the literature, as discussed in Sect.\ref{subsection:ancillary}.

 \begin{itemize}
    \setlength\itemsep{1em}
     \item[] SXDF49XGG, from \citealt{Balogh_2020}: 
        \vspace{0.2cm}
        \begin{itemize}
          \setlength\itemsep{0.5em}
            \item[-] $z_\text{lit} = 1.091$
            \item[-] Matched with Cl104, with a measured $z_\text{phot, joint} \sim 1.39$
            \item[-] Cl104 is also matched with SXDF87XXGG with $z_\text{lit} = 1.406$.
            \item[-] Cl104 is closer to SXDF87XXGG than SXDF49XGG
            \item[-]  SXDF49XGG has a velocity dispersion of $255 \pm 50 \text{ km}\cdot\text{s}^{-1}$, while SXDF87XXGG has an estimated velocity dispersion of $700 \pm 110 \text{ km}\cdot\text{s}^{-1}$ 
        \end{itemize}
    \vspace{0.5cm}
    Conclusion: the match between Cl104 and SXDF49XGG could be a spurious association along the line of sight with a low mass, foreground group of galaxies, while the most likely counterpart to Cl104 possibly is SXDF87XXGG.

    \item[] XLSSC-191, from \citealt{Adami_2018}:
        \vspace{0.2cm}
        \begin{itemize}
            \setlength\itemsep{0.5em}
            \item[-] $z_\text{lit} = 0.05$
            \item[-] Matched with Cl136, with a measured $z_\text{phot, joint} \sim 0.15$.
            \item[-] We also have $z_\text{spec, gaussian} = 0.147$, using $39$ members.
            \item[-] XLSSC-191 has a low mass ($R_\text{500, X} \sim 491 \text{ kpc}$, $M_{500, X} \sim 0.3\times 10^{14}\text{ M}_\odot$)
        \end{itemize}
    \vspace{0.5cm}
    Conclusion: the match between Cl136 and XLSSC-191 could be a spurious association along the line of sight with a low mass, foreground group of galaxies. 

    \item[] J033223.2-274943, from \citealt{Finoguenov_2015}: 
        \vspace{0.2cm}
        \begin{itemize}
            \setlength\itemsep{0.5em}
            \item[-] $z_\text{lit} = 0.578$, computed with $5$ members within $R_{200}$
            \item[-] Matched with Cl169, with a measured $z_\text{phot, joint} \sim 0.71$
            \item[-] We also have $z_\text{spec, gaussian} = 0.734$, using $30$ members. 
            \item[-] Significant angular separation between them ($\sim 2$\,arcmin, i.e, a projected physical separation of $\sim 800$ kpc at the median redshift)
        \end{itemize}
    \vspace{0.5cm}
    Conclusion: Possibly our estimation of the spectroscopic redshift of this structure is more robust, but we cannot exclude the possibility of a spurious association with a different structure along the line of sight, as their angular separation is large. 

    \item[] J033136.7-275233, from \citealt{Finoguenov_2015}:
        \vspace{0.2cm}
        \begin{itemize}
            \setlength\itemsep{0.5em}
            \item[-] $z_\text{lit} = 1.05$
            \item[-] Matched with Cl45, with a measured $z_\text{phot, joint} \sim 1.21$
            \item[-] Small angular separation between them ($\sim 0.3$ arcmin, corresponding to a projected physical separation of $\sim 160$ kpc at the median redshift) 
            \item[-] No significant concentration of galaxies in the photometric redshift distribution at $z = 1.05$ within a window of $1$\,arcmin centred at the Cl45 location. 
            \item[-] Low mass system ($\sigma_v \sim 280 \text{ km}\cdot\text{s}^{-1}$)
        \end{itemize}
    \vspace{0.5cm}
    Conclusion: it could possibly be a random association with a poor foreground group of galaxies not detected by our algorithms.         
 \end{itemize}

\end{appendices}   

\end{document}